\documentclass[natbib209]{emulateapj}
\usepackage{graphicx}
\usepackage{rotating}
\usepackage[usenames]{color}
\usepackage[colorlinks=true, pdfstartview=FitV, linkcolor=OrangeRed, citecolor=blue, urlcolor= red]{hyperref}

\shortauthors{Oh et al.}
\shorttitle{Dark and luminous matter in THINGS dwarf galaxies}

\begin{document}
\bibliographystyle{astroads}

\renewcommand{\thefootnote}{\fnsymbol{footnote}}
\newcommand{\C}{\ensuremath{\mathfrak{C}}}
\newcommand{\VNFW}{\ensuremath{V_{200}}}
\newcommand{\Vh}{\ensuremath{V_h}}
\newcommand{\kms}{\ensuremath{\mathrm{km}\,\mathrm{s}^{-1}}}
\newcommand{\kmsnospace}{\ensuremath{\mathrm{km}\,\mathrm{s}^{-1}}}
\newcommand{\barkms}{\ensuremath{\mathrm{km}\,\mathrm{s}^{-1}\,\mathrm{kpc}^{-1}}}
\newcommand{\kmsMpc}{\ensuremath{\mathrm{km}\,\mathrm{s}^{-1}\,\mathrm{Mpc}^{-1}}}
\newcommand{\etal}{et al.}
\newcommand{\LCDM}{$\Lambda$CDM}
\newcommand{\ML}{\ensuremath{\Upsilon_{\star}}}
\newcommand{\MLfix}{\ensuremath{\Upsilon_{\star}^{fix}}}
\newcommand{\MLfree}{\ensuremath{\Upsilon_{\star}^{\rm free}}}
\newcommand{\MLsps}{\ensuremath{\Upsilon_{\star}^{3.6}}}
\newcommand{\MLspl}{\ensuremath{\Upsilon_{\star}^{4.5}}}
\newcommand{\MLk}{\ensuremath{\Upsilon_{\star}^{K}}}
\newcommand{\MLb}{\ensuremath{\Upsilon_{\star}^{B}}}
\newcommand{\MLmax}{\ensuremath{\Upsilon_{max}}}
\newcommand{\Lsun}{\ensuremath{\rm{{L}_{\odot}}}}
\newcommand{\Msun}{\ensuremath{\rm{{M}_{\odot}}}}
\newcommand{\mass}{\ensuremath{\rm{{\cal M}}}}
\newcommand{\magsq}{\ensuremath{\mathrm{mag}\,\mathrm{arcsec}^{-2}}}
\newcommand{\Lsundens}{\ensuremath{\rm L_{\odot}\,\mathrm{pc}^{-2}}}
\newcommand{\surfdens}{\ensuremath{M_{\odot}\,\mathrm{pc}^{-2}}}
\newcommand{\cubedens}{\ensuremath{M_{\odot}\,\mathrm{pc}^{-3}}}

\long\def\Ignore#1{\relax}
\long\def\Comment#1{{\footnotesize #1}}

\title{Dark and luminous matter in THINGS dwarf galaxies}

\author{Se-Heon Oh\altaffilmark{1},\altaffilmark{5},
W.~J.~G.\ de Blok\altaffilmark{1},
Elias Brinks\altaffilmark{2},
Fabian Walter\altaffilmark{3} and
Robert C.\ Kennicutt, Jr.\altaffilmark{4}}

\email{seheon\_oh@ast.uct.ac.za}
\email{edeblok@ast.uct.ac.za}
\email{E.Brinks@herts.ac.uk}
\email{walter@mpia.de}
\email{robk@ast.cam.ac.uk}

\altaffiltext{1}{Department of Astronomy, University of Cape Town, Private Bag X3, Rondebosch 7701, South Africa}
\altaffiltext{2}{Centre for Astrophysics Research, University of Hertfordshire, College Lane, Hatfield, AL10 9AB, United Kingdom}
\altaffiltext{3}{Max-Planck-Institut f\"ur Astronomie, K\"onigstuhl 17, 69117 Heidelberg, Germany}
\altaffiltext{4}{Institute of Astronomy, University of Cambridge, Madingley Road, Cambridge CB3 0HA, United Kingdom}
\altaffiltext{5}{Square Kilometre Array South African Fellow}


\begin{abstract}
We present mass models for the dark matter component of 7 dwarf galaxies taken 
from ``The H{\sc i} Nearby Galaxy Survey'' (THINGS) and compare these with those from 
numerical $\Lambda$ Cold Dark Matter (\LCDM) simulations. The THINGS high-resolution data
significantly reduce observational uncertainties and thus allow us to derive accurate dark matter distributions
in these systems. We here use the bulk velocity fields when deriving 
the rotation curves of the galaxies. Compared to other types of
velocity fields, the bulk velocity field minimizes the effect of small-scale
random motions more effectively and traces the underlying kinematics of a galaxy more properly.
The ``{\it Spitzer} Infrared Nearby Galaxies Survey'' (SINGS) 3.6$\mu$m and ancillary 
optical data are used for separating the baryons from their total matter content in the galaxies.
The sample dwarf galaxies are found to be dark matter dominated over most radii. The relation
between total baryonic (stars $+$ gas) mass and maximum rotation velocity of the galaxies
is roughly consistent with the Baryonic Tully$\--$Fisher relation calibrated from
a larger sample of gas dominated low mass galaxies.
We find discrepancies between the derived dark matter distributions
of the galaxies and those of \LCDM\ simulations, even after corrections
for non-circular motions have been applied. The observed solid body-like rotation curves of the galaxies
rise too slowly to reflect the cusp-like dark matter distribution in CDM halos.
Instead, they are better described by core-like models such as pseudo-isothermal
halo models dominated by a central constant-density core.
The mean value of the logarithmic inner slopes of the mass density
profiles is $\alpha$$=$$-0.29\pm 0.07$. They are significantly different from 
the steep slope of $\sim$$-1.0$ inferred from previous dark-matter-only simulations,
and are more consistent with shallower slopes found in recent \LCDM\ simulations
of dwarf galaxies in which the effects of baryonic feedback processes are included.
\end{abstract}


\section{Introduction}
\label{Intro}

The dark matter distribution at the centers of galaxies has been
intensively debated ever since the advent of high-resolution
\LCDM\ simulations. The existence of central cusps in dark matter halos
was found in numerical simulations
(\citeauthor{Dubinski_1991} \citeyear{Dubinski_1991};
Navarro, Frenk \& White \citeyear{NFW_1996}, \citeyear{NFW_1997};
\citeauthor{Moore_1999a} \citeyear{Moore_1999a};
\citeauthor{Ghigna_2000} \citeyear{Ghigna_2000};
\citeauthor{Klypin_2001} \citeyear{Klypin_2001};
\citeauthor{Power_2002} \citeyear{Power_2002};
\citeauthor{Stoehr_2003} \citeyear{Stoehr_2003};
\citeauthor{Navarro_2004} \citeyear{Navarro_2004};
\citeauthor{Reed_2005} \citeyear{Reed_2005};
\citeauthor{Diemand_2008} \citeyear{Diemand_2008}) but challenged by the observations.
The latter support a core-like density distribution at the centers of galaxies
(\citeauthor{Flores_1994} \citeyear{Flores_1994};
\citeauthor{Moore_1994} \citeyear{Moore_1994};
\citeauthor{deBlok_2001} \citeyear{deBlok_2001};
\citeauthor{deBlok_2002} \citeyear{deBlok_2002};
\citeauthor{Bolatto_2002} \citeyear{Bolatto_2002};
\citeauthor{Weldrake_2003} \citeyear{Weldrake_2003};
\citeauthor{Simon_2003} \citeyear{Simon_2003};
\citeauthor{Swaters_2003} \citeyear{Swaters_2003};
\citeauthor{Gentile_2004} \citeyear{Gentile_2004};
\citeauthor{Oh_2008} \citeyear{Oh_2008};
\citeauthor{Trachternach_2008} \citeyear{Trachternach_2008};
\citeauthor{deBlok_2008} \citeyear{deBlok_2008}, and references therein).
A detailed observational review on where the ``cusp/core'' problem stands
is given by \cite{deBlok_2010}.

Of particular interest has been the assumption that the observations suffer from
various systematic uncertainties and that the central cusps can be ``hidden'' this way
(\citeauthor{Swaters_1999} \citeyear{Swaters_1999};
\citeauthor{van_den_Bosch_2000} \citeyear{van_den_Bosch_2000};
\citeauthor{van_den_Bosch_2001} \citeyear{van_den_Bosch_2001};
\citeauthor{Swaters_2003} \citeyear{Swaters_2003};
\citeauthor{Simon_2003} \citeyear{Simon_2003};
\citeauthor{Rhee_2004} \citeyear{Rhee_2004}).
These uncertainties consist of certain observational systematic effects
as well as the uncertainty in the stellar mass-to-light ratios (\ML) of the stellar component.
The observational systematic effects, such as beam smearing (for low-resolution radio observations),
dynamical center offsets (for slit observations) and non-circular motions affect the
derived dark matter distribution in galaxies in such a way that the apparent inner density
slopes of dark matter halos are flattened.
In addition, the fairly unconstrained \ML\ also affects the derived distribution of dark matter in galaxies
(e.g., \citeauthor{van_Albada_1986} \citeyear{van_Albada_1986}).

\begin{deluxetable*}{lccccccccccc}
\tablecaption{Properties of the THINGS dwarf galaxies}
\tablewidth{0pt}
\tablehead{
\colhead{Name} & \colhead{R.A.} & \colhead{Dec.}   & \colhead{$D$}    & \colhead{$V_{\rm sys}$}     & \colhead{$\langle \rm P.A. \rangle$}  & \colhead{$\langle \rm i^{\rm TR} \rangle$} & \colhead{$\langle \rm i^{\rm BTF} \rangle$} & \colhead{$z_{0}$} & \colhead{Metal.} & \colhead{$M_{\rm B}$} & \colhead{$\rm M_{\rm dyn}$} \\
\colhead{}   & \colhead{(h m s)}  & \colhead{($^{\circ}$ $'$ \arcsec)}  & \colhead{(Mpc)}  & \colhead{(\kms)}  & \colhead{($^{\circ}$)} & \colhead{($^{\circ}$)} & \colhead{($^{\circ}$)} & \colhead{(kpc)} & \colhead{($\rm Z_{\odot}$)} & \colhead{(mag)} & \colhead{($10^{9}\,M_{\odot}$)} \\
\colhead{} & \colhead{(1)}  & \colhead{(2)}  & \colhead{(3)}  & \colhead{(4)} & \colhead{(5)} & \colhead{(6)} & \colhead{(7)} & \colhead{(8)} & \colhead{(9)} & \colhead{(10)} & \colhead{(11)}}
\startdata
IC 2574     & 10 28 27.7 & +68 24 59 & 4.0 &53  &  53 &  55  &  46 & 0.57 & 0.20	& -18.11 & 14.62   \\
NGC 2366    & 07 28 53.4 & +69 12 51 & 3.4 &104 &  39 &  63  &  50 & 0.34 & 0.10	& -17.17 & 4.29	   \\
Holmberg I  & 09 40 32.3 & +71 11 08 & 3.8 &140 &  45 &  13  &  10 & 0.55 & 0.12	& -14.80 & 0.46	   \\
Holmberg II & 08 19 03.7 & +70 43 24 & 3.4 &156 &  175&  49  &  25 & 0.28 & 0.17	& -16.87 & 2.07    \\
M81 dwB     & 10 05 30.9 & +70 21 51 & 5.3 &346 &  311&  44  &  59 & 0.09 & 0.21	& -14.23 & 0.30	   \\
DDO 53      & 08 34 06.5 & +66 10 48 & 3.6 &18  &  131&  27  &  23 & 0.14 & 0.11	& -13.45 & 0.45	   \\
DDO 154     & 12 54 05.7 & +27 09 10 & 4.3 &375 &  229&  66  &  55 & 0.20 & 0.05 	& -14.23 & 5.40
\enddata
\\
\tablecomments{
{\bf (1)(2):} Center positions derived from a tilted-ring analysis in Section~\ref{Deriving_Rotcurs}.
The center position of DDO 154 is from \cite{Trachternach_2008};
{\bf (3):} Distance as given in \cite{Walter_2008};
{\bf (4):} Systemic velocity derived from a tilted-ring analysis in Section~\ref{Deriving_Rotcurs};
{\bf (5):} Average value of the position angle from a tilted-ring analysis in Section~\ref{Deriving_Rotcurs};
{\bf (6):} Average value of the inclination from a tilted-ring analysis in Section~\ref{Deriving_Rotcurs};
{\bf (7):} The inclination value derived from the Baryonic Tully$\--$Fisher relation (see Section~\ref{BTF_relation_THINGS});
{\bf (8):} The vertical scale height of disk derived in this paper (see Section~\ref{Stars});
{\bf (9):} Metallicities; 
{\bf (10):} Absolute B magnitude as given in \cite{Walter_2008}; 
{\bf (11):} Dynamical mass within the last measured point of the bulk rotation curve derived in this paper.}
\end{deluxetable*}

The best way to minimize these uncertainties is to use high-quality data of dark
matter-dominated objects. High-quality data ($\sim$7$\arcsec$ angular; $\leq$ 5.2 \kms\ velocity resolution)
of dwarf galaxies taken from ``The H{\sc i} Nearby Galaxy Survey''
(THINGS; \citeauthor{Walter_2008} \citeyear{Walter_2008}) significantly reduce
the systematic effects inherent in lower-quality data and thus provide
a good opportunity for addressing the dark matter distribution near the centers of galaxies.
Dwarf galaxies which are dark matter-dominated, like low surface brightness (LSB) galaxies
(\citeauthor{deBlok_1997} \citeyear{deBlok_1997}), are ideal objects for the study of
dark matter (e.g., \citeauthor{Prada_2002} \citeyear{Prada_2002}) because of the small
contribution of baryons to the total matter content. In particular the high linear
resolution of $\sim$0.2 kpc (assuming a median distance of 4 Mpc) achieved by THINGS is necessary to resolve the inner slope of the density
profile and distinguish between cusp- and core-like density profiles near the centers of
galaxies. Moreover, ``{\it Spitzer} Infrared Nearby Galaxies Survey''
(SINGS; \citeauthor{Kennicutt_2003} \citeyear{Kennicutt_2003}) data are available for our sample galaxies.
The SINGS near-IR images provide virtually dust-free pictures of the old stellar populations in galaxies.
This allows us to make reliable mass models for the stellar component of a galaxy.

We select 7 dwarf galaxies from THINGS that show a clear rotation pattern in their
velocity fields to derive their rotation curves. Although some of them have been
analyzed before, a more careful kinematic analysis is useful to derive a more accurate dark
matter distribution in these slowly rotating galaxies. In the previous analysis
(e.g., \citeauthor{Martimbeau_1994} \citeyear{Martimbeau_1994}; \citeauthor{Hunter_2001} \citeyear{Hunter_2001};
\citeauthor{Bureau_2002} \citeyear{Bureau_2002} etc.),
the intensity-weighted mean velocity field which is most likely affected by non-circular
motions in galaxies was used and the asymmetric drift correction was usually not addressed.
Both non-circular motions and pressure support tend to induce a lower observed
rotation velocity than the true one. 

In general four different types of non-circular motions in galaxies can be distinguished
on the basis of velocity fields (\citeauthor{Bosma_1978} \citeyear{Bosma_1978}):
(1) Motions associated with spiral arms. The streaming motions caused by the arms
distort the velocity field in a regular fashion (e.g., M81);
(2) Large-scale symmetric deviations. The radial change of the kinematical major axis'
position angle distorts the velocity field, while still having a central symmetry.
These velocity distortions are known as ``oval'' distortions when encountered in the inner
region, and as a ``warp'' when they occur in the outer region, respectively;
(3) Large-scale asymmetries. The tidal interaction with a neighbouring galaxy causes asymmetries
mainly in the outer regions of galaxies (e.g., M81; \citeauthor{Yun_1994} \citeyear{Yun_1994});
(4) Small-scale asymmetries. Various sources, such as supernova (SN) explosions and
stellar winds from young stars (e.g., OB associations), locally stir up the bulk motion of
gas and give rise to random motions. These are usually visible as ``kinks'' in iso-velocity
contours of velocity fields.

Of these, small-scale random motions can be classified as additional components of the velocity profiles
in the H{\sc i} data cube and result in asymmetric profiles. Therefore, a single Gaussian function cannot
properly model these (non-Gaussian) velocity profiles. To minimize the effect of these random non-circular motions
in our sample galaxies, we use the ``bulk'' velocity fields described in \cite{Oh_2008}.
We compare the bulk rotation curves with those derived from other types of velocity fields, such as the
intensity-weighted mean, peak, single Gaussian fit and hermite $h_{3}$.
In addition, we correct for the asymmetric drift
for the galaxies where the pressure support is significant with respect to the circular rotation.
We then obtain dark matter mass models of the galaxies using
\ML\ as derived in \cite{Oh_2008}. From this, we address the ``cusp/core'' problem by
comparing the derived dark matter distribution of our galaxies with that of \LCDM\ simulations.

This paper is set out as follows. In Section~\ref{Data}, we give a general description 
of the data used.
In Section~\ref{Rotation_curves}, we present the rotation curves of the THINGS dwarf galaxies.
The mass models for the baryons are presented in Section~\ref{Baryons}.
The measured dark matter fractions of the galaxies are given in Section~\ref{DM_fraction_THINGS}, and
their relation to the galaxy properties is discussed.
In Section~\ref{DM_distribution}, the derived dark matter distribution of the galaxies
is discussed with respect to the fit quality of the halo models used, the rotation curve shape and the inner
density slope. Lastly, the main results of this paper are summarized in Section~\ref{Summary}.
Data and kinematic analysis of individual galaxies are presented in the Appendix.

\section{Data}
\label{Data}

We use high-resolution H{\sc i} data of 7 nearby ($\sim$4 Mpc) dwarf galaxies from THINGS
undertaken with the NRAO\footnote[1]{The National Radio Astronomy Observatory is a facility
of the National Science Foundation operated under cooperative agreement by Associated Universities, Inc.}
Very Large Array (VLA) to derive the dark matter distribution in these systems.
Basic properties of our sample galaxies are listed in Table 1. 
See \cite{Walter_2008} for a detailed description of observations and the data reduction.
SINGS IRAC 3.6$\mu$m data\footnote[2]{NGC 2366 is not targeted in SINGS observations but retrieved from the {\it Spitzer} archive}
(with a resolution of $\sim4\arcsec$) are used to separate the contribution of
stars from the total kinematics.
In addition, ancillary optical broadband ($B$, $V$ and $R$) images of the sample galaxies
taken with the 2.1m telescope at Kitt Peak National Observatory (KPNO) as part of the SINGS survey
are used.
The data used in this paper are presented in the Appendix. IC 2574 and NGC 2366 have already been
published in \cite{Oh_2008}. However, we here make further use of the plots by extending the analysis.
For a consistency with other galaxies presented in this paper, we show the old plots again together with
some new results. Some of the galaxies (e.g., Ho I and DDO 53) have low inclinations ($<$ $30\,^{\circ}$).
The effect of inclination on the rotation curves will be discussed in Section~\ref{Deriving_Rotcurs}.

\section{Rotation curves}
\label{Rotation_curves}

\subsection{Velocity field types}
\label{VF_types}

As a first step towards deriving the rotation curve of a galaxy,
we need to extract the velocity field from the data cube.
The velocity field contains the entire 2D distribution of velocities and is therefore
less prone to systematic uncertainties in deriving rotation curves, e.g., due to
pointing offset and non-circular motions, than one-dimensional
long-slit spectra (\citeauthor{Zackrisson_2006} \citeyear{Zackrisson_2006};
\citeauthor{deBlok_2008} \citeyear{deBlok_2008}).

A velocity field can be derived in many different ways.
The most popular ones are the intensity-weighted mean (IWM),
peak, single Gaussian fit and hermite velocity fields (see \citeauthor{deBlok_2008} \citeyear{deBlok_2008}).
For a highly resolved galaxy that is not affected by non-circular motions,
these velocity fields are nearly identical to each other and
the rotation curves derived are also similar.
However, for a galaxy with dynamics severely affected by non-circular motions,
the resulting rotation curves from the different types of velocity fields
show significant differences. Therefore, we have to examine
the various types of velocity fields for a galaxy and determine
which is the least affected by non-circular motions and the most
appropriate for deriving an accurate rotation curve.
In the following sections, we briefly introduce the velocity fields mentioned above,
as well as the bulk velocity field first proposed by \cite{Oh_2008}.

\begin{figure}
\includegraphics[angle=0,width=0.49\textwidth,bb=40 164 578 715,clip=]{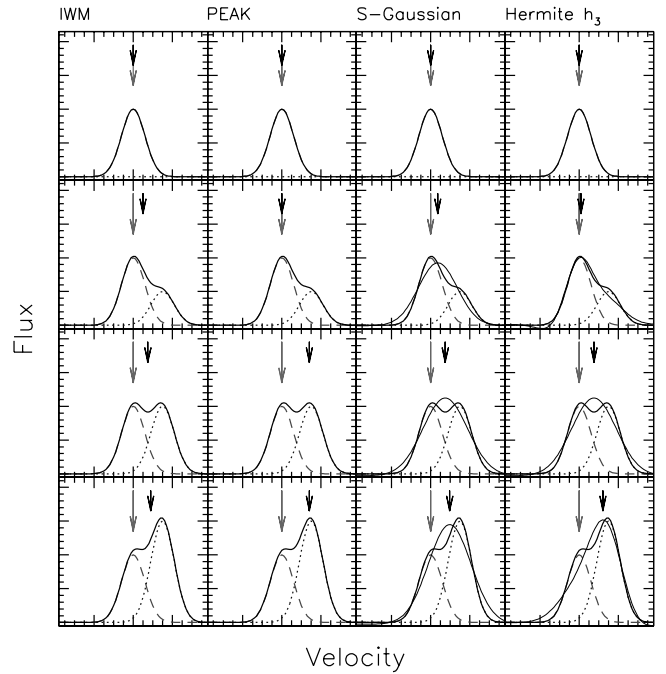}
\caption{Schematic H{\sc i} profiles with different asymmetries. The gray dashed lines represent the
bulk motion and the light-gray dotted lines indicate an additional non-circular component. The black
solid lines are the resulting profiles combining both the bulk and non-circular motions. As the non-circular
component increases from the top to bottom panels the asymmetry of the resulting profile also increases.
The long light-gray arrows in all panels indicate the central velocity of the bulk motion profile. The short
black arrows in all panels indicate the derived velocity from the IWM, peak,
single Gaussian fit, and hermite $h_{3}$ polynomial fit. The larger the asymmetry of a profile,
the larger the velocity deviation from the bulk motion. In case the non-circular motion does not
dominate the bulk motion (upper two rows), the derived velocity is close to that
of the bulk motion and peak and hermite $h_{3}$ velocities give an equally good result. However, if
the non-circular motion dominates the bulk motion (lower two rows), the derived velocity deviates significantly
from the bulk velocity even if we use the hermite $h_{3}$ polynomial fit.
\label{HI_Profiles}}
\end{figure}

\subsubsection{Intensity-weighted mean velocity field (1$^{st}$ moment map)}
\label{IWM_VF}
The IWM velocity field has been the most widely used velocity field tracing intensity-weighted
velocities along the line-of-sight through a galaxy (\citeauthor{Warner_1973} \citeyear{Warner_1973}).
The intensity-weighted mean velocity of a profile in a data cube at a given
line-of-sight for a galaxy is given as,
\begin{equation}
\label{eq:3_19}
V_{\rm{IWM}}(x, y) = \frac{\int_{-\infty}^{\infty}dv I(x, y, v) v}{\int_{-\infty}^{\infty}dv I(x, y, v)},
\end{equation}
where $I(x,y,v)$ is the flux of the profile in the data cube at a given sky position $(x, y)$ and is a function of velocity $v$.
Mapping the velocities weighted by $I(x,y,v)$ over the entire area of a galaxy gives the IWM velocity field.
As this method does not depend on profile fitting, it provides a robust estimate of velocity even for asymmetric profiles with a low S/N.
If a profile in the data cube is symmetric with respect to its central velocity,
then the IWM field properly traces the central velocity at which the peak flux is found.
However, it begins to deviate from the central velocity of a profile,
as the asymmetry of the profile increases. A schematic example of this is shown in Fig.~\ref{HI_Profiles}.

\subsubsection{Peak velocity fields}
\label{peak_VF}

Tracing the velocities at which the peak fluxes of
the profiles in a data cube are found can be an alternative way of determining
the line-of-sight velocities of a galaxy. This type of velocity field
is called a peak-intensity velocity field.
Since no fitting procedure is required, this method is simple and fast.
Unlike the IWM method, this method is able to trace the velocities at which
the highest fluxes are found, even for profiles showing significant asymmetries.
In this respect, the peak velocity is the preferred velocity compared to the ones
derived using other methods.
However, this method is sensitive to the noise in profiles with low S/N
in which case it fails to extract proper line-of-sight
velocities. See \cite{deBlok_2008} for more discussions.

\subsubsection{Single Gaussian velocity fields}
\label{SG_VF}

It is possible to fit a single Gaussian function to the velocity profiles.
A Gaussian function depends on three parameters and is given as follows:
\begin{equation}
\label{eq:3_20}
V_{\rm{Gauss}}(v) = A \exp\biggl(-\frac{(v-v_{0})^{2}}{2\sigma^{2}}\biggr ),
\end{equation}
where $v_{0}$ and $\sigma$ are the central velocity and velocity
dispersion of a profile. Due to the assumption on the shape of the profiles (i.e., Gaussian function),
this is less sensitive to the noise or (modest) asymmetries of profiles. In addition,
the least squares fit procedure provides robust estimates of velocities, even for
profiles with low S/N values. The single Gaussian velocity field
is best used in profiles where the FWHM is comparable to the
velocity resolution (\citeauthor{deBlok_2008} \citeyear{deBlok_2008}).
However, this method still suffers from significant profile asymmetries
induced by non-circular motions or projection effects of a galaxy.
As shown in Fig.~\ref{HI_Profiles}, the derived velocities from the single Gaussian fit
can deviate from the peak velocities of asymmetric profiles,
although this method provides better results than the IWM method.

\subsubsection{Hermite $h_{3}$ polynomials}
\label{Her3_VF}

It is also possible to use the Gauss-Hermite polynomial (\citeauthor{van_der_Marel1993} \citeyear{van_der_Marel1993})
to model the skewness of a non-Gaussian profile.
In addition, the Gauss-Hermite polynomial
also has a parameter called $h_{4}$, which measures the
kurtosis of a profile. However, to minimize the number of free parameters, this term
is not usually used when fitting the function.
As the skewness is built into the profile, it is efficiently applicable 
to profiles with significant asymmetries.
Compared to the peak velocity field, hermite $h_{3}$ polynomials give more
stable results, even for profiles with low S/N values.
Hermite $h_{3}$ polynomials have been used to extract
the velocity fields of the galaxies from THINGS (\citeauthor{deBlok_2008} \citeyear{deBlok_2008}).

\subsubsection{Bulk velocity fields}
\label{Bulk_VF}

A velocity profile in a data cube can consist of multiple components if there are
additional components moving at different velocities with respect to the underlying rotation of a galaxy.
Until now, we have assumed that the underlying rotation of a galaxy is the dominant motion
in a galaxy. This assumption leads us to choose the peak-intensity velocity
in a fitted or raw velocity profile as the most representative velocity.
This, however, only holds for a case where the majority of the gas moves at
this velocity. Any additional components present in a velocity profile
are then considered to be non-circular motions that deviate from the bulk rotation. But this 
is not true for a profile where non-circular motions dominate the kinematics
of a galaxy. In this case, even if a profile is decomposed successfully
with multiple components, no clues exist as to which component
is the bulk motion and which ones are the non-circular motions.
We therefore need additional constraints to distinguish the bulk motion and
non-circular motions among such decomposed components.

To this end, \cite{Oh_2008} proposed a new method to extract circularly rotating
velocity components from the H{\sc i} data cube and derive a so-called
bulk velocity field.
This type of velocity field efficiently separates small-scale random
motions from the underlying rotation of a galaxy and extracts the bulk velocity.
See \cite{deBlok_2008} for the comparison of the various types of velocity fields.
This method has been successfully used for two galaxies that are significantly affected
by non-circular motions: IC 2574 and NGC 2366 (\citeauthor{Oh_2008} \citeyear{Oh_2008}).

We extract the various types of velocity fields
(i.e., IWM, single Gaussian, hermite $h_{3}$ polynomial, and the peak and
bulk velocity fields) from the H{\sc i} data cubes
of our sample galaxies and use them to derive rotation curves, as described in
the following section. 
The natural-weighted cubes are used for this and no residual scaling,
primary beam correction or blanking\footnote[3]{Except in the determination of the bulk velocity field.
See \cite{Oh_2008} for a detailed description.}
is applied to preserve the noise characteristics. The extracted velocity fields of
the 7 THINGS dwarf galaxies are presented in the Appendix.

\subsection{Tilted-ring analysis}
\label{TRM}

Using {\sc rotcur} in GIPSY (\citeauthor{Begeman_1989} \citeyear{Begeman_1989}), 
we fit a tilted-ring model to the bulk velocity field of the galaxies to derive the ring parameters
that best describe the observed velocity fields.
We then apply these tilted-ring models obtained from the bulk velocity field to the other velocity fields
to examine the effect of the type of velocity field on the derived rotation curve.
We show the rotation curves derived from different types 
of velocity fields of the sample galaxies in the Appendix. We will compare and discuss
these rotation curves in Section~\ref{Deriving_Rotcurs}.

\subsection{Asymmetric drift correction}
\label{ADC}

Pressure support plays an important role in galaxies whose velocity
dispersions are large enough compared to their maximum rotation velocities (\citeauthor{Bureau_2002} \citeyear{Bureau_2002}).
This is the case for the galaxies in our sample whose typical maximum rotation velocities ($V_{\rm max}$) are less than $\sim$35 \kms,
except for IC 2574 ($\sim$80 \kms) and NGC 2366 ($\sim$60 \kms).
In order to obtain more reliable rotation velocities for these galaxies,
we need to correct for the asymmetric drift. 
Following the method described in \cite{Bureau_2002}, we correct for the asymmetric drift as follows,
\begin{eqnarray}
V^{2}_{\rm cor} = V^{2}_{\rm rot} + \sigma^{2}_{\rm D},
\label{V_cor}
\end{eqnarray}
where $V_{\rm rot}$ is the rotation velocity derived from the simple fit of a tilted-ring model
to the velocity field and $V_{\rm cor}$ is the asymmetric drift corrected velocity. 
The asymmetric drift correction $\sigma_{\rm D}$ is given as,
\begin{eqnarray}
\sigma^{2}_{\rm D} &=& -R\sigma^{2}\frac{\partial\rm ln(\rho \sigma^{2})}{\partial R} \nonumber \\
&=& -R\sigma^{2}\frac{\partial\rm ln(\Sigma \sigma^{2})}{\partial R},
\label{sigma_D}
\end{eqnarray}
\noindent where $\sigma$ and $\rho$ are the velocity dispersion and volume density of H{\sc i},
and $R$ is the radius of a galaxy.
In particular, $\rho$ can be converted to the H{\sc i} surface density $\Sigma$
by assuming an exponential distribution in the vertical direction and a constant
scale height (for a $1^{st}$ approximation). For the surface density $\Sigma$ and velocity dispersion $\sigma$,
we use the integrated H{\sc i} ($0^{\rm th}$ moment) and velocity dispersion ($2^{\rm nd}$ moment) maps, respectively.
Using the tilted-ring model derived
earlier from the bulk velocity field, we obtain the corrected radial profiles of $\Sigma$
and $\sigma$. To avoid large fluctuations in the derivative in Eq.~\ref{sigma_D},
we fit $\Sigma\sigma^{2}$ with an analytical function,
\begin{eqnarray}
\Sigma\sigma^{2}(R) = \frac{I_{0}(R_{0}+1)}{R_{0}+e^{\alpha R}},
\label{sigma2}
\end{eqnarray}
\noindent where $I_{0}$ and $R_{0}$ are the fitted values
in units of $\rm M_{\odot}\,{pc}^{-2}\,km^{2}\,s^{-2}$ and arcsec, respectively.
$\alpha$ is given in unit of $\rm arcsec^{-1}$.
The resulting profiles of $\sigma_{\rm D}$ and $\Sigma\sigma(R)$ for
the galaxies where the asymmetric drift corrections are needed are shown in the Appendix.

\subsection{The rotation curves of THINGS dwarf galaxies}
\label{Deriving_Rotcurs}
The resulting rotation curves of the individual galaxies derived using different
types of velocity fields and their comparison are given in the Appendix.
Below we discuss the rotation curves derived using the bulk velocity field and corrected for asymmetric drift,
where needed, in order to examine the effect of small-scale random non-circular motions. \\

\begin{figure*}
\epsscale{1.0}
\includegraphics[angle=0,width=1.0\textwidth,bb=20 402 585 685,clip=]{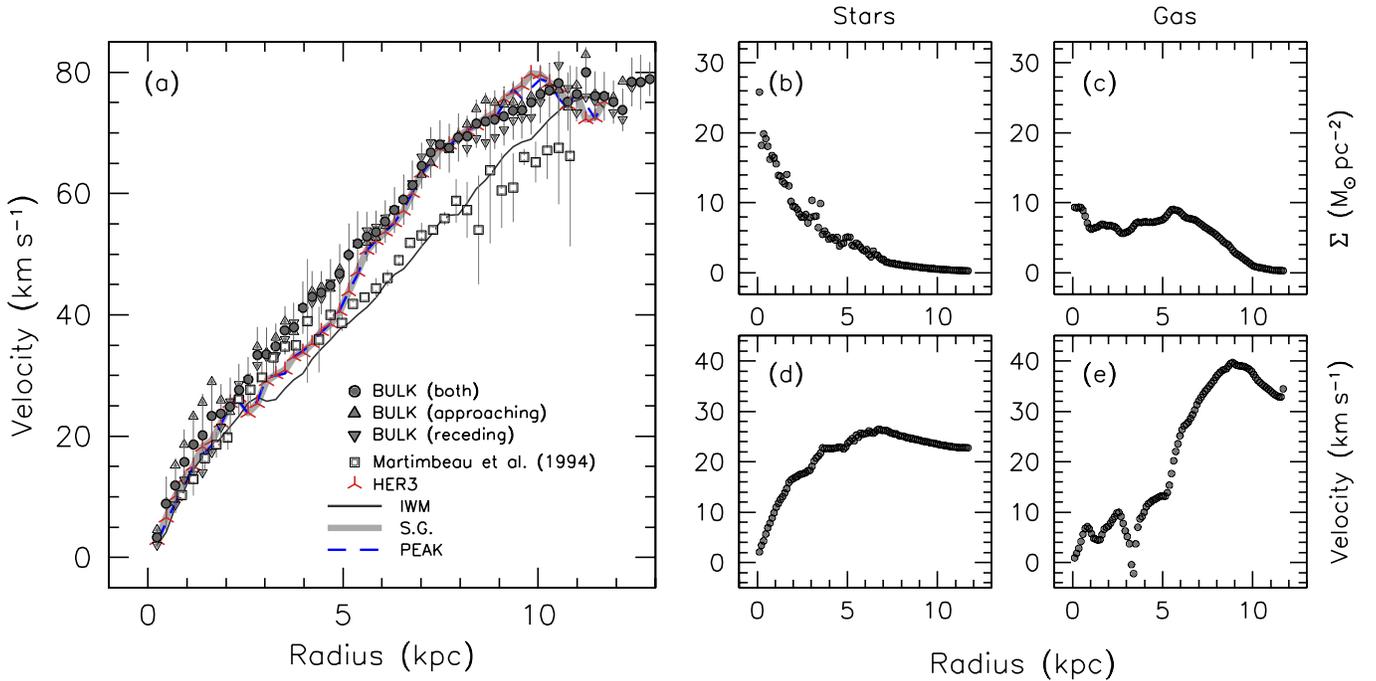}
\caption{{\bf (a):} Comparison of the bulk rotation curve of IC 2574 with rotation curves
derived using other types of velocity fields as denoted in the panel
(i.e., IWM, hermite $h_{3}$, single Gaussian and peak velocity fields).
The \cite{Martimbeau_1994} curve was derived using an IWM velocity field
with a lower resolution. They adopted a large value for the inclination ($\sim$75$^{\circ}$).
The IWM rotation curve which is most likely affected by random non-circular motions is the lowest among the others.
See Section~\ref{Deriving_Rotcurs} for more details.
{\bf (b)(c):} The radial mass surface density profiles of the stellar and gas components of IC 2574, respectively.
{\bf (d)(e):} The resulting rotation velocities of the stellar and gas components of IC 2574 derived
from the surface density profiles given in the panels (b) and (c), respectively. More details can be
found in Section~\ref{Baryons}. 
\label{ROTATION_CURVES_MASS_MODELS_IC2574}}
\end{figure*}

\noindent $\bullet$ \textbf{IC 2574:}
IC 2574 is affected by non-circular motions (\citeauthor{Walter_Brinks_1999} \citeyear{Walter_Brinks_1999})
and this is clearly seen as ``kinks'' in the iso-velocity contours of the velocity fields
as shown in Fig.~\ref{IC2574_MAPS}.
The spatial locations of these small-scale random motions are also found
in the non-circular motion velocity field (hereafter NONC velocity field) as shown in
panel (k) of Fig.~\ref{IC2574_MAPS}. As described in \cite{Oh_2008}, the NONC velocity
field only contains the velocities of the primary (i.e., strongest intensity)
components among the decomposed ones at the positions where these primary components were found
to track the non-circular motions.

For a quantitative analysis of non-circular motions, we expand the velocity fields into harmonic terms up to the 3$^{\rm{rd}}$ order,
$c_{n}$ and $s_{n}$ ($n=1, 2$ and $3$) (\citeauthor{Schoenmakers_1997} \citeyear{Schoenmakers_1997};
see also \citeauthor{Trachternach_2008} \citeyear{Trachternach_2008} for an extensive discussion of the method).
As shown in Fig.~\ref{IC2574_TR_HD}, the amplitudes of harmonic terms (e.g., $c_{2}$, $s_{1}$ and $s_{2}$; corrected for inclination)
decomposed using the hermite $h_{3}$ velocity field are $\sim$10 \kms\ in the inner regions.
However, the results from the bulk velocity fields are less than 5 \kms\ over all radii. 

In general, small-scale random motions tend to result in a lower rotation velocity than
the true one as they make the velocity gradients along the receding and approaching sides
of a galaxy less steep. This is particularly prominent for the rotation velocity derived using the IWM
velocity field which is most affected by non-circular motions. As shown in Fig.~\ref{ROTATION_CURVES_MASS_MODELS_IC2574},
the rotation curves derived from the other types of velocity field are largely consistent with each other.
At $\sim$4 kpc where non-circular motions caused by a super-giant shell are significant
(see \citeauthor{Walter_1998} \citeyear{Walter_1998}), the velocity differences between
the bulk and both the IWM and hermite $h_{3}$ curves are about $\sim$11 \kms\ and $\sim$7 \kms, respectively.
For the kinematic analysis of IC 2574, we therefore use the bulk rotation
velocity which is less affected by these random motions and thus provides a better description
of the underlying kinematics. We refer to \cite{Oh_2008} for a complete discussion on the rotation curve analysis. \\

\noindent $\bullet$ \textbf{NGC 2366:}
In Fig.~\ref{NGC2366_MAPS}, the distorted iso-velocity contours of the velocity fields indicate
that most disturbances caused by non-circular motions are present in the outer regions ($>5$ kpc), especially in the north-western part of the galaxy.
This is also confirmed by the NONC velocity field in the (l) panel of Fig.~\ref{NGC2366_MAPS}, and large
amplitudes ($\sim$10 \kms) of harmonic terms in Fig.~\ref{NGC2366_TR_HD}. 
However, these disturbances are largely removed in the bulk velocity field as shown in Fig.~\ref{NGC2366_MAPS}.

In Fig.~\ref{NGC2366_VROT_BARYONS_ALPHA}, we compare the derived rotation curves with those from the literature (\citeauthor{Swaters_1999} \citeyear{Swaters_1999};
\citeauthor{Hunter_2001} \citeyear{Hunter_2001}; \citeauthor{van_Eymeren_2009} \citeyear{van_Eymeren_2009}).
The \cite{Hunter_2001}\footnote[4]{An IWM velocity field was used.} and the THINGS IWM curves are systematically lower
than the bulk rotation curve. This is not due to different inclination assumption
since an inclination of $\sim$$65\,^{\circ}$ which is similar to our value ($\sim$$63\,^{\circ}$)
was used for the \cite{Hunter_2001} curve. Instead, the velocity difference can be due to non-circular motions in the galaxy.
This idea is supported by a significant velocity difference beyond $\sim$5 kpc where strong non-circular
motions are present as discussed above. In addition we also compare the hermite $h_{3}$ curve
with that derived using the same THINGS $h_{3}$ velocity field by \cite{van_Eymeren_2009}.
They used a slightly different center position ($\sim$20\,\arcsec\ in declination) and a lower systemic
velocity (98 \kms) but similar inclination ($\sim$$63\,^{\circ}$) and position angle ($\sim$$43\,^{\circ}$).
The \cite{van_Eymeren_2009} curve agrees well with our hermite $h_{3}$ curve but is systematically lower than the
bulk rotation curve. As in the case of IC 2574, we adopt the bulk rotation curve for the mass modeling of NGC 2366.
We refer to \cite{Oh_2008} for a complete discussion on the rotation curve analysis. \\

\noindent $\bullet$ \textbf{Ho I:}
The inclination of Ho I is the lowest among our galaxies. Therefore, the projected velocities,
$V{\sin}\,i$ (where $V$ and $i$ are the circular rotation velocity and the inclination) of the galaxy
are small, and more sensitive to the effect of non-circular motions.
As can be seen in Fig.~\ref{HoI_MAPS}, the iso-velocity contours of the velocity fields
are severely distorted, particularly in the central and north-western regions of the galaxy.
The NONC velocity field in Fig.~\ref{HoI_MAPS} also indicates the presence of
strong non-circular motions in these regions as confirmed by inspection of position-velocity
cuts along the kinematical major and minor axes (\citeauthor{Ott_2001} \citeyear{Ott_2001}).
In addition, the harmonic analysis of the bulk velocity field which is already corrected for
non-circular motions shows large amplitudes ($\sim$10 \kms) of the decomposed harmonic terms
in Fig.~\ref{HoI_TR_ADC_HD}.

To minimize the effect of the non-circular motions, we derive the rotation curve using the bulk velocity field
in Fig.~\ref{HoI_MAPS}. The derived ring parameters, such as the kinematic center, the systemic
velocity and the position angle are consistent with those found by \cite{Ott_2001}.
The rotation curve keeps increasing out to $\sim$1.5 kpc and decreases beyond that.
This also agrees well with the result by \cite{Ott_2001}, converted using the inclination of $14\,^{\circ}$.

However, the small inclination value of $14\,^{\circ}$ implies considerable uncertainty in the rotation curve.
To check this we compare the rotation curves derived using inclinations deviating $\pm10\,^{\circ}$ from our
adopted value (see the INCL panel of Fig.~\ref{HoI_TR_ADC_HD}). In the VROT panel of Fig.~\ref{HoI_TR_ADC_HD},
we find significant differences between them. In particular, the rotation curve derived using the low inclination
value ($\sim$$4\,^{\circ}$) significantly deviates from our preferred curve. However, the value of $\sim$$4\,^{\circ}$
falls at the extreme lower end of the inclinations derived from the tilted-ring fits. For reference, we show a fit
result with only the INCL left free (indicated by gray dots in the INCL panel of Fig.~\ref{HoI_TR_ADC_HD}).
The values are systematically larger than those (open circles) derived keeping all parameters free.
From this we conclude that it is unlikely that Ho I has an inclination as small as $\sim$$4\,^{\circ}$ and
we adopt a value of $14\,^{\circ}$ for the remainder of this paper. Notwithstanding the low inclination,
we will derive the rotation curves, fully keeping in mind the uncertainties due to inclination.
To avoid our conclusions being skewed by this galaxy, we will present our analysis both with and without Ho I.

The H{\sc i} velocity dispersions in Fig.~\ref{HoI_MAPS} show high values of
$\sim$12 \kms\ in the north western part (see also \citeauthor{Ott_2001} \citeyear{Ott_2001}).
Compared to the derived rotation velocity of $\sim$20 \kms\ in the outer region,
the magnitude of this velocity dispersion is significant. We therefore correct the
rotation curve for asymmetric drift as described in Section~\ref{ADC}.
The corrected curve is presented in Fig.~\ref{HoI_TR_ADC_HD}.
We adopt this corrected bulk rotation curve for the kinematic analysis of Ho I.
We again stress that in all further analysis we consider our results with and without
Ho I.
 \\

\noindent $\bullet$ \textbf{Ho II:}
We derive the rotation curve using the bulk velocity field shown in Fig.~\ref{HoII_MAPS}.
As shown in Fig.~\ref{HoII_TR_ADC_HD}, all ring parameters are well determined and are
consistent with the results by \cite{Bureau_2002}. The $2^{\rm{nd}}$ moment map
shows rather large velocity dispersions compared to the circular rotation velocity.
Therefore, we make a correction for pressure support. The asymmetric drift corrected
bulk rotation curve is presented in Fig.~\ref{HoII_TR_ADC_HD}; it is rather flat and
increases slightly beyond 4 kpc, compared to the uncorrected one. In Fig.~\ref{HoII_TR_ADC_HD},
the amplitudes of the harmonic terms derived from the hermite $h_{3}$ velocity field are less than 
5 \kms\ over most radii, although slightly larger than those from the bulk velocity field.

In Fig.~\ref{HoII_VROT_BARYONS_ALPHA}, we compare our rotation curve with that from the literature. 
The $\sim$45$\arcsec$ resolution IWM rotation curve by \cite{Bureau_2002}
not only falls below the asymmetric drift corrected bulk rotation curve but also below the THINGS
IWM one, despite the correction for asymmetric drift by the authors. The difference between the
respective tilted-ring models is not enough to explain this velocity difference.
It is likely that the Bureau \& Carignan's lower beam resolution data ($\sim$45$\arcsec$) and the
derived rotation curve with a larger ring width ($\sim$60$\arcsec$) which smooth small-scale ``wiggles''
caused by non-circular motions in the galaxy (e.g., at $\sim$2 kpc) are the main explanation for
the difference. In Fig.~\ref{HoII_VROT_BARYONS_ALPHA}, we find that the THINGS IWM curve with
a ring width of 60\arcsec\ is slightly lower than the one with 12\arcsec. We use the asymmetric
drift corrected bulk rotation curve for the mass modeling of Ho II. \\

\noindent $\bullet$ \textbf{DDO 53:}
As shown in Fig.~\ref{DDO53_MAPS}, DDO 53 shows a clear rotation pattern in its velocity field.
However, the distorted iso-velocity contours imply the presence of non-circular motions.
In particular they are prominent in the outer regions as confirmed by the extracted NONC velocity
field and the harmonic analysis as shown in Figs.~\ref{DDO53_MAPS} and \ref{DDO53_TR_ADC_HD}, respectively.
To minimize the effect of these non-circular motions, we extract the bulk velocity field
as shown in Fig.~\ref{DDO53_MAPS}. Compared to other types of velocity fields, the bulk velocity
field is noisier but the overall rotation pattern is better visible. In addition, as shown in
Fig.~\ref{DDO53_TR_ADC_HD}, the amplitudes of the harmonic terms decomposed from the bulk velocity
field are close to zero in comparison to those from the hermite $h_{3}$ velocity field.
The derived rotation curves using the bulk velocity field are shown in Fig.~\ref{DDO53_TR_ADC_HD},
and most ring parameters except inclination are well determined. The inclination shows a large scatter
as a function of radius, especially in the inner regions.

We examine the effect of inclination on the rotation velocity
by changing it by $\pm10\,^{\circ}$ and performing tilted-ring fits while keeping other
ring parameters the same. Although the rotation curve derived using the lower inclination
value ($\sim$$17\,^{\circ}$) is higher by up to $\sim$10 \kms, this low inclination value
seems not plausible for DDO 53. Like the case of Ho I, we show a fit result with only the
INCL free as indicated by gray dots in the INCL panel of Fig.~\ref{DDO53_TR_ADC_HD}.
They are larger than those (open circles) from the very first run with all ring parameters free.
The lower inclination value ($\sim$$17\,^{\circ}$) can be regarded as a lower limit.

The maximum bulk rotation velocity is $\sim$18 \kms\ which is comparable to the values
found for the velocity dispersion in the outer regions of the galaxy (see Fig.~\ref{DDO53_MAPS}).
This demands an asymmetric drift correction, and the corrected curve is shown
in Fig.~\ref{DDO53_TR_ADC_HD}. The corrected curve keeps increasing to $\sim$34 \kms\ at 2 kpc.
We use this rotation velocity for the mass modeling of DDO 53. \\

\noindent $\bullet$ \textbf{M81dwB:}
The extracted velocity fields themselves show little difference with respect to each other
as shown in Fig~\ref{M81DWB_MAPS}. The NONC velocity field shows no significant non-circular
motions in the galaxy, except in the very outer regions. Moreover, as shown in Fig.~\ref{M81DWB_TR_ADC_HD},
the amplitudes of the harmonic terms (corrected for inclination) decomposed from both the IWM
and hermite $h_{3}$ velocity fields are less than 5 \kms\ over most radii. In addition, the difference
between the harmonic terms derived from the IWM and hermite $h_{3}$ velocity fields is also negligible.
This implies that the effect of non-circular motions on the velocity fields is not significant.
Therefore, we can use the hermite $h_{3}$ velocity field to derive the rotation curve.
As already discussed in Section~\ref{VF_types}, the hermite $h_{3}$ velocity field
gives a robust estimate for the underlying circular rotation of a galaxy in which
non-circular motions are insignificant.
In Fig.~\ref{M81DWB_TR_ADC_HD}, the derived hermite $h_{3}$
rotation curve keeps increasing out to 0.5 kpc and then stays flat to 1 kpc.
Beyond that, the curve rapidly declines but this is mainly due to the small number of pixels
that contain signal and the large uncertainties in the velocities of the outer rings.

The large velocity dispersions ($\sim$15 \kms) in the outer regions are significant
compared to the maximum rotation velocity of $\sim$28 \kms. We therefore perform the
asymmetric drift correction for the circular rotation after which the corrected curve keeps
increasing out to 1 kpc as shown in Fig.~\ref{M81DWB_TR_ADC_HD}. \\

\noindent $\bullet$ \textbf{DDO 154:}
The complete description of the data and the mass modeling including the tilted-ring analysis
is given in detail in \cite{deBlok_2008}.
It shows a regular rotation pattern in the hermite $h_{3}$ velocity field in Fig.~\ref{DDO154_MAPS}
(see also Fig. 81 in \citeauthor{deBlok_2008} \citeyear{deBlok_2008}).
In addition, no significant non-circular motions were found in the galaxy
from a harmonic analysis of the velocity field (\citeauthor{Trachternach_2008} \citeyear{Trachternach_2008}).
We therefore use the hermite $h_{3}$ rotation curve as in the case of M81dwB.
As described in \cite{deBlok_2008}, the resulting rotation curve resembles that of a galaxy with solid-body rotation
but increases more steeply in the inner regions than previous determinations
(e.g., \citeauthor{Carignan_1988} \citeyear{Carignan_1988}; \citeauthor{Carignan_1998} \citeyear{Carignan_1998})
for which the IWM velocity fields with lower beam resolutions were used.
In this paper, we use the hermite $h_{3}$ rotation curve derived in \cite{deBlok_2008}
for the mass modeling of DDO 154, and refer to their paper for a complete discussion. \\

In summary, the rotation velocities derived from the bulk velocity fields of the THINGS dwarf galaxies
(except M81dwB and DDO 154 where a hermite $h_{3}$ velocity field was used) generally
show the most rapid increase compared to those from the other types of velocity
fields, such as the IWM, peak, single Gaussian fit and hermite $h_{3}$. The IWM
velocities show the slowest increase, especially in the inner region of the galaxy.
The rotation velocities derived from the peak, single Gaussian fit and hermite $h_{3}$ velocity
fields show an increasingly steeper gradient than the IWM velocity, but somewhat less steep than
the bulk velocity. This is due to their different abilities to take asymmetries of profiles affected
by non-circular motions into account.
The IWM velocity field is the one most affected by non-circular motions. The random non-circular motions induce a smaller
velocity gradient across the IWM velocity field, which results in a rotation velocity that increases more slowly.
In contrast, the bulk velocity field minimizes the effect of random motions, and properly extracts
the underlying circular rotation. 

We also examine the sensitivity of the rotation curves to the exact value of inclination. Of our galaxies, the rotation curves of
Ho I and DDO 53 (whose inclination values are $\sim$14\,$^{\circ}$ and $\sim$27\,$^{\circ}$, respectively)
are most sensitive to changes in inclination. However, the adopted inclination values from the
tilted-ring fits appear to be plausible. In addition, they also agree well with those
derived independently from the Baryonic Tully$\--$Fisher relation as will be discussed in
Section~\ref{BTF_relation_THINGS} later.

\section{Mass models of baryons}
\label{Baryons}

The rotation curve reflects the dynamics of the total
matter content in a galaxy, including the baryons and the dark matter.
We therefore subtract the dynamical contribution of baryons from the
total dynamics to determine the dark matter component only.
To this end, we first derive radial distributions of the baryons in our
galaxies and derive mass models for them.

\subsection{Stellar component}
\label{Stars}

We derive the mass models for the stellar components of our sample galaxies
following the method described in \citeauthor{Oh_2008} (\citeyear{Oh_2008}; see also \citeauthor{deBlok_2008} \citeyear{deBlok_2008}).
Firstly, we derive the luminosity profiles of the galaxies by applying the tilted-ring
models derived in Section~\ref{TRM} to the IRAC 3.6$\mu \rm m$ images from SINGS
to derive radially averaged surface brightness profiles.
These are shown in the Appendix.
We then convert the luminosity profiles to mass density profiles in units of
\surfdens\ using an empirical \ML\ relation derived from population synthesis
models, as described in \cite{Oh_2008}.
The empirical relation derives \ML\ in the IRAC 3.6$\mu \rm m$ band (\MLsps)
from the \ML\ in $K$ band (\MLk) which in turn is determined using optical
colors and metallicity of a galaxy, as given in \cite{Bell_2001}.
The optical ($B$, $V$ and $R$) surface brightness profiles and colors ($B-V$ and $B-R$)
used for determining \MLk\ are shown in the Appendix.
Here, we use constant average colors for our galaxies, except for IC 2574 where
the radial distribution of colors can be derived. We also show the metallicity of
the sample galaxies in Table 1. From this we compute \MLsps\ for the sample galaxies,
as shown in the Appendix.
\cite{Leroy_2008} use an empirical $K$-to-3.6$\mu$m calibration to derive $K$-band
fluxes from the IRAC 3.6$\mu$m images for a number of THINGS galaxies. They then derive
the stellar disk masses adopting a fixed $\MLk$=$0.5$. \cite{deBlok_2008} make a comparison
between the stellar disk masses derived using our method and the approach by \cite{Leroy_2008},
and find that they agree well with each other. 

Using the \MLsps\ values, we derive the mass density profiles
of stellar components of the sample galaxies (presented in the Appendix).
We then calculate the rotation velocities for the stellar components from
the mass density profiles, assuming a vertical $\rm {sech}^{2}(z)$ scale height distribution.
We use $h/z_{0}$=$5$, the ratio between the vertical scale height $z_{0}$ and the radial scale length $h$
of disk, as determined in \citeauthor{van_der_Kruit_1981} (\citeyear{van_der_Kruit_1981};
see also \citeauthor{Kregel_2002} \citeyear{Kregel_2002}).
The derived $z_{0}$ values are given in Table 1. The average
value is $z_{0}\simeq0.32$ kpc. We construct the final mass models for the
stellar components of the sample galaxies using the {\sc rotmod} task in GIPSY,
the results of which are shown in the Appendix.

\begin{figure}
\epsscale{1.0}
\includegraphics[angle=0,width=0.49\textwidth,bb=20 141 578 688,clip=]{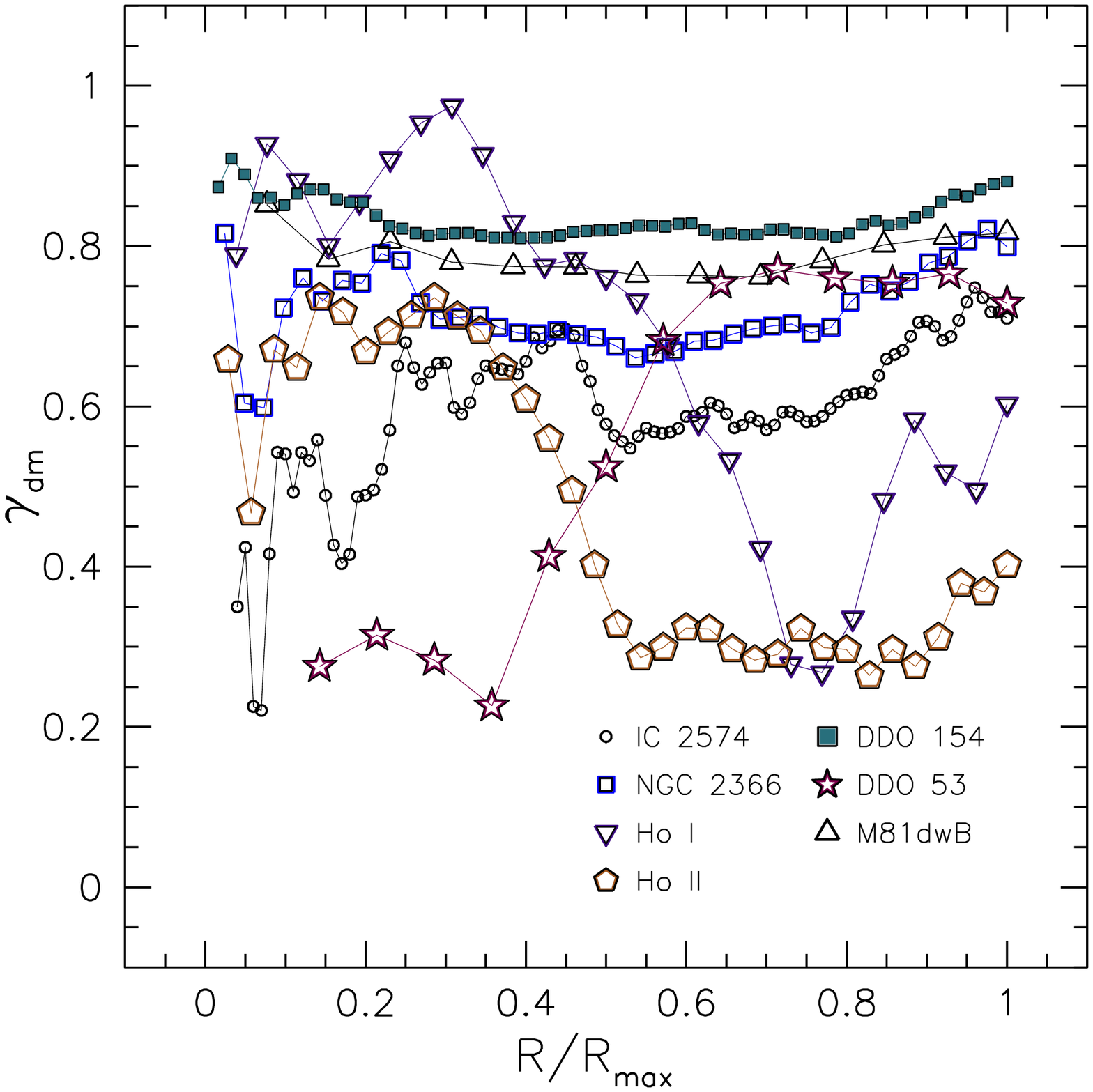}
\caption{The dark matter fraction $\gamma_{\rm dm}$ (as described in Eq.~\ref{gamma_dm}) of the 7 THINGS dwarf galaxies.
Most galaxies are dark matter dominated across all radii except the inner region of
DDO 53 and the outer region of Ho II, respectively. This is discussed in detail in Section~\ref{DM_fraction_THINGS_properties}
\label{THINGS_DM_fract}}
\end{figure}

\begin{figure*} 
\epsscale{1.0}
\includegraphics[angle=0,width=1.0\textwidth,bb=20 400 565 685,clip=]{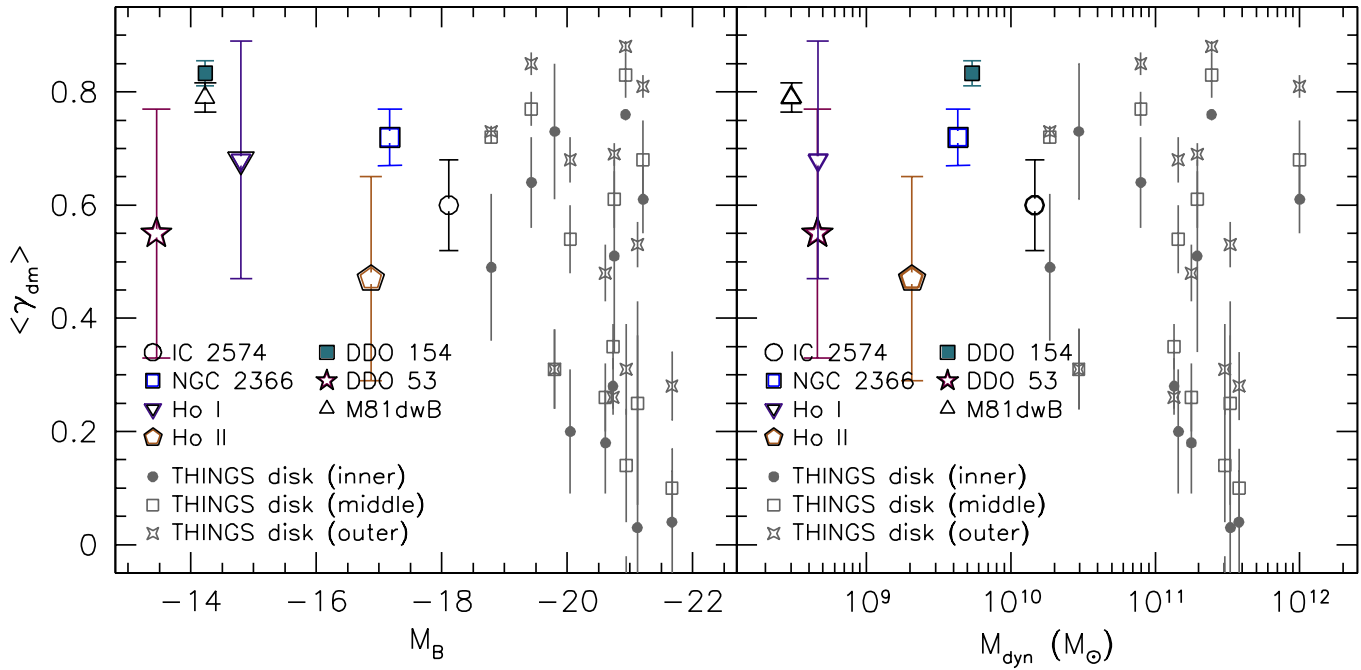}
\caption{{\bf Left:} The relationship between the dark matter fraction and the absolute B magnitude
of 19 THINGS dwarf and spiral galaxies. $<$$\gamma_{\rm dm}$$>$ is determined by radially averaging
$\gamma_{\rm dm}$ values of each galaxy. For the spiral galaxies, $<$$\gamma_{\rm dm}$$>$ values
are calculated over three regions, splitting a galaxy into three annuli (inner, middle and outer)
as indicated by different symbols. {\bf Right:} The relationship between the mean dark matter fraction
$<$$\gamma_{\rm dm}$$>$ and the dynamical mass of the same galaxies. See Section~\ref{DM_fraction_THINGS_properties}
for more discussions.
\label{DM_fract_Bmag_Mdyn}}
\end{figure*}

\subsection{Gas component}
\label{Gas}
The H{\sc i} surface density profile in \surfdens\ units can be directly
derived from the observed H{\sc i} column density. In order to calculate the
radial H{\sc i} distribution of the sample galaxies, we apply the tilted-ring
models derived in Section~\ref{TRM} to the integrated H{\sc i} maps to derive
azimuthally averaged radial H{\sc i} profiles.
We scale the derived H{\sc i} surface density profile by a factor of 1.4
(\citeauthor{deBlok_2008} \citeyear{deBlok_2008})
to account for helium and metals
and calculate the rotation velocities for the gas component.
For this we assume an infinitely thin disk and
use the task {\sc rotmod} implemented in GIPSY. The gas surface density profiles
and the gas rotation velocities of our galaxies are presented in the Appendix.

\section{Dark matter halo and luminous matter}
\label{DM_fraction_THINGS}

In general, dwarf galaxies are dark matter-dominated throughout due to the small
contribution of baryons to the total dynamics, as is the case in LSB galaxies
(e.g., \citeauthor{deBlok_1997} \citeyear{deBlok_1997}; \citeauthor{Prada_2002} \citeyear{Prada_2002}).
Therefore, dwarf galaxies have been considered to be ideal objects for
studying dark matter properties in galaxies.
Of particular interest is testing the dark matter distribution as predicted from cosmological simulations.
In this section, we calculate the dark matter fraction of our galaxies
and verify if dark matter indeed dominates the total
dynamics of these systems. Furthermore, we also examine the relationship between
the dark matter fraction and other galaxy properties, such as the dynamical mass,
the absolute $B$ magnitude and the Baryonic Tully$\--$Fisher (BTF) relation 
(\citeauthor{Bell_2001} \citeyear{Bell_2001}; \citeauthor{Verheijen_2001} \citeyear{Verheijen_2001};
\citeauthor{McGaugh_2004} \citeyear{McGaugh_2004}; \citeauthor{De_Rijcke_2007} \citeyear{De_Rijcke_2007}).

\subsection{Dark matter fraction and galaxy properties}
\label{DM_fraction_THINGS_properties}

We derive the radial dark matter fraction of our galaxies using,
\begin{equation}
\gamma_{\rm{dm}} = \frac{M_{\rm{DM}}}{M_{\rm{tot}}} = \frac{V_{\rm{tot}}^{2} - V_{\rm{star}}^{2} - V_{\rm{gas}}^{2}}{V_{\rm {tot}}^{2}},
\label{gamma_dm}
\end{equation}
\noindent where $V_{\rm{tot}}$ is the observed total rotation velocity, and
$V_{\rm{star}}$ and $V_{\rm{gas}}$ are the rotation velocities of stars and gas, respectively.
For this measurement, we use $V_{\rm{tot}}$, $V_{\rm{star}}$ and $V_{\rm{gas}}$ of our galaxies
as derived in Sections~\ref{Deriving_Rotcurs} and \ref{Baryons} (see the Appendix).

In Fig.~\ref{THINGS_DM_fract}, we plot $\gamma_{\rm{dm}}$ as a function of radius.
The radii are normalized to the maximum radius ($R_{\rm{max}}$) at which the last data
point is measured. Most of our galaxies show large values of $\gamma_{\rm{dm}}$ of
about 0.7 over the radial range. This implies that they are indeed dark matter-dominated
over most of their radial range. Ho II and DDO 53 show
radial gradients. The value of $\gamma_{\rm{dm}}$ for Ho II is $\sim$0.7 within 0.4$R_{\rm{max}}$,
but decreases to $\sim$0.3 in the outer parts. The $\gamma_{\rm{dm}}$ value
of DDO 53 is $\sim$0.3 in the inner parts ($<$0.4$R_{\rm{max}}$), but increases
to $\sim$0.7 in the outer parts. Note that the contribution of the gas component
to the total dynamics is larger than that of stars in the outer parts of Ho II and the
inner parts of DDO 53.

We proceed by examining the relationship between the dark matter fraction, the absolute $B$ magnitude and
the dynamical mass of 12 spiral galaxies from THINGS and the 7 dwarf galaxies from the current sample.
In the left panel of Fig.~\ref{DM_fract_Bmag_Mdyn}, we plot the radial average of $\gamma_{\rm{dm}}$
against the absolute $B$ magnitude of individual galaxies. For the spiral galaxies where the inner and
outer regions are totally dominated by baryons and dark matter, respectively, we calculate average
$\gamma_{\rm{dm}}$ values over three regions, splitting a galaxy into three annuli (inner, middle and outer).
For this, we choose an inner radius at which the rotation curve reaches its flat part, and split
the region beyond it into two equal-size radial bins for the middle and outer annuli. The calculated $\gamma_{\rm{dm}}$
values within the annuli for each spiral galaxy are indicated by different symbols in Fig.~\ref{DM_fract_Bmag_Mdyn}.
As expected, the outer region of the spiral galaxies is more dark matter dominated than the inner region.
In addition it is likely that the dark matter fraction in the outer region of the spiral galaxies is similar
to that of the dwarf galaxies.

In the right panel of Fig.~\ref{DM_fract_Bmag_Mdyn}, we show the relationship between the dark matter fraction
and the dynamical mass of the galaxies.
Likewise, for the spiral galaxies we calculate average $\gamma_{\rm{dm}}$ values over
three regions, splitting a galaxy into three annuli (inner, middle and outer).
Considering that more luminous galaxies are in general more massive
(e.g., \citeauthor{Guo_2010} \citeyear{Guo_2010}; \citeauthor{Dutton_2010} \citeyear{Dutton_2010}),
this is largely consistent with the relationship between $\gamma_{\rm{dm}}$
and absolute $B$ magnitude, as shown above.

\subsection{The baryonic Tully$\--$Fisher relation}
\label{BTF_relation_THINGS}

\begin{figure}
\epsscale{1.0}
\includegraphics[angle=0,width=0.49\textwidth,bb=23 165 535 655,clip=]{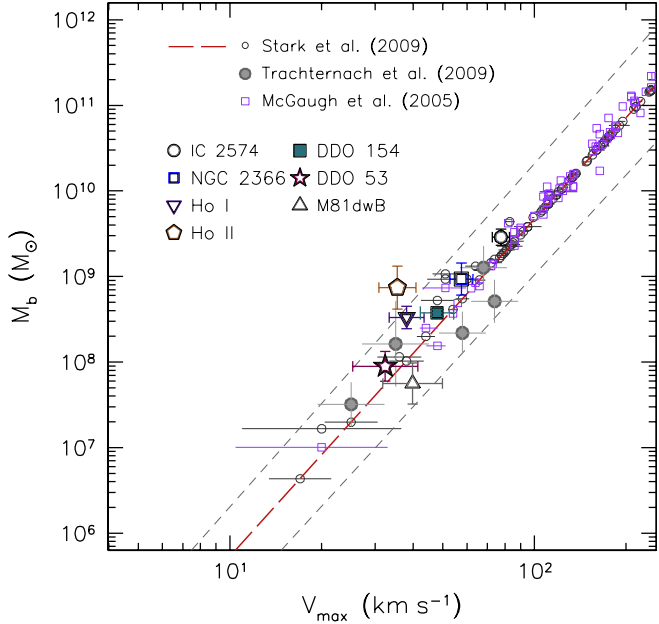}
\caption{The baryonic Tully$\--$Fisher relation of the THINGS dwarf galaxies. The baryonic mass includes the
stellar and gas components derived in Section~\ref{Baryons}. The long dashed-line indicates the BTF relation
calibrated using a sample of gas dominated galaxies in \cite{Stark_2009}, and the short dashed-lines indicate
the uncertainty in the relation. See Section~\ref{BTF_relation_THINGS} for more details.
\label{BTF_relation}}
\end{figure}

We also examine whether the THINGS dwarf sample galaxies follow the Baryonic
Tully$\--$Fisher (BTF) relation. There have been several efforts to calibrate the
BTF relation using a sample of gas dominated low mass systems
(\citeauthor{McGaugh_2000} \citeyear{McGaugh_2000}; \citeauthor{McGaugh_2005} \citeyear{McGaugh_2005};
\citeauthor{Stark_2009} \citeyear{Stark_2009}).
These found that the broken continuity of the classical Tully$\--$Fisher relation
of low-mass galaxies can be restored by using their total baryonic mass (i.e., including
not only stars but also the gas component). As shown in Fig.~\ref{BTF_relation},
we plot the baryonic (stars + gas) mass of our galaxies derived in Section~\ref{Baryons}
against the maximum rotation velocity at the last measured point.
They are roughly consistent with the BTF relation (indicated as the dashed line) from \cite{Stark_2009}
within the uncertainty but systematically slightly higher than the line except M81dwB. This could be owing to
the underestimated maximum rotation velocities of the galaxies. Some of the rotation
curves derived in Section~\ref{Deriving_Rotcurs} still keep increasing at the last
measured point, which implies a larger maximum rotation velocity.

As already discussed in Section~\ref{Deriving_Rotcurs}, the rotation curves of some of
our sample galaxies are sensitive to the exact value of inclination (e.g., Ho I and DDO 53).
As a sanity check, we therefore derive inclinations based on the BTF relation. 
The observed line-of-sight velocity of a galaxy at the last
measured point $R_{\rm max}$ can be expressed by the following equation
(if we only consider the azimuthal velocity component),
\begin{equation}
\label{eq:Vlos_BTF}
\displaystyle V_{\rm{obs}}(R_{\rm max}) = V_{\rm{sys}} + V^{\rm TR}_{\rm max}\times\sin\,i^{\rm TR}\times\cos\theta,
\end{equation}
\noindent where $V_{\rm{sys}}$ is the systemic velocity, $\theta$ is the position angle,
$i^{\rm TR}$ is the inclination and $V^{\rm TR}_{\rm max}$ is the maximum rotation velocity
derived from tilted-ring analysis. The BTF relation yields an estimate
of maximum rotation velocity $V^{\rm BTF}_{\rm max}$ at a given baryonic mass.
Therefore, in Eq.~\ref{eq:Vlos_BTF} $V^{\rm TR}_{\rm max}$ can be substituted with $V^{\rm BTF}_{\rm max}$
and the corresponding inclination value $i^{\rm BTF}$ which gives the same $V_{\rm{obs}}(R_{\rm max})$ can
be calculated using the following formula,
\begin{equation}
\label{BTF_inclination}
i^{\rm BTF} = \arcsin\biggl(\frac{V^{\rm TR}_{\rm max}}{V^{\rm BTF}_{\rm max}}\times\sin\,i^{\rm TR}\biggr).
\end{equation}
The derived $i^{\rm BTF}$ values of the THINGS dwarf galaxies sample are given in Table 1, and they
are smaller than those derived from tilted-ring analysis except for M81dwB. This is because the
inferred $V^{\rm BTF}_{\rm max}$ values of our galaxies are larger than the $V^{\rm TR}_{\rm max}$
values as shown in Fig.~\ref{BTF_relation}.
Given the uncertainties in the estimates, the inclination values derived from both the BTF relation
and the tilted-ring analysis are not significantly different from each other except for Ho II. 
However, as can be seen from not only the tilted-ring analysis (including the position-velocity diagram)
but also the comparison of rotation velocities in the Appendix it is unlikely that Ho II has
an inclination ($\sim25^{\circ}$) as low as that inferred from the BTF relation.

\section{Dark matter distribution}
\label{DM_distribution}

In this section, we compare the derived dark matter distribution of the THINGS dwarf galaxies
with that inferred from structure formation N-body simulations based on the \LCDM\ paradigm.
For this we use an NFW halo model (\citeauthor{NFW_1996} \citeyear{NFW_1996}, \citeyear{NFW_1997})
which is given as,
\begin{equation}
\label{eq:3_1}
\rho_{\rm{NFW}}(R) = \frac{\rho_{i}}{(R/R_{s})(1+R/R_{s})^{2}},
\end{equation}
where $\rho_{i}$ is the initial density of the Universe at the
time of the collapse of the halo and $R_{s}$ is the characteristic
radius of the dark matter halo. This gives a ``cusp'' feature having
a power law mass density distribution $\rho \sim R^{-1}$ towards
the centers of galaxies. The corresponding rotation velocity induced by this potential has the following form,
\begin{equation}
\label{eq:3_2}
V_{\rm{NFW}}(R) = V_{200}\sqrt{\frac{{\rm{ln}}(1+cx)-cx/(1+cx)}{x[{\rm{ln}}(1+c)-c/(1+c)]}},
\end{equation}
where $c$ is the concentration parameter defined as $R_{200}/R_{s}$.
$V_{200}$ is the rotation velocity at radius $R_{200}$ where the mass
density contrast exceeds 200 and $x$ is defined as $R/R_{200}$.

In addition, we also use an observationally motivated pseudo-isothermal halo model
as an extreme representation of ``core-like'' halo models (e.g., \citeauthor{Begeman_1991} \citeyear{Begeman_1991}).
It has the following form,
\begin{equation}
\label{eq:3_4}
\rho_{\rm{ISO}}(R) = \frac{\rho_{0}}{1+(R/R_C)^{2}},
\end{equation}
where $\rho_{0}$ and $R_C$ are the core-density and core-radius
of the halo, respectively. This gives rise to a mass distribution
with a sizable constant density-core ($\rho \sim R^{0}$) at the centers of galaxies.
The rotation velocity induced by the mass distribution is given as,
\begin{equation}
\label{eq:3_5}
V_{\rm{ISO}}(R) = \sqrt{4\pi G\rho_{0}R_C^{2}\Biggl[1-\frac{R_C}{R}{\rm{atan}}\Biggl(\frac{R}{R_C}\Biggr)\Biggr]}.
\end{equation}

Using these two halo models, we examine which model is preferable to describe
the observed dark matter distribution of our galaxies.

\begin{sidewaysfigure*}
\epsscale{1.0}
\includegraphics[angle=0,width=1.0\textwidth,bb=20 395 565 670,clip=]{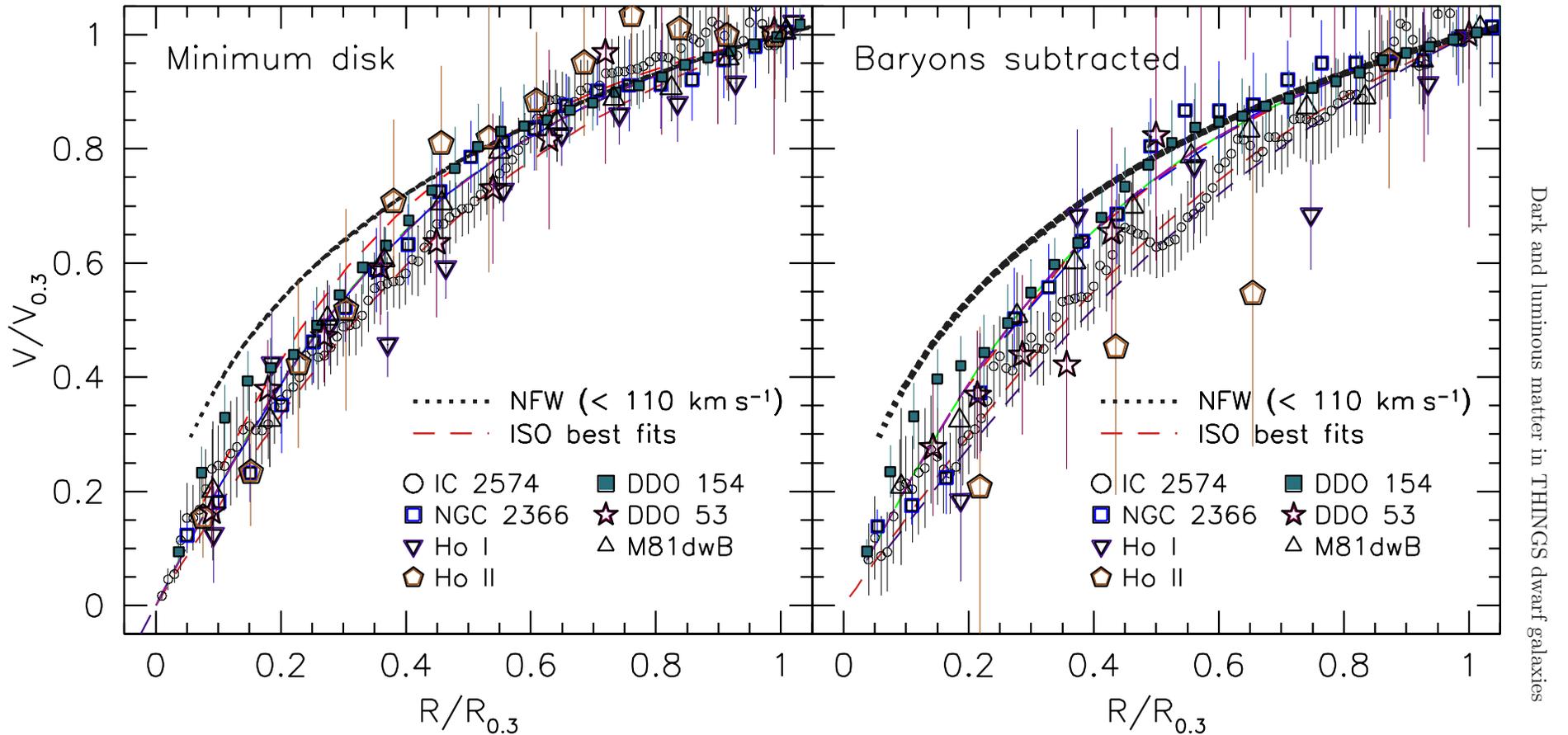}
\caption{{\bf Left:} The shape of the total rotation curves (not corrected for baryons) of the 7 THINGS dwarf galaxies.
The rotation curves are scaled with respect to the rotation velocity $V_{0.3}$ at $R_{0.3}$ where the logarithmic slope
of the curve is $d{\rm log}V/d{\rm log}R=0.3$ (\citeauthor{Hayashi_2006} \citeyear{Hayashi_2006}).
We overplot the rotation curves of the NFW models with $V_{200}$ ranging from 10 to 110 \kms.
The dotted lines indicate the NFW models with $V_{200}$ less than 110 \kms.
The scaled rotation curves of the best fit pseudo-isothermal halo models
(denoted as ISO) are also overplotted. See Section~\ref{Rotcur_shape} for more details.
{\bf Right:} The shape of the dark matter rotation curves of the 7 THINGS dwarf galaxies. These are corrected
for baryons and are scaled with respect to the rotation velocity $V_{0.3}$ at $R_{0.3}$.
Same legends as in the left panel. Compared to the total rotation curves in the left panel, the dark matter rotation curves
increase less steeply but they are similar due to the low baryonic fraction of the galaxies as discussed
in Section~\ref{DM_fraction_THINGS_properties}. See Section~\ref{Rotcur_shape} for more discussions.
\label{THINGS_NFW_comp_min_opt_disk}}
\end{sidewaysfigure*}

\begin{figure*}
\epsscale{1.0}
\includegraphics[angle=0,width=1.0 \textwidth,bb=0 150 580 690,clip=]{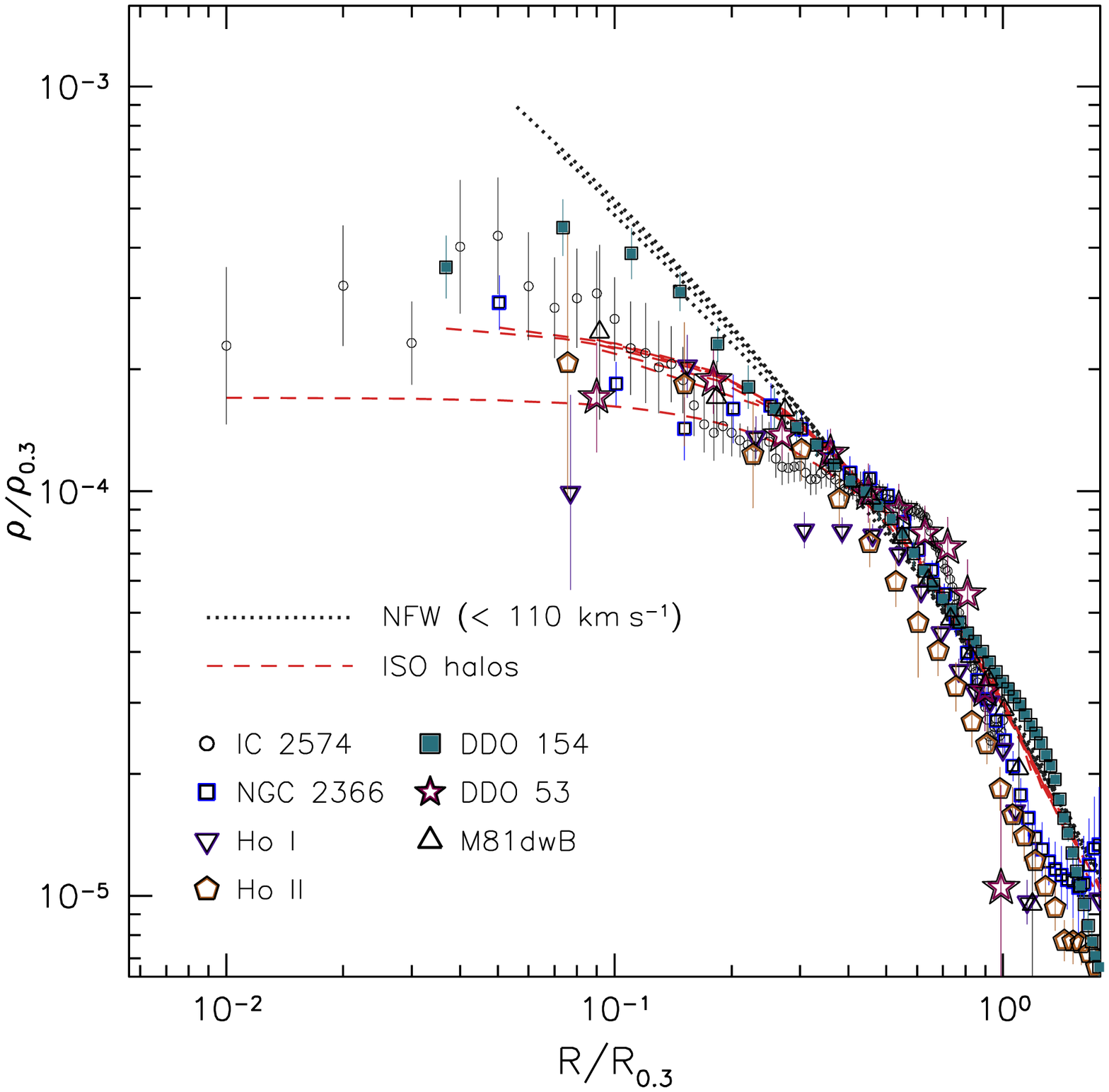}
\caption{The dark matter density profiles of the 7 THINGS dwarf galaxies. The profiles are derived using
the scaled rotation curves (assuming minimum disk) as described in Section~\ref{Rotcur_shape} (see also Fig.~\ref{THINGS_NFW_comp_min_opt_disk}).
The dotted lines represent the mass density profiles of NFW models ($\alpha\sim$$-1.0$) with
$V_{200}$ ranging from 10 to 110 \kms. The dashed lines indicate the mass
density profiles of the best fit pseudo-isothermal halo models ($\alpha\sim0.0$). See Section~\ref{DM_profiles} for more details.
\label{THINGS_DM_density_profile}}
\end{figure*}

\subsection{Dark matter mass modeling}
\label{DM_modeling}

We subtract the dynamical contribution of the baryons from the total kinematics and construct
mass models of the dark matter halos of our galaxies.
We fit the two halo models, i.e., the NFW and pseudo-isothermal models (see e.g. \citeauthor{Oh_2008} \citeyear{Oh_2008}),
to the bulk rotation curves derived in Section~\ref{Rotation_curves}, taking into account the mass
models of the baryons. When performing the fits, we use various assumptions for \ML,
such as ``maximum disk'', ``minimum disk'' and ``minimum$+$gas disk''
as well as the model \MLsps\ value as described in Section~\ref{Stars}.
The maximum disk assumes that the observed rotation curve in the inner regions of a galaxy
is almost entirely due to the stellar component (\citeauthor{van_Albada_1986} \citeyear{van_Albada_1986}).
Therefore, the dark matter properties derived using this assumption will provide a
lower limit to its mass distribution. In contrast, the minimum disk hypothesis ignores the contribution of baryons
and attributes the rotation curve to the dark matter component only (\citeauthor{van_Albada_1986} \citeyear{van_Albada_1986}).
This yields a robust upper limit on the properties of dark matter.
The minimum$+$gas disk ignores the stellar component but includes the gas component.

The fit results of individual galaxies are presented in the figures (``Mass modeling results'') and Tables in the Appendix.
We find that in terms of fit-quality (i.e., $\chi^{2}_{red}$) pseudo-isothermal
halo models are mostly preferred over NFW halo models in describing the dark matter
distribution of our galaxies. 
In addition the mean value of the logarithmic central halo surface
density $\log (\rho_{0} R_C)$ in units of \surfdens\ of our sample galaxies is $\sim 1.62 \pm 0.14$.
This is smaller than the relation ($\log \rho_{0} R_{0} = 2.15 \pm 0.2$) found by \cite{Donato_2009}
from a sample of galaxies. However, considering that the relation found by \cite{Donato_2009}
was derived assuming the Burkert profile (\citeauthor{Burkert_1995} \citeyear{Burkert_1995}),
they are not significantly different from each other.

In most galaxies, the NFW halo model fails to fit the rotation curves, irrespective of
the assumptions of \ML\ (i.e., maximum, minimum, minimum$+$gas and model \MLsps\ disk),
giving negative (or close to zero) $c$ values. Even if the fits are feasible (e.g., Ho II),
pseudo-isothermal halo models are still slightly better in describing the rotation curves
irrespective of the all assumptions on \ML. We also fit the NFW model to the rotation curves
with only $V_{200}$ as a free parameter after fixing $c$ to 9 which is similar to typical values
(e.g., 8--10; \citeauthor{McGaugh_2003} \citeyear{McGaugh_2003}) predicted from \LCDM\ cosmology.
However, as shown in the Tables in the Appendix, the best-fit $\chi^{2}_{red}$ values are even larger than those from the fits
with both $c$ and $V_{200}$ as free parameters. Moreover, at the inner regions of the rotation
curves, the fitted NFW halo models are too steep. This will be further discussed in the following section.

It is also interesting how well the ``minimum disk'' assumption on \ML\ provides
a good description of the baryonic mass distributions of the galaxies. As shown
in the Appendix, the best-fit $\chi^{2}_{red}$ values are close to the ones
obtained assuming the model \MLsps. This confirms that the THINGS dwarf galaxies
are indeed dark-matter-dominated.

\subsection{Rotation curve shape}
\label{Rotcur_shape}

The divergent density profiles (e.g., $\sim$$R^{\alpha}$ where $\alpha$$\sim$$-0.8$ at 120 pc; \citeauthor{Stadel_2009} \citeyear{Stadel_2009};
see also \citeauthor{Graham_2006} \citeyear{Graham_2006})
found in \LCDM\ simulations are expected towards the centers of galaxies and thus ought to be observed
(see also \citeauthor{Navarro_2010} \citeyear{Navarro_2010}). Dark matter slopes at smaller radius
are often steeper although recent simulations for dwarf galaxies that include the detailed description
and effect of baryonic feedback processes show shallower slopes (\citeauthor{Governato_2010} \citeyear{Governato_2010}).
A comparison of these simulations with THINGS dwarf galaxies will be discussed in a separate paper (Oh et al.).

The cusp-like dark matter distribution in turn forces a unique shape to the rotation curve.
Therefore, a comparison of the shape of the rotation curves between
observation and simulations provides an additional test of \LCDM\ cosmology.
Moreover, the low baryon fraction of our galaxies, as found in Section~\ref{DM_fraction_THINGS_properties},
allows us to directly compare the rotation curve shapes with those
of \LCDM\ simulations even if they are partially affected by the baryons.

For this comparison, we model CDM halos covering a wide range of
$V_{200}$ from 10 to 110 \kms. The concentration parameter $c$
corresponding to a particular value of $V_{200}$ is determined by the following empirical $c-V_{200}$ relation
from the WMAP\footnote[5]{The Wilkinson Microwave Anisotropy Probe (\citeauthor{Spergel_2003} \citeyear{Spergel_2003};
\citeauthor{Spergel_2007} \citeyear{Spergel_2007})} observations in \citeauthor{McGaugh_2007} (\citeyear{McGaugh_2007};
see also \citeauthor{deBlok_2003} \citeyear{deBlok_2003}),
\begin{equation}
\log \VNFW = 2 \C - \log[g(c)] - \log(\frac{h}{2}),
\end{equation}
where
\begin{equation}
g(c) = \frac{c^2}{\ln(1+c)-c/(1+c)},
\end{equation}
$h=H_{0}/100 \kmsMpc$ and $\C=1.61$ for the 3 year WMAP parameters (\citeauthor{Spergel_2007} \citeyear{Spergel_2007}).
We adopt $h=0.75$. We refer the reader to \cite{McGaugh_2007} for more details.

To be able to compare any discrepancies in shape, we scale both the rotation curves of our galaxies
and those of the adopted CDM halos to the velocity $V_{0.3}$ at the radius $R_{0.3}$,
where $R_{0.3}$ is the radius where the logarithmic slope of the curve is $d{\rm{log}}V/d{\rm{log}}R$$=$0.3.
As discussed in \cite{Hayashi_2006}, the NFW curves are well resolved at the scaling
radius $R_{0.3}$ (corresponding to $\sim$$0.4R_{\rm s}$ where $R_{\rm s}$ is given in Eq.~\ref{eq:3_1}) as their asymptotic
slopes are about $d{\rm{log}}V/d{\rm{log}}R$$=$0.5.
In addition, this scaling radius is also well determined in observed rotation curves
since it lies between the inner linear ($d{\rm{log}}V/d{\rm{log}}R$$=$1) and the
outer flat ($d{\rm{log}}V/d{\rm{log}}R$$=$0) regions of the rotation curves of most disk galaxies
(\citeauthor{Hayashi_2006} \citeyear{Hayashi_2006}).
This also holds for our sample galaxies, except for IC 2574 where the outermost logarithmic slope
is still larger than 0.3. In the case of IC 2574, we scale the rotation curve to the maximum radius $R_{\rm{max}}$ where
the last data point is measured, and corresponding maximum rotation velocity.

We plot the scaled rotation curves in the left panel of Fig.~\ref{THINGS_NFW_comp_min_opt_disk}. These rotation curves
are not corrected for baryons, and assume the minimum disk model as described in Section~\ref{DM_modeling}.
Similarly, in the right panel of Fig.~\ref{THINGS_NFW_comp_min_opt_disk}, we plot the scaled rotation curves corrected
for baryons derived from Section~\ref{Baryons}. Although the rotation curves corrected for baryons
increase less steeply than the curves assuming a minimum disk, they are
very similar. This directly shows that using the minimum disk assumption gives
a good description of the dark matter distribution in our galaxies.
In Fig.~\ref{THINGS_NFW_comp_min_opt_disk},
the CDM rotation curves with $V_{200}$ less than 110 \kms\ are represented
by dotted lines. Of our sample galaxies, IC 2574 has the largest maximum
rotation velocity of about 80 \kms. Therefore, the CDM rotation curve
with $V_{200}$=$110$ \kms\ is a hard upper limit for our galaxies assuming that $V_{\rm{max}}\sim V_{200}$.
We also overplot the best fits of pseudo-isothermal models (dashed lines) derived using
the minimum disk assumption and derived \MLsps\ in Fig.~\ref{THINGS_NFW_comp_min_opt_disk}.

As can be seen in Fig.~\ref{THINGS_NFW_comp_min_opt_disk},
the rotation curve shapes of the galaxies are similar and consistent with those of pseudo-isothermal
halo models. However, they are inconsistent with those of \LCDM\ simulations.
The rotation curves of \LCDM\ simulations rise too steeply to match the observations.
The difference in rotation curve shapes between our galaxies and
\LCDM\ simulations is particularly prominent in the inner regions of galaxies
i.e., at radii less than $R_{0.3}$. This difference is further enhanced for
the CDM rotation curves with $V_{200}$ less than 110 \kms.
In conclusion, the solid body-like rotation curves of our galaxies
rise too slowly to reflect the cusp-like dark matter distribution in CDM halos.

\subsection{Dark matter density profile}
\label{DM_profiles}
Direct conversion of the galaxy rotation curve to the mass density profile
allows us to examine the radial matter distribution in the galaxy. In particular,
the measured inner slope of the density profile is critical for resolving the
``cusp/core'' problem at the galaxy center.
The Poisson equation ($\nabla^{2}\Phi$ = $4\pi G\rho$, where $\Phi = -GM/R$) can be used
for the conversion under the assumption of a spherical mass distribution. The mass density
$\rho$ is directly derived from the rotation curve $V(R)$, as follows (see \citeauthor{deBlok_2001} \citeyear{deBlok_2001} for more details),
\begin{equation}
\rho(R) = \frac{1}{4\pi G}\Biggl[2\frac{V}{R}\frac{\partial V}{\partial R} + \Biggl(\frac{V}{R}\Biggr)^{2}\Biggr].
\label{rho_poisson}
\end{equation}

Using Eq.~\ref{rho_poisson}, we directly convert the total rotation curves into mass density profiles.
Here, we use the minimum disk hypothesis (i.e., ignores baryons). As already discussed in Section~\ref{DM_fraction_THINGS_properties}, our
galaxies are mostly dark matter-dominated and this ``minimum disk'' assumption
is a good approximation in describing their dynamics.
Particularly useful is the fact that it gives a hard upper limit to the dark matter density.

In this way, we derive the mass density profiles of the 7 THINGS dwarf galaxies and
present them in the Appendix. We also derive the mass density profiles using
the scaled rotation curves derived assuming minimum disk in
Fig.~\ref{THINGS_NFW_comp_min_opt_disk}, and plot them in Fig.~\ref{THINGS_DM_density_profile}.
The best fits of the NFW and pseudo-isothermal models are also overplotted.
Despite the scatter, the derived mass density profiles are more consistent with the
pseudo-isothermal models as shown in Fig.~\ref{THINGS_DM_density_profile}.

\begin{figure}
\epsscale{1.0}
\includegraphics[angle=0,width=0.49\textwidth,bb=27 159.5 578 690,clip=]{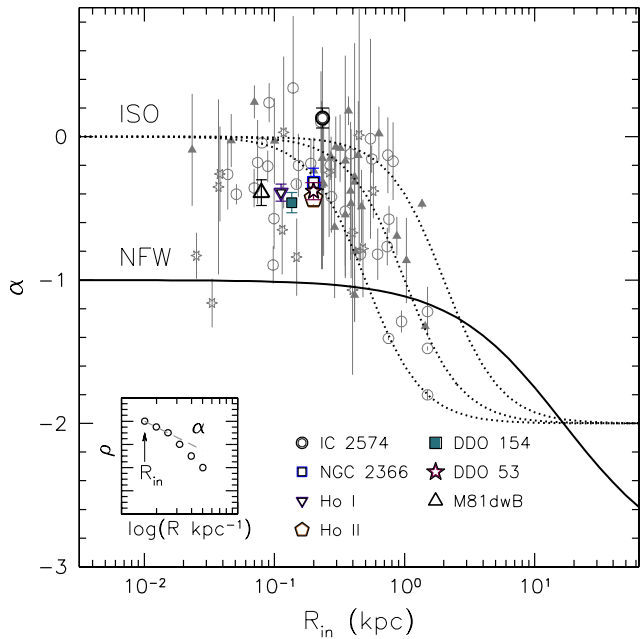}
\caption{The inner slope of the dark matter density profile plotted against the
radius of the innermost point. The inner density slope $\alpha$ is measured by a least squares fit
to the inner data point as described in the small figure.
The inner-slopes of the mass density profiles of the 7 THINGS dwarf galaxies
are overplotted with earlier papers and they are consistent with previous measurements of LSB galaxies.
The pseudo-isothermal model is preferred over the NFW model to explain the observational data.
Gray symbols: open circles (\citeauthor{deBlok_2001} \citeyear{deBlok_2001});
triangles (\citeauthor{deBlok_2002} \citeyear{deBlok_2002});
open stars (\citeauthor{Swaters_2003} \citeyear{Swaters_2003}).
See Section~\ref{DM_profiles} for more discussions.
\label{Alpha_Rinner}}
\end{figure}

To quantify the degree of concentration of the dark matter distribution towards the galaxy center,
we measure the logarithmic inner slope of the density profile. For this
measurement, we first need to determine a break-radius where the slope
changes most rapidly.
The inner density slope is then measured by performing a
least squares fit to the data points within the break-radius. For the uncertainty,
we re-measure the slope twice, including the first data point outside the break-radius
and excluding the data point at the break radius. The mean difference between
these two slopes is adopted as the slope uncertainty $\Delta \alpha$. The measured
slope $\alpha$ and slope uncertainty $\Delta \alpha$ of the galaxies
are shown in the Appendix. In addition, we overplot
the mass density profiles of NFW and pseudo-isothermal halo models which are
best fitted to the rotation curves of the galaxies.
From this, we find that the mean value of the inner density slopes for the galaxies
is $\alpha$$=$$-0.29\pm 0.07$ (and $-0.27\pm 0.07$ without Ho I which has a low inclination.
See Section~\ref{Deriving_Rotcurs} for details). 
These rather flat slopes are in very good agreement with the value of $\alpha= -0.2\pm 0.2$
found in the earlier work of \citeauthor{deBlok_2001} (\citeyear{deBlok_2001}; see also \citeauthor{deBlok_2002} \citeyear{deBlok_2002}) for
a larger number of LSB galaxies. They are, however, in contrast with the steep slope
of $\sim$$-0.8$ predicted by \LCDM\ simulations (e.g., \citeauthor{Stadel_2009} \citeyear{Stadel_2009};
\citeauthor{Navarro_2010} \citeyear{Navarro_2010}) as well as those by the classical simulations
(e.g., Navarro, Frenk \& White \citeyear{NFW_1996}, \citeyear{NFW_1997}).
This implies that the sample galaxies show slightly increasing or even constant density profiles towards their centers.

We also examine how the mass model differs when it is based on the hermite $h_{3}$ rotation curve
instead of the bulk one. For this, we use IC 2574 which shows strong non-circular motions close
to the center. As shown in the ``Mass density profile'' panel of Fig.~\ref{IC2574_VROT_BARYONS_ALPHA},
the mass density profile derived using the hermite $h_{3}$ rotation curve is found to be slightly
lower than that from the bulk rotation curve at the central regions. This is mainly due to the
lower hermite $h_{3}$ rotation velocity, resulting in smaller velocity gradients
${\partial V}/{\partial R}$ in Eq.~\ref{rho_poisson} and thus smaller densities. The measured inner
density slope is $\alpha$$=$$0.00\pm 0.19$ which is similar, within the error, to that ($\alpha$$=$$0.13\pm 0.07$) based on the bulk rotation curve.
This supports earlier studies that suggest that the effect of systematic non-circular motions in dwarf galaxies is not
enough to hide the central cusps (e.g., \citeauthor{Gentile_2004} \citeyear{Gentile_2004};
\citeauthor{Trachternach_2008} \citeyear{Trachternach_2008};
\citeauthor{van_Eymeren_2009} \citeyear{van_Eymeren_2009}).

In Fig.~\ref{Alpha_Rinner}, we plot the logarithmic inner density slope $\alpha$ against
resolution of a rotation curve. At high resolutions ($R_{\rm in}<$ 1 kpc) the slopes of
the NFW and pseudo-isothermal halo models can be clearly distinguished but at low
resolutions ($R_{\rm in}\sim$1 kpc) the slopes of the two models are approximately equal
(\citeauthor{deBlok_2001} \citeyear{deBlok_2001}).
Because of their proximity ($\sim$4 Mpc) and their highly-resolved
rotation curves, the innermost radius of the rotation curves that can be
probed for our galaxies is about 0.1-0.2 kpc.
We also overplot the theoretical $\alpha - R_{\rm{in}}$ relations of NFW and pseudo-isothermal halo models
as solid and dotted lines, respectively. The highly-resolved rotation curves of our galaxies
(i.e., $R_{\rm{in}}\sim$0.2 kpc) deviate significantly from the prediction of NFW CDM models.
In particular, around $R_{\rm{in}} \sim$0.1 kpc where the predictions of the two halo models
are clearly distinct, the $\alpha - R_{\rm{in}}$ trend of our galaxies is more consistent with
those of pseudo-isothermal halo models.

\section{Conclusions}
\label{Summary}

In this paper we have presented high-resolution mass models of the 7 dwarf galaxies,
IC 2574, NGC 2366, Ho I, Ho II, DDO 53, DDO 154 and M81dwB from the THINGS survey,
and examined their dark matter distribution by comparison with classical \LCDM\ simulations.
The THINGS high-resolution data significantly reduce observational systematic effects,
such as beam smearing, center offset and non-circular motions. 
When deriving the rotation curves, we used various types of velocity fields,
such as intensity-weighted mean, peak, single Gaussian,
hermite $h_{3}$ and bulk velocity fields, and compared the results.
In particular the bulk velocity field was able to efficiently remove
small-scale random motions and allowed us to better determine the total kinematics
of the galaxies.

We also found that the relation between the total baryonic mass (stars $+$ gas) and
the maximum rotation velocity of the galaxies is roughly consistent with the
Baryonic Tully$\--$Fisher (BTF) relation calibrated from a larger sample of low mass galaxies. 
Especially, the inclination values derived if one takes the BTF relation at face value are not significantly
different from those derived from a tilted-ring analysis. This implies that the BTF
relation can be used as an alternative way for deriving inclinations of galaxies
for which it is difficult to apply a tilted-ring analysis.

We derived the mass models of baryons, and subtracted them from the total kinematics.
For the stellar component, we used SINGS 3.6$\mu$m and optical data determined by
the stellar mass--to--light ratio \ML\ for the 3.6$\mu$m band.
For the purpose of our study we use the 3.6$\mu$m {\it Spitzer} images to estimate the mass of
the old stellar population in our target galaxies. Even though this band may contain some
dust emission features, we consider it to be the best consistent tracer of the stellar masses
(see discussion in \citeauthor{Leroy_2008} \citeyear{Leroy_2008}, \citeauthor{deBlok_2008} \citeyear{deBlok_2008}
and \citeauthor{Oh_2008} \citeyear{Oh_2008}).
These therefore allow us to estimate the old stellar population that
dominates the stellar continuum emission in the infrared regime. Although our sample
dwarf galaxies are dark matter dominated as indicated by their low baryonic fraction,
the population synthesis \MLsps\ values gave slightly better or similar fits than not only
the maximum disk but also the minimum ($+$gas) disk assumptions in describing the stellar
component.

With the help of the well determined total kinematics and the mass models of baryons,
we were able to accurately constrain the dark matter distribution
in the galaxies. From this, we found a significant discrepancy in the dark matter distribution
between the THINGS dwarf galaxies and classical dark-matter-only cosmological simulations
both in the rotation curve shape and the inner slope $\alpha$ of the mass density
profiles. The rotation curves of the galaxies rise less steeply to be consistent with the cusp
feature at the centers. In addition the mean value of the inner slopes of the mass density
profiles is $\alpha$$=$$-0.29\pm 0.07$ (and $-0.27\pm 0.07$ without Ho I which has a low inclination),
significantly deviating from $\sim$$-1$ inferred from dark-matter-only simulations. Considering the fact
that the bulk rotation curves which show the most rapid increase compared to the others
(particularly in the inner regions) were used, the results provide good evidence that the
central dark matter distribution in dwarf galaxies is not cusp-like, as suggested by earlier studies.

It is most likely that both the lack of resolution and the absence of baryonic physics in the
older simulations play the dominant role for the discrepancy.
In order to distinguish the core and cusp-like models clearly, it is indispensable for the
simulations to resolve scales smaller than 1 kpc (\citeauthor{deBlok_2008} \citeyear{deBlok_2008}).
In addition baryons are dynamically important in the central regions and their feedback
like gas outflows driven by star formation or supernovae may affect the dark matter
distribution indirectly through gravitational interaction on galaxy scales. Therefore
improvements on both resolution and description of baryonic feedback processes in simulations
will provide a major contribution toward a solution for the ``cusp/core'' problem.
We refer a companion paper (\citeauthor{Oh_2011} \citeyear{Oh_2011}) where we compare the results from the latest high-resolution
cosmological $\rm N-body+Smoothed$ Particle Hydrodynamic (SPH) simulations by \cite{Governato_2010},
of dwarf galaxies that include effects of baryonic feedback processes which result in shallower
slopes of $\alpha$.

\acknowledgements
SHOH acknowledges financial support from the South African Square
Kilometre Array Project. The work of WJGdB is based upon research supported by the South
African Research Chairs Initiative of the Department of Science and
Technology and National Research Foundation.
This research has made use of the NASA/IPAC
Extragalactic Database (NED) which is operated by the Jet Propulsion
Laboratory, California Institute of Technology, under contract with the
National Aeronautics and Space Administration.
This publication makes use of data products from the Two Micron All
Sky Survey, which is a joint project of the University of
Massachusetts and the Infrared Processing and Analysis
Center/California Institute of Technology, funded by the National
Aeronautics and Space Administration and the National Science
Foundation.

\bibliography{THINGS_Dwarfs_DM_profiles_2col}

\begin{thebibliography}{77}
\expandafter\ifx\csname natexlab\endcsname\relax\def\natexlab#1{#1}\fi
\expandafter\ifx\csname href\endcsname\relax
  \def\href#1#2{}\fi
\expandafter\ifx\csname urllinklabel\endcsname\relax
  \def\urllinklabel{[LINK]}\fi
\expandafter\ifx\csname adsurllinklabel\endcsname\relax
  \def\adsurllinklabel{[ADS]}\fi

\bibitem[{{Begeman}(1989)}]{Begeman_1989}
{Begeman}, K. 1989, \aap, 223, 47


\bibitem[{{Begeman} {et~al.}(1991){Begeman}, {Broeils}, \&
  {Sanders}}]{Begeman_1991}
{Begeman}, K., {Broeils}, A.~H., \& {Sanders}, R.~H. 1991, \mnras, 249, 523


\bibitem[{{Bell} \& {de Jong}(2001)}]{Bell_2001}
{Bell}, E.~F. \& {de Jong}, R.~S. 2001, \apj, 550, 212


\bibitem[{{Bolatto} {et~al.}(2002){Bolatto}, {Simon}, {Leroy}, \&
  {Blitz}}]{Bolatto_2002}
{Bolatto}, A.~D., {Simon}, J.~D., {Leroy}, A., \& {Blitz}, L. 2002, \apj, 565,
  238


\bibitem[{Bosma(1978)}]{Bosma_1978}
Bosma, A. 1978 (PhD Thesis, University of Groningen)


\bibitem[{{Bruzual} \& {Charlot}(2003)}]{Bruzual_Charlot_2003}
{Bruzual}, G. \& {Charlot}, S. 2003, \mnras, 344, 1000


\bibitem[{{Bureau} \& {Carignan}(2002)}]{Bureau_2002}
{Bureau}, M. \& {Carignan}, C. 2002, \aj, 123, 1316


\bibitem[{{Burkert}(1995)}]{Burkert_1995}
{Burkert}, A. 1995, \apjl, 447, L25


\bibitem[{{Carignan} \& {Freeman}(1988)}]{Carignan_1988}
{Carignan}, C. \& {Freeman}, K.~C. 1988, \apj, 332, L33


\bibitem[{{Carignan} \& {Purton}(1998)}]{Carignan_1998}
{Carignan}, C. \& {Purton}, C. 1998, \apj, 506, 125


\bibitem[{{de Blok}(2010)}]{deBlok_2010}
{de Blok}, W.~J.~G. 2010, AdAst, 2010, 1


\bibitem[{{de Blok} \& {Bosma}(2002)}]{deBlok_2002}
{de Blok}, W.~J.~G. \& {Bosma}, A. 2002, \aap, 385, 816


\bibitem[{{de Blok} {et~al.}(2003){de Blok}, {Bosma}, \&
  {McGaugh}}]{deBlok_2003}
{de Blok}, W.~J.~G., {Bosma}, A., \& {McGaugh}, S.~S. 2003, \mnras, 340, 657


\bibitem[{{de Blok} \& {McGaugh}(1997)}]{deBlok_1997}
{de Blok}, W.~J.~G. \& {McGaugh}, S.~S. 1997, \mnras, 290, 533


\bibitem[{{de Blok} {et~al.}(2001){de Blok}, {McGaugh}, {Bosma}, \&
  {Rubin}}]{deBlok_2001}
{de Blok}, W.~J.~G., {McGaugh}, S.~S., {Bosma}, A., \& {Rubin}, V.~C. 2001,
  \apj, 552, 23


\bibitem[{{de Blok} {et~al.}(2008){de Blok}, {Walter}, {Brinks},
  {Trachternach}, {Oh}, \& {Kennicutt}}]{deBlok_2008}
{de Blok}, W.~J.~G., {Walter}, F., {Brinks}, E., {Trachternach}, C., {Oh},
  S.-H., \& {Kennicutt}, R.~C. 2008, \aj, 136, 2648


\bibitem[{{De Rijcke} \& \etal(2007)}]{De_Rijcke_2007}
{De Rijcke}, S. \& \etal. 2007, \apj, 659, 1172


\bibitem[{{Diemand} {et~al.}(2008){Diemand}, {Kuhlen}, {Madau}, {Zemp},
  {Moore}, \& {Potter}}]{Diemand_2008}
{Diemand}, J., {Kuhlen}, M., {Madau}, P., {Zemp}, M., {Moore}, B., \& {Potter},
  D.~{Stadel}, J. 2008, Nature, 454, 735


\bibitem[{{Donato} {et~al.}(2009){Donato}, {Gentile}, {Salucci}, {Frigerio
  Martins}, {Wilkinson}, {Gilmore}, {Grebel}, {Koch}, \& {Wyse}}]{Donato_2009}
{Donato}, F., {Gentile}, G., {Salucci}, P., {Frigerio Martins}, C.,
  {Wilkinson}, M.~I., {Gilmore}, G., {Grebel}, E.~K., {Koch}, A., \& {Wyse}, R.
  2009, \mnras, 397, 1169


\bibitem[{{Dubinski} \& {Carlberg}(1991)}]{Dubinski_1991}
{Dubinski}, J. \& {Carlberg}, R.~G. 1991, \apj, 378, 496


\bibitem[{{Dutton} {et~al.}(2010){Dutton}, {Conroy}, {van den Bosch}, {Prada},
  \& {More}}]{Dutton_2010}
{Dutton}, A.~A., {Conroy}, C., {van den Bosch}, F.~C., {Prada}, F., \& {More},
  S. 2010, arXiv:1004.4626v1


\bibitem[{{Flores} \& {Primack}(1994)}]{Flores_1994}
{Flores}, R.~A. \& {Primack}, J.~R. 1994, \apj, 427, L1


\bibitem[{{Gentile} {et~al.}(2004){Gentile}, {Salucci}, {Klein}, {Vergani}, \&
  {Kalberla}}]{Gentile_2004}
{Gentile}, G., {Salucci}, P., {Klein}, U., {Vergani}, D., \& {Kalberla}, P.
  2004, \mnras, 351, 903


\bibitem[{{Ghigna} {et~al.}(2000){Ghigna}, {Moore}, {Governato}, {Lake},
  {Quinn}, \& {Stadel}}]{Ghigna_2000}
{Ghigna}, S., {Moore}, B., {Governato}, F., {Lake}, G., {Quinn}, T., \&
  {Stadel}, J. 2000, \apj, 544, 616


\bibitem[{{Governato} {et~al.}(2010){Governato}, {Brook}, {Mayer}, {Brooks},
  {Rhee}, {Wadsley}, {Jonsson}, {Willman}, {Stinson}, {Quinn}, \&
  {Madau}}]{Governato_2010}
{Governato}, F., {Brook}, C., {Mayer}, L., {Brooks}, A., {Rhee}, G., {Wadsley},
  J., {Jonsson}, P., {Willman}, B., {Stinson}, G., {Quinn}, T., \& {Madau}, P.
  2010, Nature, 463, 203


\bibitem[{{Graham} {et~al.}(2006){Graham}, {Merritt}, {Moore}, {Diemand}, \&
  {Terzi{\'c}}}]{Graham_2006}
{Graham}, A.~W., {Merritt}, D., {Moore}, B., {Diemand}, J., \& {Terzi{\'c}}, B.
  2006, \aj, 132, 2701


\bibitem[{{Guo} {et~al.}(2010){Guo}, {White}, {Li}, \&
  {Boylan-Kolchin}}]{Guo_2010}
{Guo}, Q., {White}, S.~D.~M., {Li}, C., \& {Boylan-Kolchin}, M. 2010, \mnras,
  404, 1111


\bibitem[{{Hayashi} \& {Navarro}(2006)}]{Hayashi_2006}
{Hayashi}, E. \& {Navarro}, J.~F. 2006, \mnras, 373, 1117


\bibitem[{{Hunter} {et~al.}(2001){Hunter}, {Elmegreen}, \& {van
  Woerden}}]{Hunter_2001}
{Hunter}, D., {Elmegreen}, B., \& {van Woerden}, H. 2001, \apj, 556, 773


\bibitem[{{Kennicutt} {et~al.}(2003){Kennicutt}, {Armus}, {Bendo}, {Calzetti},
  {Dale}, {Draine}, {Engelbracht}, {Gordon}, {Grauer}, \&
  {Helou}}]{Kennicutt_2003}
{Kennicutt}, R.~C., {Armus}, L., {Bendo}, G., {Calzetti}, D., {Dale}, D.~A.,
  {Draine}, B.~T., {Engelbracht}, C.~W., {Gordon}, K.~D., {Grauer}, A.~D., \&
  {Helou}, G. 2003, \pasp, 115, 928


\bibitem[{{Klypin} {et~al.}(2001){Klypin}, {Kravtsov}, {Bullok}, \&
  {Primack}}]{Klypin_2001}
{Klypin}, A., {Kravtsov}, A.~V., {Bullok}, J.~S., \& {Primack}, J.~R. 2001,
  \apj, 554, 903


\bibitem[{{Kregel} {et~al.}(2002){Kregel}, {van der Kruit}, \& {de
  Grijs}}]{Kregel_2002}
{Kregel}, M., {van der Kruit}, P.~C., \& {de Grijs}, R. 2002, \mnras, 334, 646


\bibitem[{{Leroy} {et~al.}(2008){Leroy}, {Walter}, {Brinks}, {Bigiel}, {de
  Blok}, {Madore}, \& {Thornley}}]{Leroy_2008}
{Leroy}, A.~K., {Walter}, F., {Brinks}, E., {Bigiel}, F., {de Blok}, W.~J.~G.,
  {Madore}, B., \& {Thornley}, M.~D. 2008, \aj, 136, 2782


\bibitem[{{Martimbeau} {et~al.}(1994){Martimbeau}, {Carignan}, \&
  Roy}]{Martimbeau_1994}
{Martimbeau}, N., {Carignan}, C., \& Roy, J.-R. 1994, \aj, 107, 543


\bibitem[{{McGaugh}(2004)}]{McGaugh_2004}
{McGaugh}, S.~S. 2004, \apj, 609, 652


\bibitem[{{McGaugh} {et~al.}(2003){McGaugh}, {Barker}, \& {de
  Blok}}]{McGaugh_2003}
{McGaugh}, S.~S., {Barker}, M.~K., \& {de Blok}, W.~J.~G. 2003, \apj, 584, 566


\bibitem[{{McGaugh} {et~al.}(2007){McGaugh}, {de Blok}, {Schombert}, {Kuzio de
  Naray}, \& {Kim}}]{McGaugh_2007}
{McGaugh}, S.~S., {de Blok}, W.~J.~G., {Schombert}, J.~M., {Kuzio de Naray},
  R., \& {Kim}, J.~H. 2007, \apj, 659, 149


\bibitem[{{McGaugh} {et~al.}(2000){McGaugh}, {Schombert}, {Bothun}, \& {de
  Blok}}]{McGaugh_2000}
{McGaugh}, S.~S., {Schombert}, J.~M., {Bothun}, G.~D., \& {de Blok}, W.~J.~G.
  2000, \apj, 533, L99


\bibitem[{{McGaugh} {et~al.}(2005){McGaugh}, {Schombert}, {Bothun}, \& {de
  Blok}}]{McGaugh_2005}
---. 2005, \apj, 632, 859


\bibitem[{{Moore}(1994)}]{Moore_1994}
{Moore}, B. 1994, Nature, 370, 629


\bibitem[{{Moore} {et~al.}(1999){Moore}, {Quinn}, {Governato}, {Stadel}, \&
  {Lake}}]{Moore_1999a}
{Moore}, B., {Quinn}, T., {Governato}, F., {Stadel}, J., \& {Lake}, G. 1999,
  \mnras, 310, 1147


\bibitem[{{Navarro} {et~al.}(1996){Navarro}, {Frenk}, \& {White}}]{NFW_1996}
{Navarro}, J.~F., {Frenk}, C.~S., \& {White}, S.~D.~M. 1996, \apj, 462, 563


\bibitem[{{Navarro} {et~al.}(1997){Navarro}, {Frenk}, \& {White}}]{NFW_1997}
---. 1997, \apj, 490, 493


\bibitem[{{Navarro} {et~al.}(2004){Navarro}, {Hayashi}, {Power}, {Jenkins},
  {Frenk}, {White}, {Springel}, {Stadel}, \& {Quinn}}]{Navarro_2004}
{Navarro}, J.~F., {Hayashi}, E., {Power}, C., {Jenkins}, A.~R., {Frenk}, C.~S.,
  {White}, S.~D.~M., {Springel}, V., {Stadel}, J., \& {Quinn}, T.~R. 2004,
  \mnras, 349, 1039


\bibitem[{{Navarro} {et~al.}(2010){Navarro}, {Ludlow}, {Springel}, {Wang},
  {Vogelsberger}, {White}, {Jenkins}, {Frenk}, \& {Helmi}}]{Navarro_2010}
{Navarro}, J.~F., {Ludlow}, A., {Springel}, V., {Wang}, J., {Vogelsberger}, M.,
  {White}, S.~D.~M., {Jenkins}, A., {Frenk}, C.~S., \& {Helmi}, A. 2010,
  \mnras, 402, 21


\bibitem[{{Oh} {et~al.}(2011){Oh}, {Brook}, {Governato}, {Brinks}, {Mayer}, {de
  Blok}, {Brooks}, \& {Walter}}]{Oh_2011}
{Oh}, S., {Brook}, C., {Governato}, F., {Brinks}, E., {Mayer}, L., {de Blok},
  W.~J.~G., {Brooks}, A., \& {Walter}, F. 2011, ArXiv e-prints, AJ accepted


\bibitem[{{Oh} {et~al.}(2008){Oh}, {de Blok}, {Walter}, {Brinks}, \&
  {Kennicutt}}]{Oh_2008}
{Oh}, S.-H., {de Blok}, W.~J.~G., {Walter}, F., {Brinks}, E., \& {Kennicutt},
  R.~C. 2008, \aj, 136, 2761


\bibitem[{{Ott} {et~al.}(2001){Ott}, {Walter}, {Brinks}, D., {Dirsch}, \&
  {Klein}}]{Ott_2001}
{Ott}, J., {Walter}, F., {Brinks}, E., D., V.~S., {Dirsch}, B., \& {Klein}, U.
  2001, \aj, 122, 3070


\bibitem[{{Power} {et~al.}(2002){Power}, {Navarro}, {Jenkins}, {Frenk},
  {White}, {Springel}, {Stadel}, \& {Quinn}}]{Power_2002}
{Power}, C., {Navarro}, J.~F., {Jenkins}, A.~R., {Frenk}, C.~S., {White},
  S.~D.~M., {Springel}, V., {Stadel}, J., \& {Quinn}, T. 2002, \mnras, 338, 14


\bibitem[{{Prada} \& {Burkert}(2002)}]{Prada_2002}
{Prada}, F. \& {Burkert}, A. 2002, \apjl, 564, 73


\bibitem[{{Reed} {et~al.}(2005){Reed}, {Governato}, {Verde}, {Gardner},
  {Quinn}, {Stadel}, \& {Lake}}]{Reed_2005}
{Reed}, D., {Governato}, F., {Verde}, L., {Gardner}, J., {Quinn}, T., {Stadel},
  J.~{Merritt}, D., \& {Lake}, G. 2005, \mnras, 357, 82


\bibitem[{{Rhee} {et~al.}(2004){Rhee}, {Valenzuela}, {Klypin}, {Holtzman}, \&
  {Moorthy}}]{Rhee_2004}
{Rhee}, G., {Valenzuela}, O., {Klypin}, A., {Holtzman}, J., \& {Moorthy}, B.
  2004, \apj, 617, 1059


\bibitem[{{Schoenmakers}(1999)}]{Schoenmakers_1999}
{Schoenmakers}, R.~H.~M. 1999


\bibitem[{{Schoenmakers} {et~al.}(1997){Schoenmakers}, {Franx}, \& {de
  Zeeuw}}]{Schoenmakers_1997}
{Schoenmakers}, R.~H.~M., {Franx}, M., \& {de Zeeuw}, P.~T. 1997, \mnras, 292,
  349


\bibitem[{{Simon} {et~al.}(2003){Simon}, {Bolatto}, {Leroy}, \&
  {Blitz}}]{Simon_2003}
{Simon}, J.~D., {Bolatto}, A.~D., {Leroy}, A., \& {Blitz}, L. 2003, \apj, 596,
  957


\bibitem[{{Spergel} {et~al.}(2007){Spergel}, {Bean}, {Dor\'e}, {Nolta},
  {Bennett}, {Dunkley}, {Hinshaw}, N., {Komatsu}, {Page}, {Peiris}, {Verde},
  {Halpern}, {Hill}, {Kogut}, {Limon}, {Meyer}, {Odegard}, {Tucker}, {Weiland},
  {Wollack}, \& {Wright}}]{Spergel_2007}
{Spergel}, D.~N., {Bean}, R., {Dor\'e}, O., {Nolta}, M.~R., {Bennett}, C.~L.,
  {Dunkley}, J., {Hinshaw}, G., N., J., {Komatsu}, E., {Page}, L., {Peiris},
  H.~V., {Verde}, L., {Halpern}, M., {Hill}, R.~S., {Kogut}, A., {Limon}, M.,
  {Meyer}, S.~S., {Odegard}, N., {Tucker}, G.~S., {Weiland}, J.~L., {Wollack},
  E., \& {Wright}, E.~L. 2007, \apjs, 170, 377


\bibitem[{{Spergel} {et~al.}(2003){Spergel}, {Verde}, {Peiris}, {Komatsu},
  {Nolta}, {Bennett}, {Halpern}, {Hinshaw}, {Jarosik}, {Kogut}, {Limon},
  {Meyer}, {Page}, {Tucker}, {Weiland}, {Wollack}, \& {Wright}}]{Spergel_2003}
{Spergel}, D.~N., {Verde}, L., {Peiris}, H.~V., {Komatsu}, E., {Nolta}, M.~R.,
  {Bennett}, C.~L., {Halpern}, M., {Hinshaw}, G., {Jarosik}, N., {Kogut}, A.,
  {Limon}, M., {Meyer}, S.~S., {Page}, L., {Tucker}, G.~S., {Weiland}, J.~L.,
  {Wollack}, E., \& {Wright}, E.~L. 2003, \apjs, 148, 175


\bibitem[{{Stadel} {et~al.}(2009){Stadel}, {Potter}, {Moore}, {Diemand},
  {Madau}, {Zemp}, {Kuhlen}, \& {Quilis}}]{Stadel_2009}
{Stadel}, J., {Potter}, D., {Moore}, B., {Diemand}, J., {Madau}, P., {Zemp},
  M., {Kuhlen}, M., \& {Quilis}, V. 2009, \mnras, 398, L21


\bibitem[{{Stark} {et~al.}(2009){Stark}, {McGaugh}, \& {Swaters}}]{Stark_2009}
{Stark}, D.~V., {McGaugh}, S.~S., \& {Swaters}, R.~A. 2009, \aj, 139, 312


\bibitem[{{Stoehr} {et~al.}(2003){Stoehr}, {White}, {Springel}, {Tormen}, \&
  {Yoshida}}]{Stoehr_2003}
{Stoehr}, F., {White}, S.~D.~M., {Springel}, V., {Tormen}, G., \& {Yoshida}, N.
  2003, \mnras, 345, 1313


\bibitem[{{Swaters}(1999)}]{Swaters_1999}
{Swaters}, R.~A. 1999 (PhD Thesis, University of Groningen)


\bibitem[{{Swaters} {et~al.}(2003){Swaters}, {Madore}, {van den Bosch}, \&
  {Balcells}}]{Swaters_2003}
{Swaters}, R.~A., {Madore}, B.~F., {van den Bosch}, F.~C., \& {Balcells}, M.
  2003, \apj, 583, 732


\bibitem[{{Trachternach} {et~al.}(2008){Trachternach}, {de Blok}, {Brinks},
  {Walter}, \& {Kennicutt}}]{Trachternach_2008}
{Trachternach}, C., {de Blok}, W.~J.~G., {Brinks}, E., {Walter}, F., \&
  {Kennicutt}, R.~C. 2008, \aj, 136, 2720


\bibitem[{{van Albada} \& {Sancisi}(1986)}]{van_Albada_1986}
{van Albada}, T.~S. \& {Sancisi}, R. 1986, Philos. Trans. R. Soc. London A,
  320, 447


\bibitem[{{van den Bosch} {et~al.}(2000){van den Bosch}, {Robertson},
  {Dalcanton}, \& {de Blok}}]{van_den_Bosch_2000}
{van den Bosch}, F.~C., {Robertson}, B.~E., {Dalcanton}, J.~J., \& {de Blok},
  W.~J.~G. 2000, \aj, 119, 1579


\bibitem[{{van den Bosch} \& {Swaters}(2001)}]{van_den_Bosch_2001}
{van den Bosch}, F.~C. \& {Swaters}, R.~A. 2001, \mnras, 325, 1017


\bibitem[{{van der Kruit} \& {Searle}(1981)}]{van_der_Kruit_1981}
{van der Kruit}, P.~C. \& {Searle}, L. 1981, \aap, 95, 105


\bibitem[{{van der Marel} \& {Franx}(1993)}]{van_der_Marel1993}
{van der Marel}, R.~P. \& {Franx}, M. 1993, \apj, 407, 525


\bibitem[{{van Eymeren} {et~al.}(2009){van Eymeren}, {Trachternach},
  {Koribalski}, \& {Dettmar}}]{van_Eymeren_2009}
{van Eymeren}, J., {Trachternach}, C., {Koribalski}, B.~S., \& {Dettmar}, R.
  2009, \aap, 505, 1


\bibitem[{{Verheijen}(2001)}]{Verheijen_2001}
{Verheijen}, M.~A.~W. 2001, \apj, 563, 694


\bibitem[{{Walter} \& {Brinks}(1999)}]{Walter_Brinks_1999}
{Walter}, F. \& {Brinks}, E. 1999, \aj, 118, 273


\bibitem[{{Walter} {et~al.}(2008){Walter}, {Brinks}, {de Blok}, {Bigiel},
  {Kennicutt}, {Thornley}, \& {Leroy}}]{Walter_2008}
{Walter}, F., {Brinks}, E., {de Blok}, W.~J.~G., {Bigiel}, F., {Kennicutt},
  R.~C., {Thornley}, M., \& {Leroy}, A. 2008, \aj, 136, 2563


\bibitem[{{Walter} {et~al.}(1998){Walter}, {Kerp}, {Duric}, {Brinks}, \&
  {Klein}}]{Walter_1998}
{Walter}, F., {Kerp}, J., {Duric}, N., {Brinks}, E., \& {Klein}, U. 1998, \apj,
  502, 143


\bibitem[{{Warner} {et~al.}(1973){Warner}, {Wright}, \&
  {Baldwin}}]{Warner_1973}
{Warner}, P.~J., {Wright}, M.~C.~H., \& {Baldwin}, J.~E. 1973, \mnras, 163, 163


\bibitem[{{Weldrake} {et~al.}(2003){Weldrake}, {de Blok}, \&
  {Walter}}]{Weldrake_2003}
{Weldrake}, D.~T.~F., {de Blok}, W.~J.~G., \& {Walter}, F. 2003, \mnras, 340,
  12


\bibitem[{{Yun} {et~al.}(1994){Yun}, {Ho}, \& {Lo}}]{Yun_1994}
{Yun}, M.~S., {Ho}, P.~T.~P., \& {Lo}, K.~Y. 1994, \nat, 372, 530


\bibitem[{{Zackrisson} {et~al.}(2006){Zackrisson}, {Bergvall}, {Marquart}, \&
  {\"Ostlin}}]{Zackrisson_2006}
{Zackrisson}, E., {Bergvall}, N., {Marquart}, T., \& {\"Ostlin}, G. 2006, \aap,
  452, 857


\end{thebibliography}

\clearpage

\clearpage

\renewcommand{\thesection}{Appendix}
\label{Appendix}
\section{Data and kinematic analysis}

\newcounter{appendix_section}
\renewcommand{\thesection}{A.\arabic{appendix_section}}
\renewcommand{\theequation}{A.\arabic{equation}}
\renewcommand{\thefigure}{A.\arabic{figure}}

In the following we present the data and kinematic analysis of 7 dwarf galaxies from ``The H{\sc i} Nearby Galaxy Survey (THINGS)''.
The kinematic analysis includes (1) the tilted ring model, (2) the harmonic analysis,
(3) the mass models of baryons and dark matter and (4) the dark matter density profile.
The following are general descriptions of the figures. \\

\noindent $\bullet$ {\bf Data$\--$} We show the total intensity maps in {\it Spitzer} IRAC 3.6$\mu$m, optical {\it B}, {\it R}\--bands
and H{\sc i} 21 cm. The latter can be used to directly derive the H{\sc i} surface density. The stellar surface density is based on
the {\it Spitzer} 3.6$\mu$m map and information about the optical colors (see main text for details). 
The $\rm 2^{nd}$ moment map showing the velocity dispersions of the H{\sc i}
profiles is also given. We then compare the five types of velocity fields
extracted from the H{\sc i} data cube: The intensity-weighted mean (IWM),
the Peak-intensity (PEAK), the Single Gaussian profile (SGFIT), the hermite {\it $h_{3}$} (HER3) and the
bulk velocity fields (BULK). Additionally, we show the extracted velocity field of
strong non-circular motions (NONC). We also show a major-axis position-velocity (P-V) diagram overlayed with
the derived bulk rotation curve corrected for inclination. For the extraction of the integrated H{\sc i} map,
the intensity-weighted mean velocity field, and the $\rm 2^{nd}$ moment map,
the natural-weighted data cube is used. See \cite{Walter_2008} for a detailed description of the data cubes. \\

\noindent $\bullet$ {\bf Rotation curves$\--$} The tilted ring model derived from the bulk
(or hermite $h_{3}$ for M81dwB) velocity field. Note that the black solid lines are not the
fits to the gray open circles. The open gray circles indicate the fit made with all ring parameters
``free''. The final rotation curves (black solid lines) are derived after several iterations. \\

\noindent $\bullet$ {\bf Asymmetric drift correction$\--$} For galaxies where
the velocity dispersion is comparable to the maximum rotation velocity, we correct for the asymmetric
drift following the method described in \cite{Bureau_2002}.
See Section~\ref{ADC} for a detailed description. \\

\noindent $\bullet$ {\bf Harmonic analysis$\--$} Harmonic expansion of the hermite $h_{3}$
and bulk velocity fields. Gray circles and black dots represent the results
from the hermite $h_{3}$ and the bulk velocity fields, respectively.
$c_0$ and $c_1$ are the systemic and the rotation velocities.
$c_2,c_3,s_1,s_2$, and $s_3$ components quantify non-circular motion components.
In the bottom-rightmost panel, we show a global elongation of the potential, $\epsilon_{pot}\,{\rm sin}\,2\phi_{2}$
calculated at each radius as described in \cite{Schoenmakers_1997} and \cite{Schoenmakers_1999}.
This measurement can be used as an additional test
for CDM halos (e.g., \citeauthor{Trachternach_2008} \citeyear{Trachternach_2008}).
The black solid and gray dashed lines indicate the average values
of the potential derived using the bulk and hermite $h_{3}$ velocity fields, respectively. \\

\noindent $\bullet$ {\bf Mass models of baryons$\--$}
{\bf (a):} Azimuthally averaged surface brightness
profiles in the 3.6$\mu$m, $R$, $V$, and $B$ bands (top to bottom) derived
assuming the tilted-ring parameters derived as above. These are not
corrected for inclination except for the 3.6$\mu$m. The lines shown are
fits to the data which are partly filled.
{\bf (b)(f):} Derived values of \ML\ in the $K$ and 3.6$\mu$m bands from \cite{Bruzual_Charlot_2003} population
synthesis models. The dotted and dashed lines are computed
using optical colors ($B-R$ and $B-V$) in {\bf (e)} and the mean value (solid line) is adopted
as the final \ML. The relationships between \MLk\ and
optical colors (e.g., $B-R$, $B-V$) are adopted from the models of Bell \& de Jong (2001).
For the conversion of \MLk\ to \MLsps, Eq. 6 in \cite{Oh_2008} is used.
{\bf (e):} The optical colors ($B-R$ and $B-V$) derived from the surface brightness
profiles in {\bf (a)}. Where $B-R$ was not available, only $B-V$ is given.
{\bf (c)(d):} Mass models for the stellar component.
The stellar mass surface density is derived from the 3.6$\mu$m surface brightness
(inclination corrected) in {\bf (a)} using the \MLsps\ values shown in {\bf (f)}.
The resulting expected rotation velocity for H{\sc i} if it were to move in circular
orbits in the potential corresponding to the optical mass density only is then derived from this.
{\bf (g)(h):} The mass model for the gas component.
The radial mass surface density distribution of
neutral gas is scaled by 1.4 to account for He and
metals. \\

\noindent $\bullet$ {\bf Comparison of rotation curves$\--$} Comparison of the rotation velocity derived
from the bulk velocity field with those from the other types of velocity fields
(i.e., IWM, hermite $h_{3}$, single Gaussian fit and peak velocity fields) and the
literature in case other measurements are available. For the bulk rotation velocity, we
derive rotation velocities for receding and approaching side only, by keeping the ring
parameters the same. These are indicated as the gray (inverse) triangles. We also show the
bulk rotation velocity corrected with $i^{\rm BTF}$ derived from the Baryonic Tully$\--$Fisher (BTF) relation.  \\

\noindent $\bullet$ {\bf Mass density profile$\--$} The derived mass density profile.
The dashed and solid lines show the best fits of the NFW halo model and
the pseudo-isothermal halo model to the rotation curve, respectively.
The measured inner slope $\alpha$ is shown in the panel. \\

\noindent $\bullet$ {\bf Mass modeling results$\--$} Disk-halo decomposition of the bulk rotation curve
(asymmetric drift corrected where needed) is made under various \ML\ assumptions (\MLsps, maximum disk,
minimum disk $+$ gas and minimum disk). For M81dwB, the asymmetric drift corrected hermite $h_{3}$ rotation
curve is used.
\clearpage

\begin{figure}
\epsscale{1.0}
\figurenum{A.1}
\includegraphics[angle=0,width=1.0\textwidth,bb=60 80 540 680,clip=]{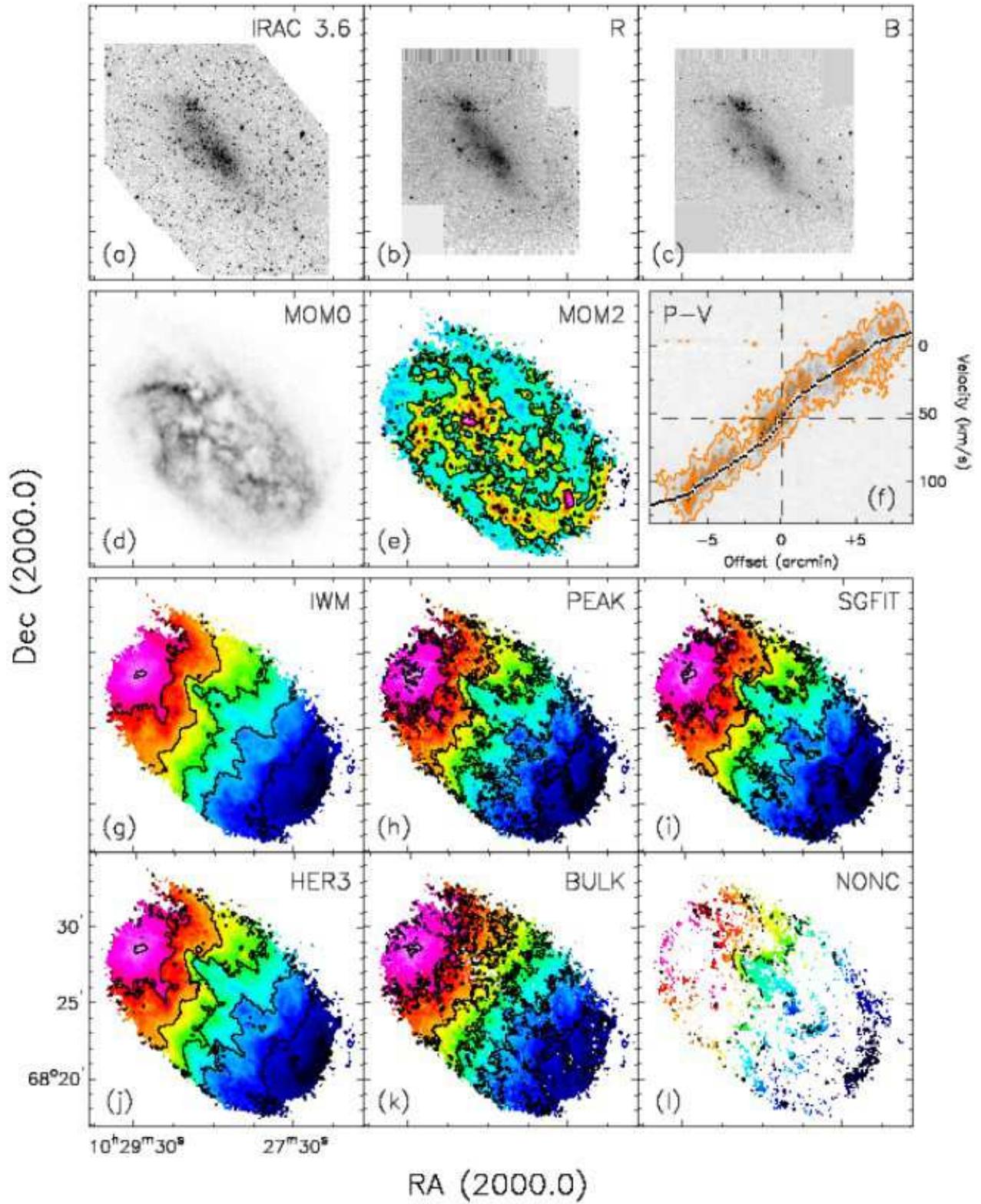}
\caption{{\bf Data:} Total intensity maps and velocity fields of IC 2574.
{\bf (a)(b)(c):}: Total intensity maps in {\it Spitzer} IRAC 3.6 $\mu$m, optical {\it R} and {\it B} bands.
{\bf (d):} Integrated H{\sc i} map (moment 0). The gray-scale levels run from 0 to 600 $\rm mJy\,beam^{-1}\,\kms$.
{\bf (e):} Velocity dispersion map (moment 2). Velocity contours run from $0$ to $25$\,\kms\,with a spacing of $5$\,\kms. 
{\bf (f):} Position-velocity diagram taken along the average position angle of the major axis as listed in Table 1. 
Contours start at $+2\sigma$ in steps of $8\sigma$. The dashed lines indicate the systemic velocity and position
of the kinematic center derived in this paper. Overplotted is the bulk rotation curve corrected for the average
inclination from the tilted-ring analysis as listed in Table 1. 
{\bf (g)(h)(i)(j)(k)(l):} Velocity fields. Contours run from $-10$\,\kms\,to $110$\,\kms\,with a spacing of $20$\,\kms.
\label{IC2574_MAPS}}
\end{figure}

\begin{figure}
\epsscale{1.0}
\figurenum{A.2}
\includegraphics[angle=0,width=1.0\textwidth,bb=45 165 570 695,clip=]{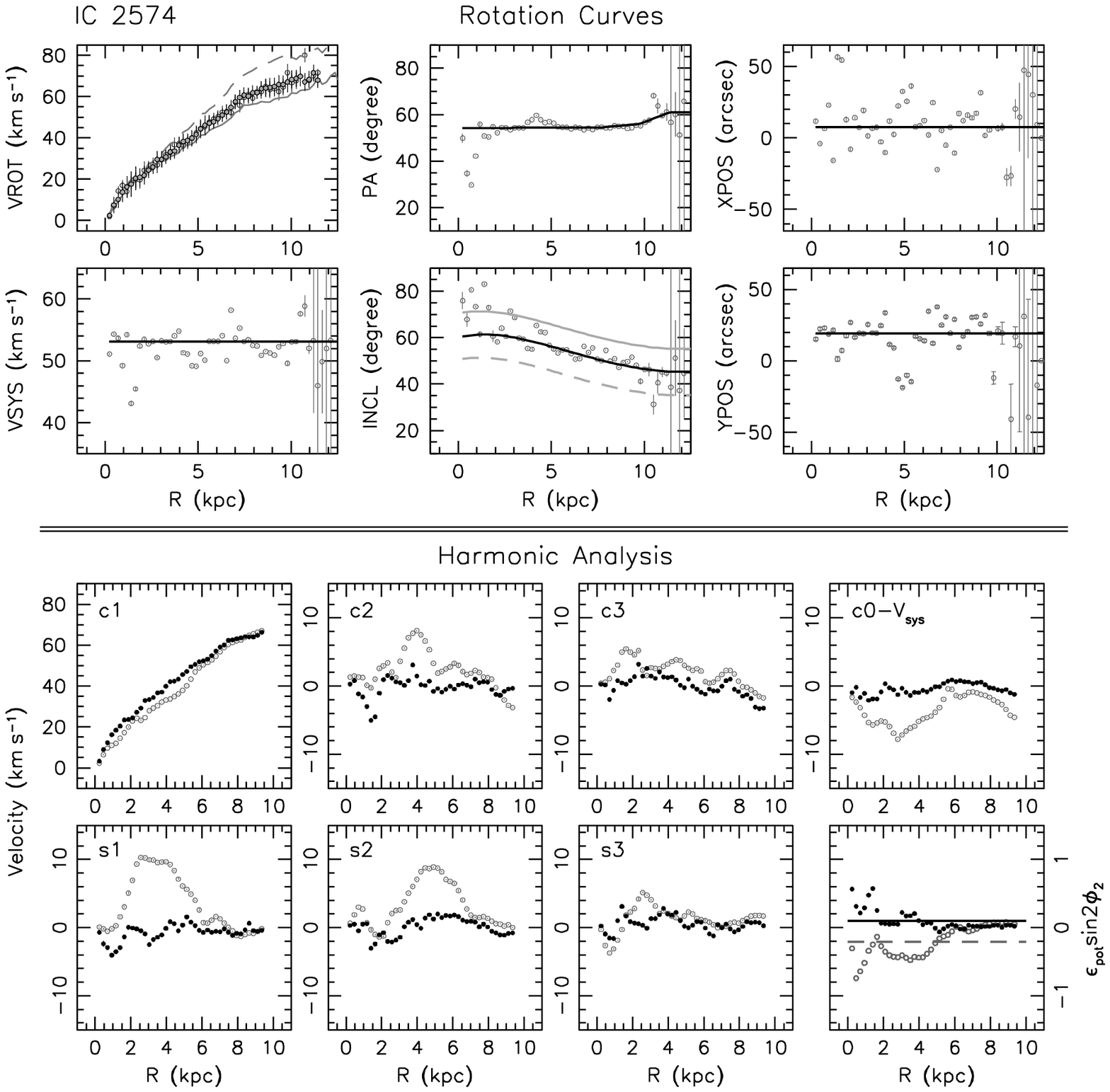}
\caption{{\bf Rotation curves:} The tilted ring model derived from the bulk velocity field of IC 2574.
The open gray circles in all panels indicate the fit made with all ring parameters free. 
The gray dots in the VROT panel were derived using the entire velocity field after fixing other ring parameters
to the values (black solid lines) as shown in the panels. To examine the sensitivity of the rotation curve
to the inclination, we vary the inclination by $+10$ and $-10\,^{\circ}$ as indicated by the gray
solid and dashed lines, respectively, in the bottom-middle panel. 
We derive the rotation curves using these inclinations while keeping other ring
parameters the same. The resulting rotation curves are indicated
by gray solid (for $+10\,^{\circ}$ inclination) and dashed (for $-10\,^{\circ}$ inclination) lines in the VROT panel.
{\bf Harmonic analysis:} Harmonic expansion of the velocity fields for IC 2574.
The black dots and gray open circles indicate the results from the bulk and hermite $h_{3}$ velocity fields, respectively.
In the bottom-rightmost panel, the solid and dashed lines indicate global elongations of the potential
measured using the bulk and hermite $h_{3}$ velocity fields.
\label{IC2574_TR_HD}}
\end{figure}
{\clearpage}

\begin{figure}
\epsscale{1.0}
\figurenum{A.3}
\includegraphics[angle=0,width=1.0\textwidth,bb=43 175 543 695,clip=]{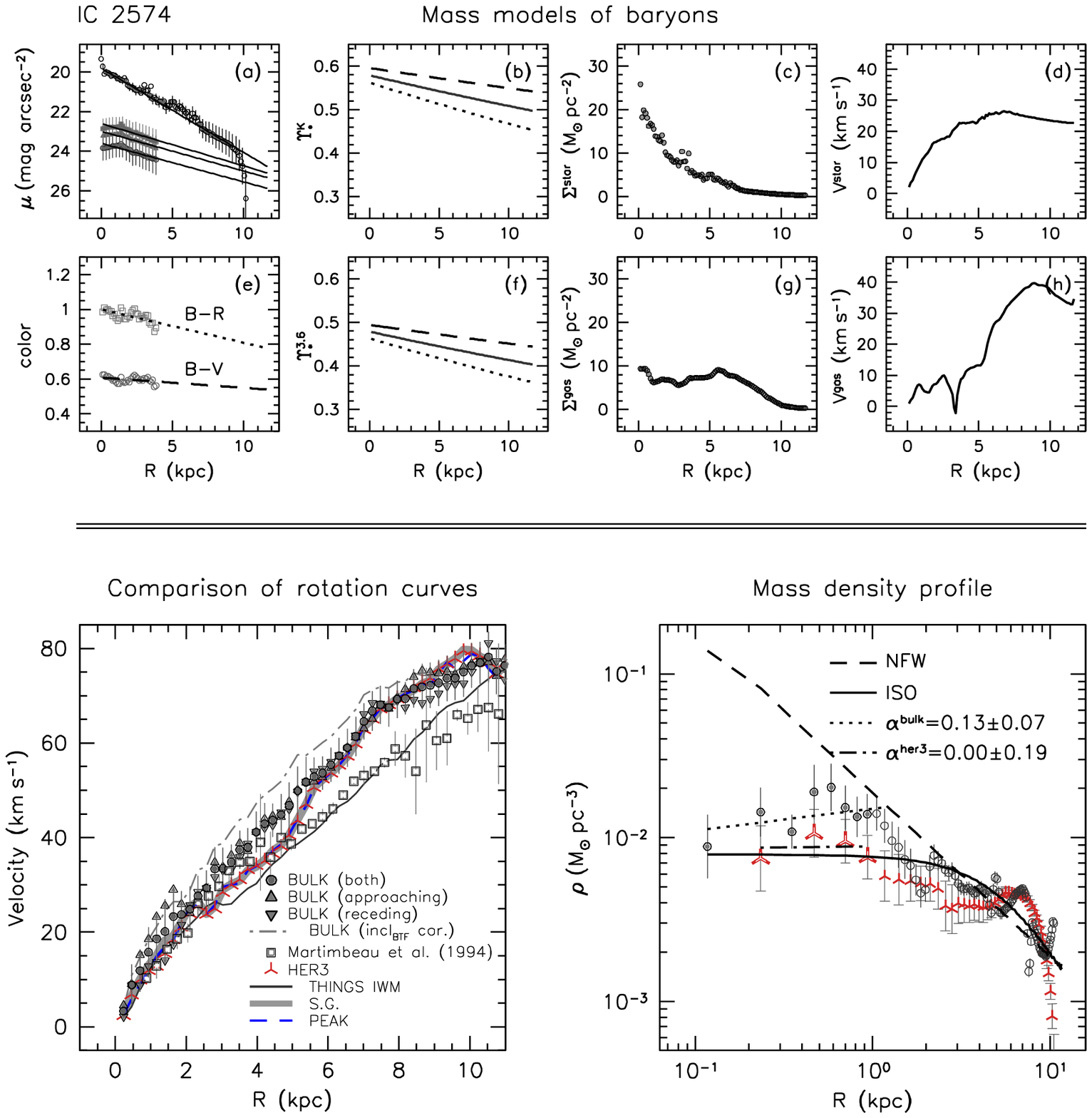}
\caption{{\bf Mass models of baryons:} Mass models for the gas and stellar components of IC 2574.
{\bf (a):} Azimuthally averaged surface brightness profiles in the 3.6$\mu$m, $R$, $V$ and $B$ bands (top to bottom).
{\bf (b)(f):} The stellar mass-to-light values in the $K$ and 3.6$\mu$m bands derived from stellar population synthesis models.
{\bf (c)(d):} The mass surface density and the resulting rotation velocity for the stellar component. 
{\bf (e):} Optical colors.
{\bf (g)(h):} The mass surface density (scaled by 1.4 to account for He and metals) and the resulting
rotation velocity for the gas component.
{\bf Comparison of rotation curves:} Comparison of the H{\sc i} rotation curves derived using different types
of velocity fields (i.e., bulk, IWM, hermite $h_{3}$, single Gaussian and peak velocity fields as denoted
in the panel) for IC 2574. This figure is the same as the panel (a) of Fig.~\ref{ROTATION_CURVES_MASS_MODELS_IC2574}.
{\bf Mass density profile:} The derived mass density profile of IC 2574.
The open circles and tripod-like symbols represent the mass density profiles derived from the bulk and hermite $h_{3}$
rotation curves assuming minimum disk, respectively. The inner density slopes $\alpha$ are measured by
least squares fits (dotted and dot-dashed lines) to the data points indicated by gray dots and larger
tripod-like symbols, and shown in the panel.
\label{IC2574_VROT_BARYONS_ALPHA}}
\end{figure}
{\clearpage}

\begin{figure}
\epsscale{1.0}
\figurenum{A.4}
\includegraphics[angle=0,width=1.0\textwidth,bb=20 150 580 720,clip=]{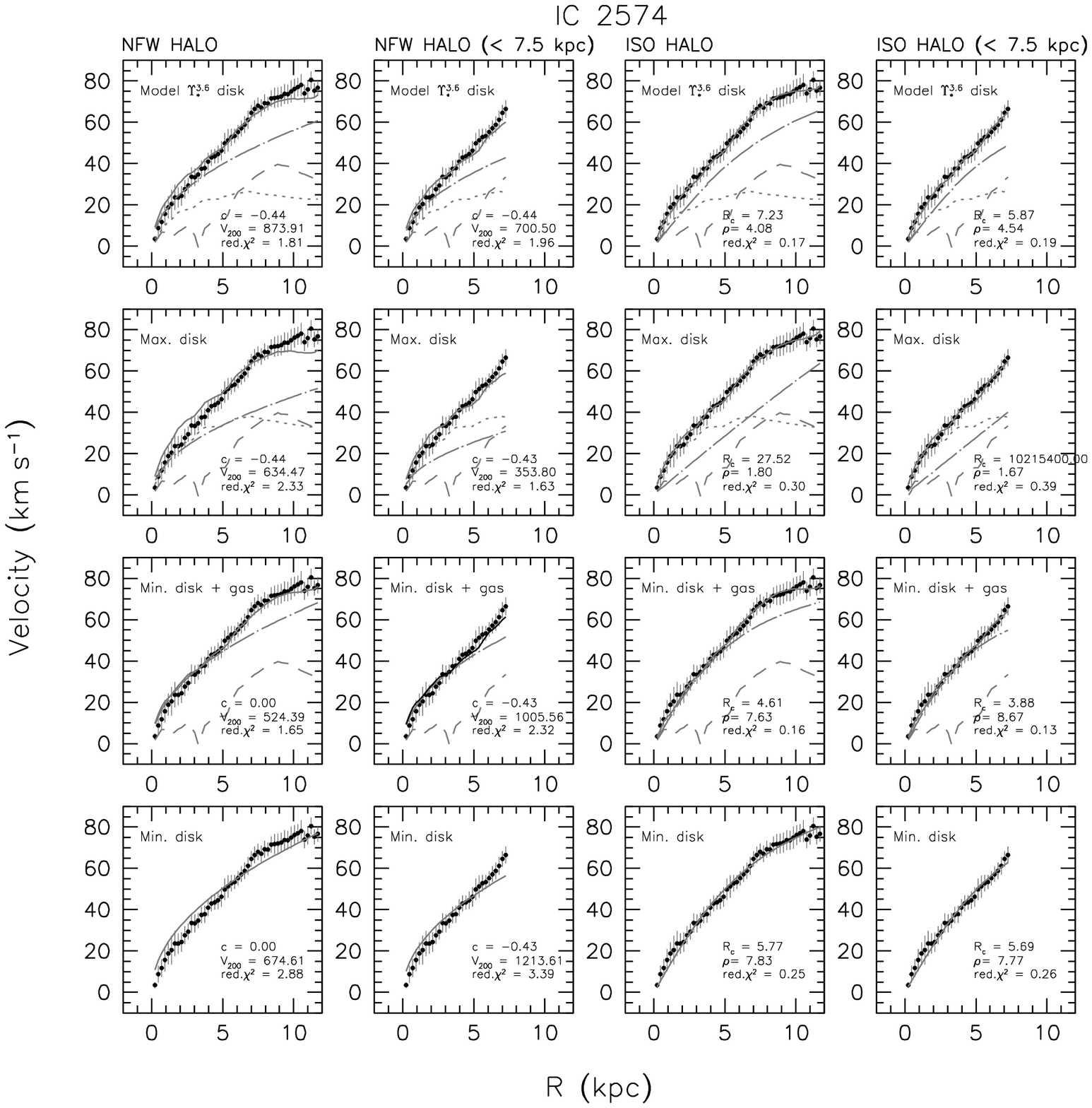}
\caption{{\bf Mass modeling results:} Disk-halo decomposition of the IC 2574 rotation curve 
under various \ML\ assumptions (\MLsps, maximum disk, $\rm minimum disk + gas$ and minimum disks).
The black dots indicate the bulk rotation curve, and the short and long dashed lines show the
rotation velocities of the stellar and gas components, respectively. The fitted parameters
of NFW and pseudo-isothermal halo models (long dash-dotted lines) are denoted on each panel.
\label{IC2574_MD}}
\end{figure}
{\clearpage}

\begin{table*}
\scriptsize
\caption{Parameters of dark halo models for IC 2574}
\label{NFWHALO_IC2574}
\begin{center}
\begin{tabular}{@{}lccccccrc}
\hline
\hline
\noalign{\vskip 2pt}
& \multicolumn{3}{c}{NFW halo (entire region)} && \multicolumn{3}{c}{NFW halo ($<$ 7.5 kpc)}\\
\cline{2-4} \cline{6-8}\\
\multicolumn{1}{c}{\ML\ assumption}   &  \multicolumn{1}{c}{$c$}   &  \multicolumn{1}{c}{$V_{200}$}  &  \multicolumn{1}{c}{$\chi_{red.}^2$}    &&  \multicolumn{1}{c}{$c$}   &  \multicolumn{1}{c}{$V_{200}$}  &  \multicolumn{1}{c}{$\chi_{red.}^2$}\\
\multicolumn{1}{c}{(1)} & \multicolumn{1}{c}{(2)}   & \multicolumn{1}{c}{(3)}      & \multicolumn{1}{c}{(4)}    && \multicolumn{1}{c}{(5)} & \multicolumn{1}{c}{(6)} & \multicolumn{1}{c}{(7)}\\
\hline
Min. disk               & $<0.1$ ({\bf 9.0})  &  674.6 $\pm$ 18.3 ({\bf 54.9 $\pm$ 1.7})      & 2.88 ({\bf 11.64})  && $<0.1$   & 1213.6 $\pm$ ...  & 3.39\\
Min. disk+gas           & $<0.1$ ({\bf 9.0}) &   524.3 $\pm$ 51.7 ({\bf 45.9 $\pm$ 1.5})     & 1.65 ({\bf 7.85})  &&   $<0.1$   & 1005.5 $\pm$ ... & 2.32\\
Max. disk               & $<0.1$ ({\bf 9.0}) &   634.4 $\pm$ ...  ({\bf 30.8 $\pm$ 1.8})      & 2.33 ({\bf 7.71})  &&  $<0.1$   & 353.8 $\pm$ ... & 1.63\\
Model $\Upsilon_{*}^{3.6}$ disk & $<0.1$ ({\bf 9.0}) & 873.9 $\pm$ ... ({\bf 39.3 $\pm$ 1.7}) & 1.81 ({\bf 7.86})  &&  $<0.1$   & 700.5 $\pm$ ... & 1.96\\
\hline\\
\noalign{\vskip 2pt}
 & \multicolumn{3}{c}{Pseudo-isothermal halo (entire region)} && \multicolumn{3}{c}{Pseudo-isothermal halo ($<$ 7.5 kpc)}\\
\cline{2-4} \cline{6-8} \\
\multicolumn{1}{c}{\ML\ assumption}    &  \multicolumn{1}{c}{$R_{C}$} &   \multicolumn{1}{c}{$\rho_0$}    & \multicolumn{1}{c}{$\chi_{red.}^2$}     && \multicolumn{1}{c}{$R_{C}$}      & \multicolumn{1}{c}{$\rho_{0}$}  & \multicolumn{1}{c}{$\chi_{red.}^2$} \\
\multicolumn{1}{c}{(8)} & \multicolumn{1}{c}{(9)}   & \multicolumn{1}{c}{(10)}      & \multicolumn{1}{c}{(11)}   && \multicolumn{1}{c}{(12)} & \multicolumn{1}{c}{(13)} & \multicolumn{1}{c}{(14)} \\
\hline
Min. disk               & 5.77 $\pm$ 0.16       & 7.8 $\pm$ 0.2    & 0.25   && 5.69 $\pm$ 0.35       & 7.8 $\pm$ 0.3     & 0.26\\
Min. disk+gas           & 4.61 $\pm$ 0.12       & 7.6 $\pm$ 0.2   & 0.16   &&  3.88 $\pm$ 0.16       & 8.7 $\pm$ 0.3     & 0.13\\
Max. disk               & 27.52 $\pm$ 10.22       & 1.8 $\pm$ 0.1    & 0.30   && $...$               & 1.7 $\pm$ 0.9     & 0.39\\
Model $\Upsilon_{*}^{3.6}$ disk & 7.23 $\pm$ 0.30       & 4.1 $\pm$ 0.1    & 0.17   && 5.87 $\pm$ 0.55       & 4.5 $\pm$ 0.3     & 0.19\\
\hline
\end{tabular}
\medskip\noindent
\begin{minipage}{161mm}
\noindent
\\
{\bf Note.$\--$}
{\bf (1)(8):} The stellar mass-to-light ratio \ML\ assumptions. ``Model \MLsps\ disk'' uses the values derived from the population synthesis models in Section~\ref{Stars}.
{\bf (2)(5):} Concentration parameter c of NFW halo model (NFW 1996, 1997). We also fit the NFW model to
the rotation curves with only $V_{200}$ as a free parameter after fixing $c$ to 9. The corresponding
best-fit $V_{200}$ and $\chi^{2}_{red}$ values are given in the brackets in (3) and (4), respectively.
{\bf (3)(6):} The rotation velocity (\kms)\,at radius $R_{200}$ where the density constrast exceeds 200 (Navarro \etal\ 1996).
{\bf (4)(7)(11)(14):} Reduced $\chi^{2}$ value.
{\bf (9)(12):} Fitted core-radius of pseudo-isothermal halo model (kpc).
{\bf (10)(13):} Fitted core-density of pseudo-isothermal halo model ($10^{-3}$ \cubedens).
{\bf ($...$):} blank due to unphysically large value or not well-constrained uncertainties.
\end{minipage}
\end{center}
\end{table*}

\begin{figure}
\figurenum{A.5}
\epsscale{1.0}
\includegraphics[angle=0,width=1.0\textwidth,bb=60 80 540 680,clip=]{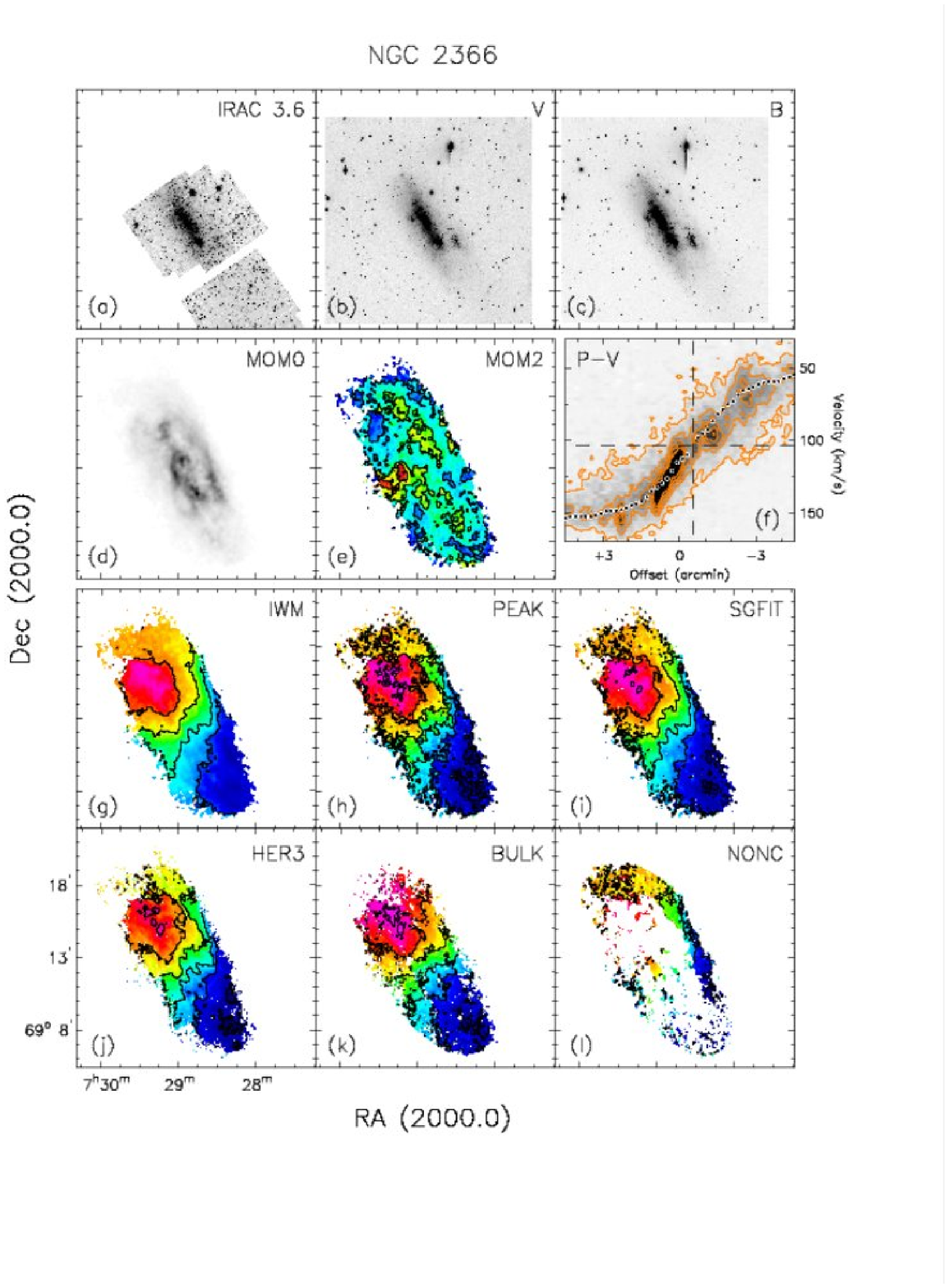}
\caption{{\bf Data:} Total intensity maps and velocity fields of NGC 2366.
{\bf (a)(b)(c):}: Total intensity maps in {\it Spitzer} IRAC 3.6 $\mu$m, optical {\it V} and {\it B} bands.
{\bf (d):} Integrated H{\sc i} map (moment 0). The gray-scale levels run from 0 to 1000 $\rm mJy\,beam^{-1}\,\kms$.
{\bf (e):} Velocity dispersion map (moment 2). Velocity contours run from $0$ to $25$\,\kms\,with a spacing of $5$\,\kms. 
{\bf (f):} Position-velocity diagram taken along the average position angle of the major axis as listed in Table 1. 
Contours start at $+2\sigma$ in steps of $8\sigma$. The dashed lines indicate the systemic velocity and position
of the kinematic center derived in this paper. Overplotted is the bulk rotation curve corrected for the average
inclination from the tilted-ring analysis as listed in Table 1. 
{\bf (g)(h)(i)(j)(k)(l):} Velocity fields. Contours run from $30$\,\kms\,to $180$\,\kms\,with a spacing of $20$\,\kms.
\label{NGC2366_MAPS}}
\end{figure}
{\clearpage}

\begin{figure}
\figurenum{A.6}
\epsscale{1.0}
\includegraphics[angle=0,width=1.0\textwidth,bb=45 165 570 695,clip=]{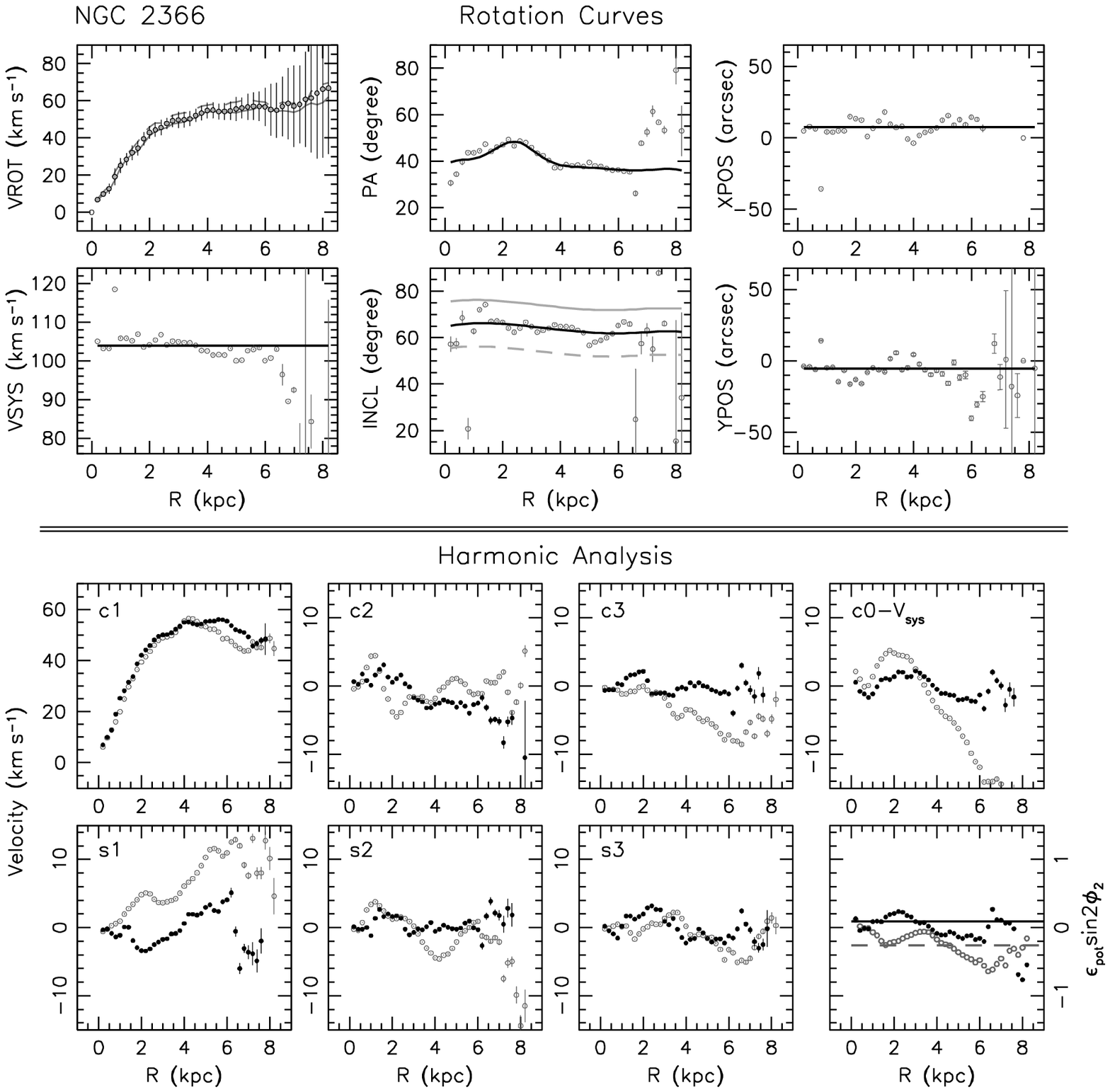}
\caption{{\bf Rotation curves:} The tilted ring model derived from the bulk velocity field of NGC 2366.
The open gray circles in all panels indicate the fit made with all ring parameters free. 
The gray dots in the VROT panel were derived using the entire velocity field after fixing other ring parameters
to the values (black solid lines) as shown in the panels. To examine the sensitivity of the rotation curve
to the inclination, we vary the inclination by $+10$ and $-10\,^{\circ}$ as indicated by the gray
solid and dashed lines, respectively, in the right-middle panel. 
We derive the rotation curves using these inclinations while keeping other ring
parameters the same. The resulting rotation curves are indicated
by gray solid (for $+10\,^{\circ}$ inclination) and dashed (for $-10\,^{\circ}$ inclination) lines in the VROT panel.
{\bf Harmonic analysis:} Harmonic expansion of the velocity fields for NGC 2366.
The black dots and gray open circles indicate the results from the bulk and hermite $h_{3}$ velocity fields, respectively.
In the bottom-rightmost panel, the solid and dashed lines indicate global elongations of the potential
measured using the bulk and hermite $h_{3}$ velocity fields.
\label{NGC2366_TR_HD}}
\end{figure}
{\clearpage}

\begin{figure}
\epsscale{1.0}
\figurenum{A.7}
\includegraphics[angle=0,width=1.0\textwidth,bb=43 175 543 695,clip=]{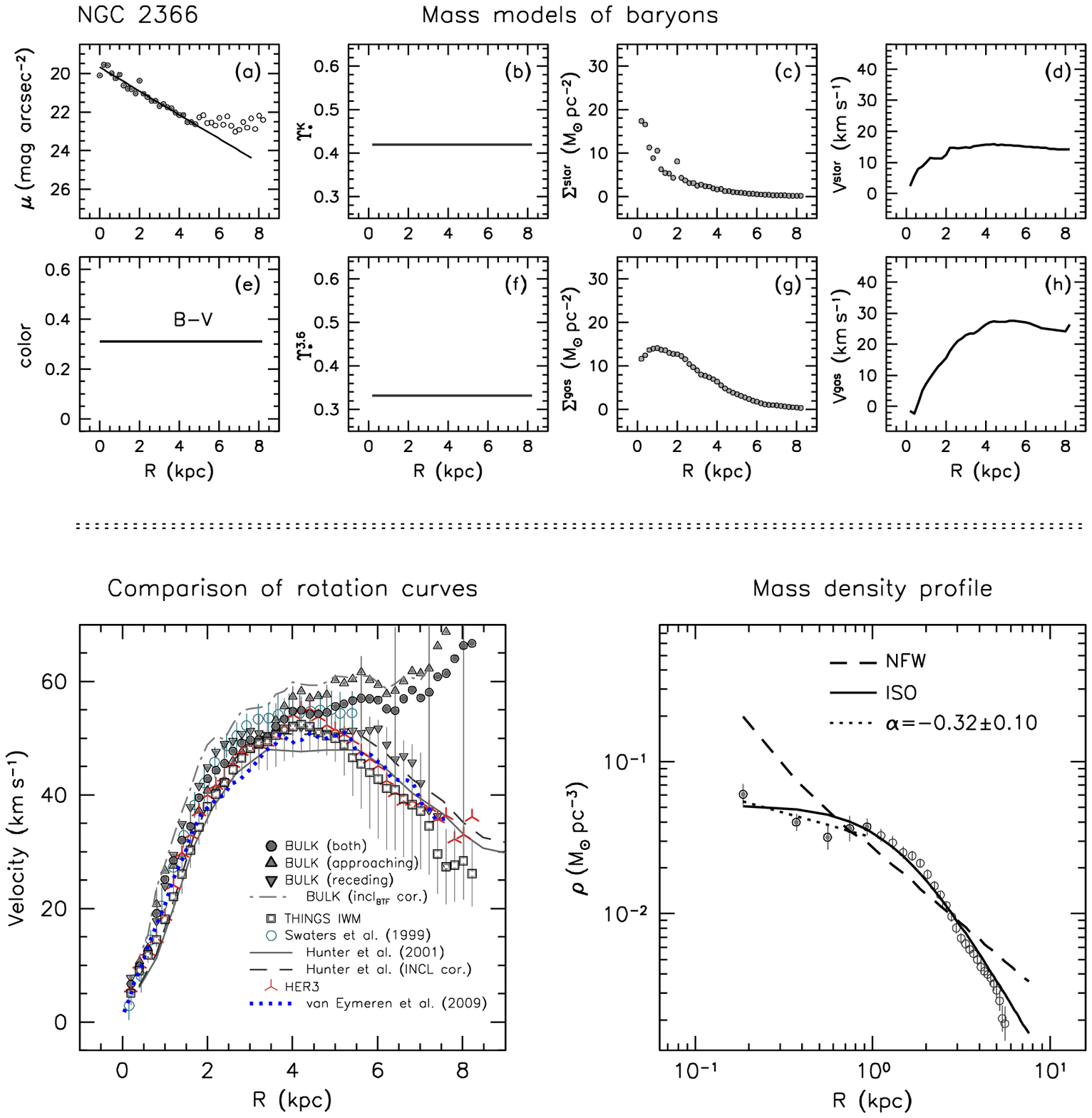}
\caption{{\bf Mass models of baryons:} Mass models for the gas and stellar components of NGC 2366.
{\bf (a):} The azimuthally averaged 3.6$\mu$m surface brightness profile.
{\bf (b)(f):} The stellar mass-to-light values in the $K$ and 3.6$\mu$m bands derived from stellar population synthesis models.
{\bf (c)(d):} The mass surface density and the resulting rotation velocity for the stellar component.
{\bf (e):} Optical color.
{\bf (g)(h):} The mass surface density (scaled by 1.4 to account for He and metals) and the resulting
rotation velocity for the gas component.
{\bf Comparison of rotation curves:} Comparison of the H{\sc i} rotation curves for NGC 2366. See Section~\ref{Deriving_Rotcurs}
for a detailed discussion. These rotation curves have also been discussed in detail \cite{Oh_2008}. 
{\bf Mass density profile:} The derived mass density profile of NGC 2366.
The open circles represent the mass density profile derived from the bulk rotation curve
assuming minimum disk. The inner density slope $\alpha$ is measured by a least squares fit (dotted line)
to the data points indicated by gray dots, and shown in the panel.
\label{NGC2366_VROT_BARYONS_ALPHA}}
\end{figure}
{\clearpage}

\begin{figure}
\figurenum{A.8}
\epsscale{1.0}
\includegraphics[angle=0,width=1.0\textwidth,bb=20 150 580 720,clip=]{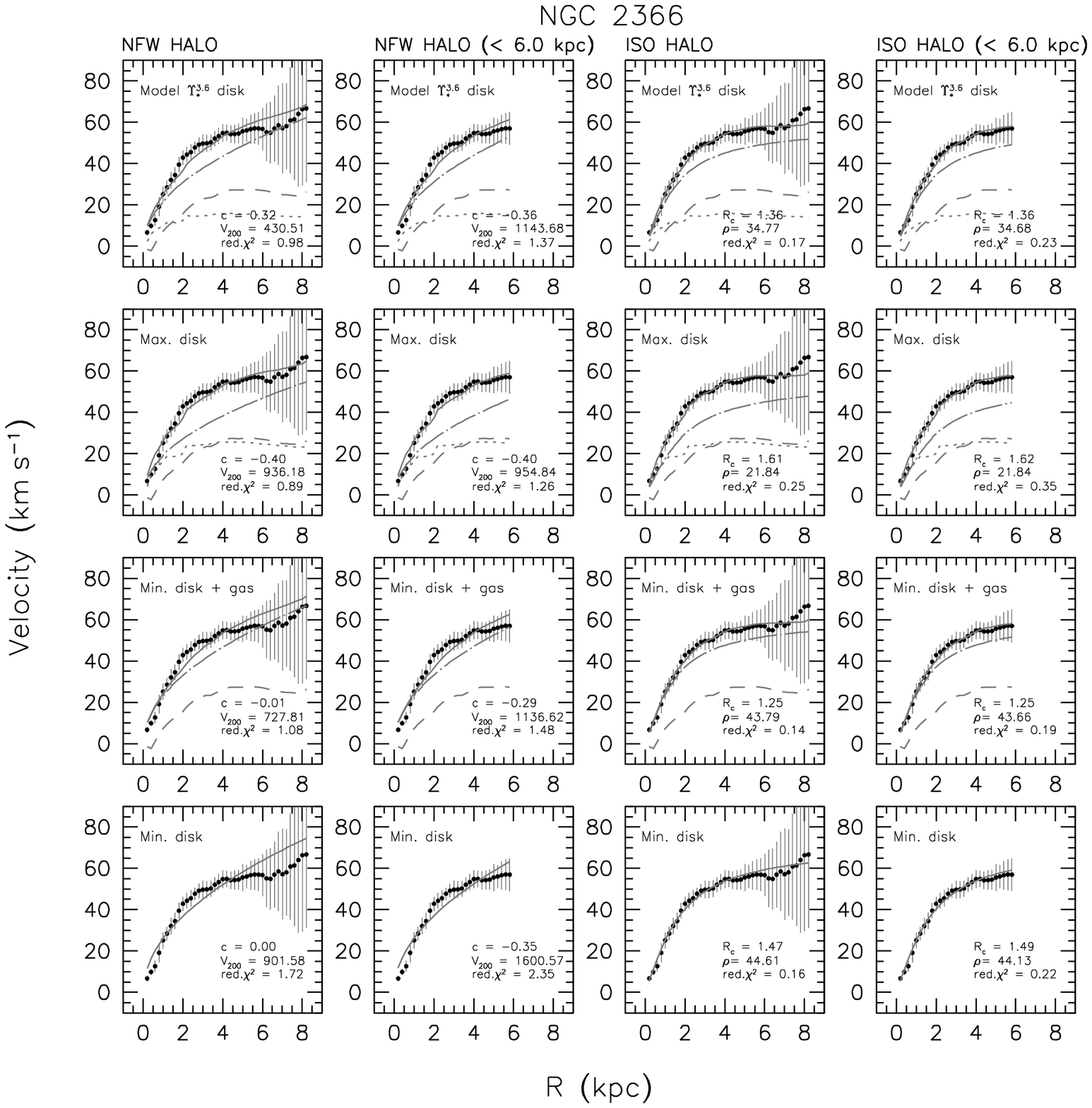}
\caption{{\bf Mass modeling results:} Disk-halo decomposition of the NGC 2366 rotation curve 
under various \ML\ assumptions (\MLsps, maximum disk, $\rm minimum disk + gas$ and minimum disks).
The black dots indicate the bulk rotation curve, and the short and long dashed lines show the
rotation velocities of the stellar and gas components, respectively. The fitted parameters
of NFW and pseudo-isothermal halo models (long dash-dotted lines) are denoted on each panel.
\label{NGC2366_MD}}
\end{figure}
{\clearpage}

\begin{table*}
\scriptsize
\caption{Parameters of dark halo models for NGC 2366}
\label{NFWHALO_NGC2366}
\begin{center}
\begin{tabular}{@{}lccccccrc}
\hline
\hline
\noalign{\vskip 2pt}
& \multicolumn{3}{c}{NFW halo (entire region)} && \multicolumn{3}{c}{NFW halo ($<$ 6.0 kpc)}\\
\cline{2-4} \cline{6-8}\\
\multicolumn{1}{c}{\ML\ assumption}   &  \multicolumn{1}{c}{$c$}   &  \multicolumn{1}{c}{$V_{200}$}  &  \multicolumn{1}{c}{$\chi_{red.}^2$}    &&  \multicolumn{1}{c}{$c$}   &  \multicolumn{1}{c}{$V_{200}$}   &  \multicolumn{1}{c}{$\chi_{red.}^2$}\\
\multicolumn{1}{c}{(1)} & \multicolumn{1}{c}{(2)}   & \multicolumn{1}{c}{(3)}      & \multicolumn{1}{c}{(4)}    && \multicolumn{1}{c}{(5)}  & \multicolumn{1}{c}{(6)} & \multicolumn{1}{c}{(7)} \\
\hline
Min. disk        & $<0.1$ ({\bf 9.0})     & 901.5 $\pm$ 478.4 ({\bf 50.7 $\pm$ 2.7})  & 1.72  ({\bf 4.54})      &&  $<0.1$       & 1600.5 $\pm$ $...$    & 2.35\\
Min. disk+gas    & $<0.1$ ({\bf 9.0})     & 727.8 $\pm$ $...$ ({\bf 42.5 $\pm$ 2.3})  & 1.08  ({\bf 3.02})      &&  $<0.1$        & 1136.6 $\pm$ $...$    & 1.48\\
Max. disk        & $<0.1$ ({\bf 9.0})     & 936.1 $\pm$ $...$ ({\bf 31.7 $\pm$ 2.4})  & 0.89  ({\bf 2.86})      &&  $<0.1$        & 954.8 $\pm$ $...$    & 1.26\\
Model $\Upsilon_{*}^{3.6}$ disk & $<0.1$ ({\bf 9.0})  & 630.7 $\pm$ $...$ ({\bf 38.6 $\pm$ 2.3}) & 0.98 ({\bf 2.96}) &&  $<0.1$   & 1143.6 $\pm$ $...$    & 1.37\\
\hline\\
\noalign{\vskip 2pt}
 & \multicolumn{3}{c}{Pseudo-isothermal halo (entire region)} && \multicolumn{3}{c}{Pseudo-isothermal halo ($<$ 6.0 kpc)}\\
\cline{2-4} \cline{6-8} \\
\multicolumn{1}{c}{\ML\ assumption}    & \multicolumn{1}{c}{$R_C$} &   \multicolumn{1}{c}{$\rho_0$}    & \multicolumn{1}{c}{$\chi_{red.}^2$}     && \multicolumn{1}{c}{$R_C$}      & \multicolumn{1}{c}{$\rho_0$}   & \multicolumn{1}{c}{$\chi_{red.}^2$}\\
\multicolumn{1}{c}{(8)} & \multicolumn{1}{c}{(9)}   & \multicolumn{1}{c}{(10)}      & \multicolumn{1}{c}{(11)}    && \multicolumn{1}{c}{(12)}  & \multicolumn{1}{c}{(13)} & \multicolumn{1}{c}{(14)}  \\
\hline
Min. disk        & 1.47 $\pm$ 0.06       & 44.6 $\pm$ 2.2         & 0.16         && 1.49 $\pm$ 0.07       & 44.1 $\pm$ 2.6         & 0.21\\
Min. disk+gas    & 1.25 $\pm$ 0.05       & 43.8 $\pm$ 2.4         & 0.13         && 1.25 $\pm$ 0.06       & 43.7 $\pm$ 2.8         & 0.18\\
Max. disk        & 1.61 $\pm$ 0.15       & 21.8 $\pm$ 2.5         & 0.25         && 1.62 $\pm$ 0.18       & 21.8 $\pm$ 3.0         & 0.34\\
Model $\Upsilon_{*}^{3.6}$ disk & 1.36 $\pm$ 0.07 & 34.8 $\pm$ 2.4         & 0.17         && 1.36 $\pm$ 0.09       & 34.7 $\pm$ 2.9         & 0.23\\
\hline
\end{tabular}
\medskip\noindent
\begin{minipage}{158mm}
\noindent
\\
{\bf Note.$\--$}
{\bf (1)(8):} The stellar mass-to-light ratio \ML\ assumptions. ``Model \MLsps\ disk'' uses the values derived from the population synthesis models in Section~\ref{Stars}.
{\bf (2)(5):} Concentration parameter c of NFW halo model (NFW 1996, 1997). We also fit the NFW model to
the rotation curves with only $V_{200}$ as a free parameter after fixing $c$ to 9. The corresponding
best-fit $V_{200}$ and $\chi^{2}_{red}$ values are given in the brackets in (3) and (4), respectively.
{\bf (3)(6):} The rotation velocity (\kms)\,at radius $R_{200}$ where the density constrast exceeds 200 (Navarro \etal\ 1996).
{\bf (4)(7)(11)(14):} Reduced $\chi^{2}$ value.
{\bf (9)(12):} Fitted core-radius of pseudo-isothermal halo model (kpc).
{\bf (10)(13):} Fitted core-density of pseudo-isothermal halo model ($10^{-3}$ \cubedens).
{\bf ($...$):} blank due to unphysically large value or not well-constrained uncertainties.
\end{minipage}
\end{center}
\end{table*}

\begin{figure}
\figurenum{A.9}
\epsscale{1.0}
\includegraphics[angle=0,width=1.0\textwidth,bb=60 80 540 680,clip=]{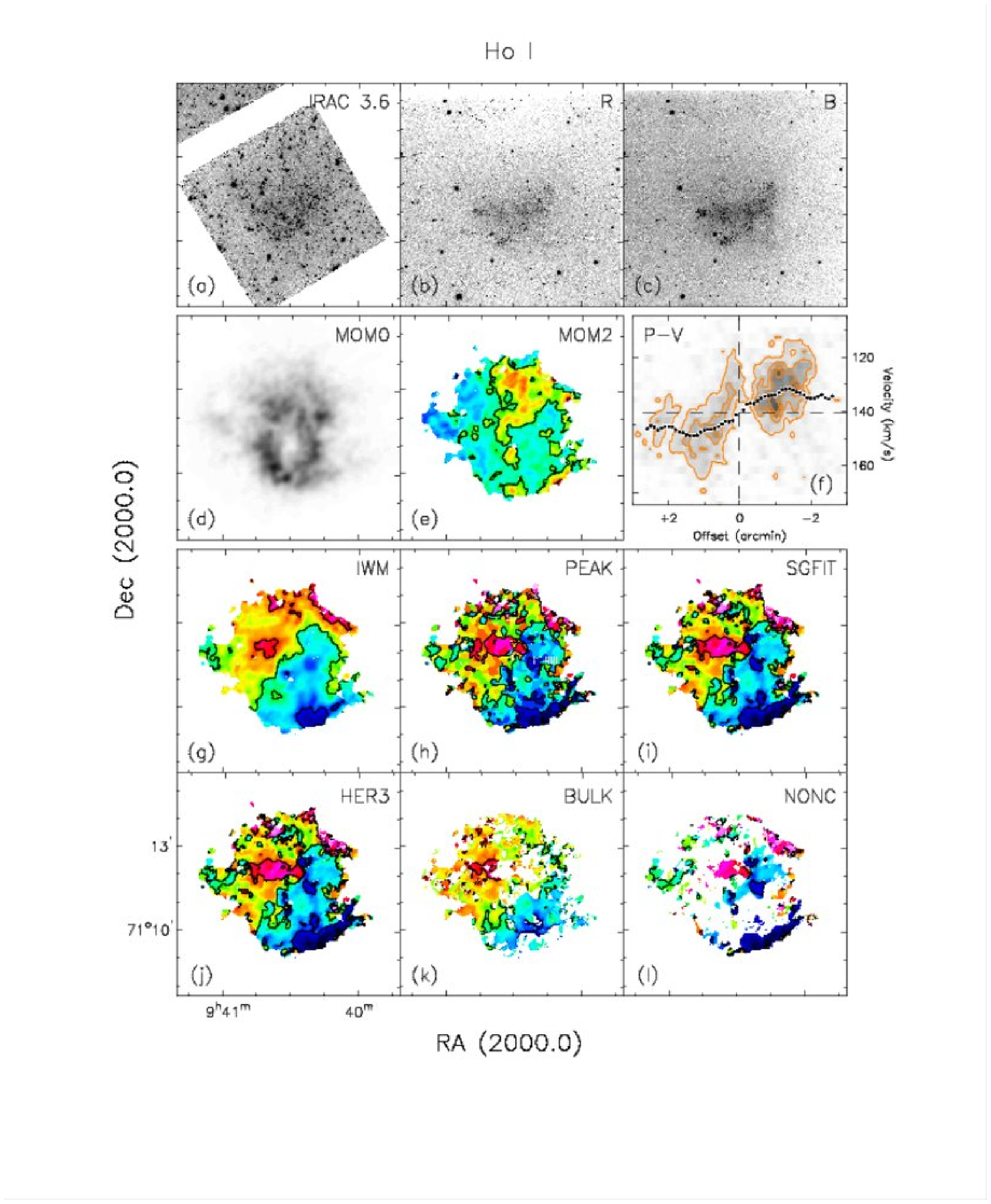}
\caption{{\bf Data:} Total intensity maps and velocity fields of Ho I.
{\bf (a)(b)(c):}: Total intensity maps in {\it Spitzer} IRAC 3.6 $\mu$m, optical {\it R} and {\it B} bands.
{\bf (d):} Integrated H{\sc i} map (moment 0). The gray-scale levels run from 0 to 400 $\rm mJy\,beam^{-1}\,\kms$.
{\bf (e):} Velocity dispersion map (moment 2). Velocity contours run from $0$ to $25$\,\kms\,with a spacing of $5$\,\kms. 
{\bf (f):} Position-velocity diagram taken along the average position angle of the major axis as listed in Table 1. 
Contours start at $+2\sigma$ in steps of $3\sigma$. The dashed lines indicate the systemic velocity and position
of the kinematic center derived in this paper. Overplotted is the bulk rotation curve corrected for the average
inclination from the tilted-ring analysis as listed in Table 1. 
{\bf (g)(h)(i)(j)(k)(l):} Velocity fields. Contours run from $120$\,\kms\,to $160$\,\kms\,with a spacing of $10$\,\kms.
\label{HoI_MAPS}}
\end{figure}
{\clearpage}

\begin{figure}
\figurenum{A.10}
\epsscale{1.0}
\includegraphics[angle=0,width=1.0\textwidth,bb=45 140 570 720,clip=]{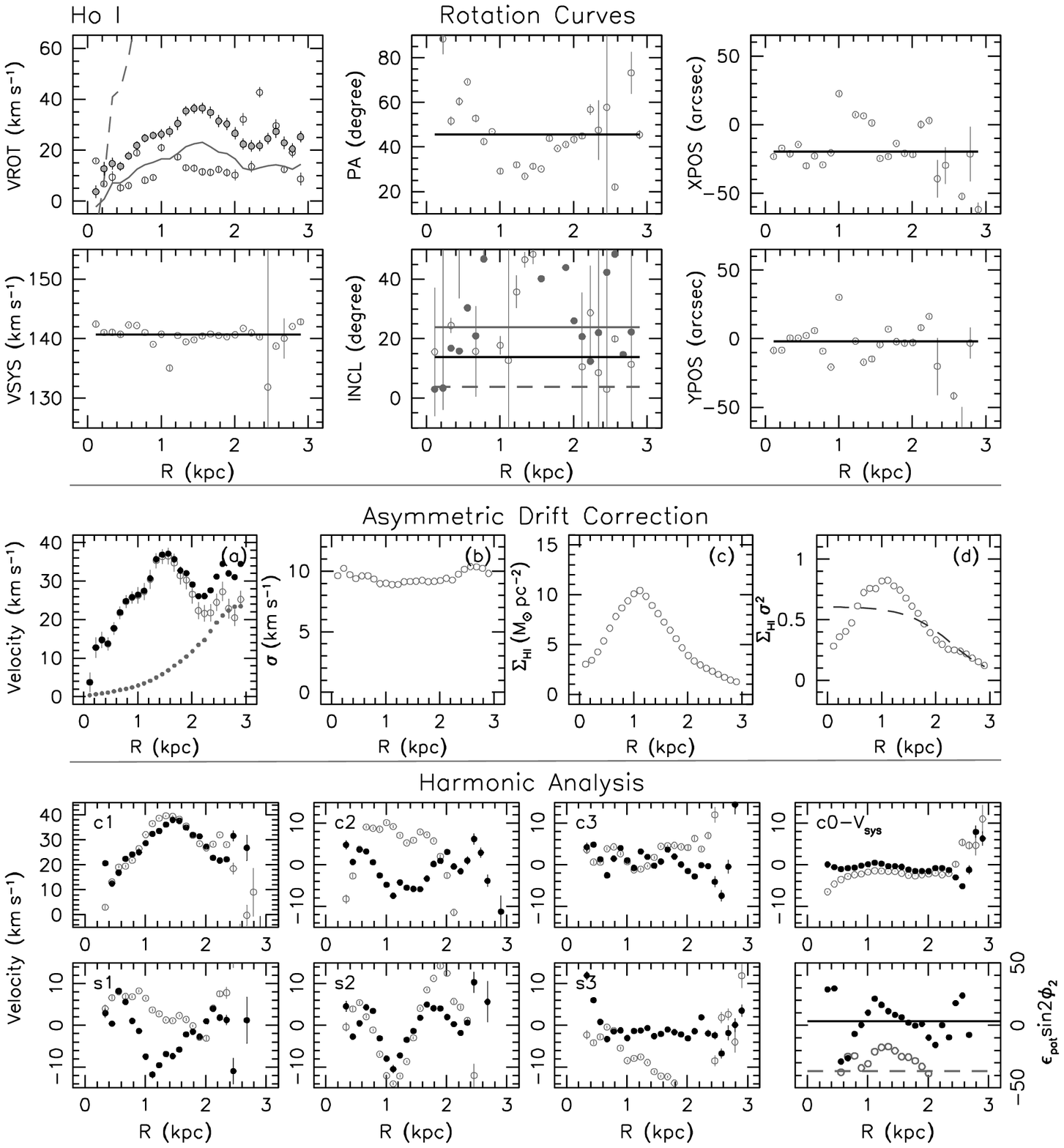}
\caption{{\bf Rotation curves:} The tilted ring model derived from the bulk velocity field of Ho I.
The open gray circles in all panels indicate the fit made with all ring parameters free. 
The gray dots in the VROT panel were derived using the entire velocity field after fixing other ring parameters
to the values (black solid lines) as shown in the panels. To examine the sensitivity of the rotation curve
to the inclination, we vary the inclination by $+10$ and $-10\,^{\circ}$ as indicated by the gray
solid and dashed lines, respectively, in the right-middle panel. 
We derive the rotation curves using these inclinations while keeping other ring
parameters the same. The resulting rotation curves are indicated
by gray solid (for $+10\,^{\circ}$ inclination) and dashed (for $-10\,^{\circ}$ inclination) lines in the VROT panel.
{\bf Asymmetric drift correction:}
{\bf a:} Gray filled dots indicate the derived radial velocity
correction for the asymmetric drift $\sigma_{\rm D}$. Black open and filled
dots represent the uncorrected and corrected curves for the asymmetric drift, respectively.
{\bf b:} Azimuthally averaged H{\sc i} velocity dispersion.
{\bf c:} Azimuthally averaged H{\sc i} surface density.
{\bf d:} The dashed line indicates a fit to $\Sigma\sigma^{2}$ with an analytical function.
{\bf Harmonic analysis:} Harmonic expansion of the velocity fields for Ho I.
The black dots and gray open circles indicate the results from the bulk and hermite $h_{3}$ velocity fields, respectively.
In the bottom-rightmost panel, the solid and dashed lines indicate global elongations of the potential
measured using the bulk and hermite $h_{3}$ velocity fields.
\label{HoI_TR_ADC_HD}}
\end{figure}
{\clearpage}

\begin{figure}
\epsscale{1.0}
\figurenum{A.11}
\includegraphics[angle=0,width=1.0\textwidth,bb=43 175 543 695,clip=]{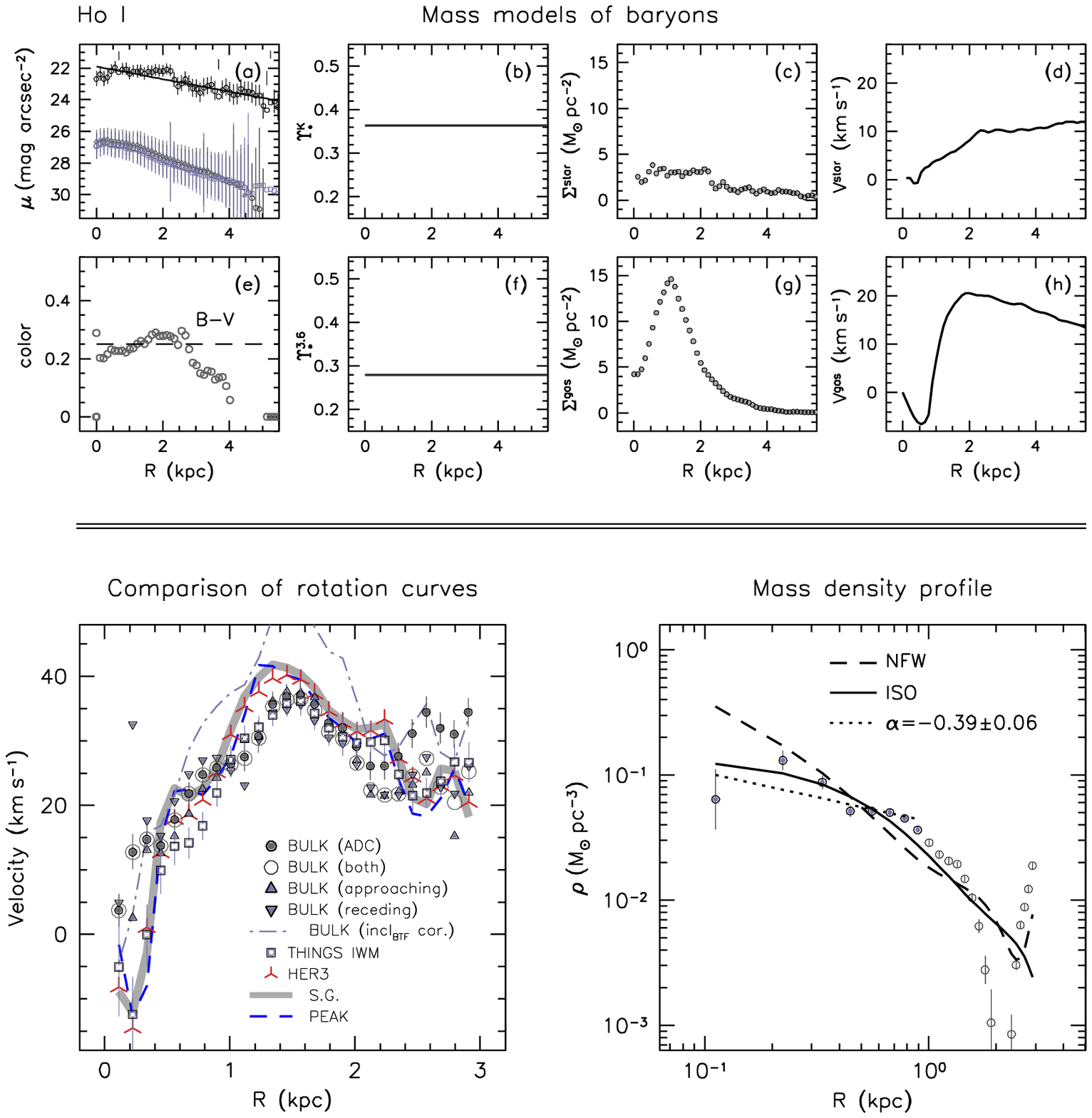}
\caption{
{\bf Mass models of baryons:} Mass models for the gas and stellar components of Ho I.
{\bf (a):} Azimuthally averaged surface brightness profiles in the 3.6$\mu$m, $R$, $V$ and $B$ bands (top to bottom).
{\bf (b)(f):} The stellar mass-to-light values in the $K$ and 3.6$\mu$m bands derived from stellar population synthesis models.
{\bf (c)(d):} The mass surface density and the resulting rotation velocity for the stellar component.
{\bf (e):} Optical color.
{\bf (g)(h):} The mass surface density (scaled by 1.4 to account for He and metals) and the resulting
rotation velocity for the gas component.
{\bf Comparison of rotation curves:} Comparison of the H{\sc i} rotation curves derived using different types
of velocity fields (i.e., bulk, IWM, hermite $h_{3}$, single Gaussian and peak velocity fields as denoted in 
the panel) for Ho I. See Section~\ref{Deriving_Rotcurs} for more information.
{\bf Mass density profile:} The derived mass density profile of Ho I.
The open circles represent the mass density profile derived from the bulk rotation curve
assuming minimum disk. The inner density slope $\alpha$ is measured by a least squares fit (dotted line)
to the data points indicated by gray dots, and shown in the panel.
\label{HoI_VROT_BARYONS_ALPHA}}
\end{figure}
{\clearpage}

\begin{figure}
\figurenum{A.12}
\epsscale{1.0}
\includegraphics[angle=0,width=1.0\textwidth,bb=20 150 580 720,clip=]{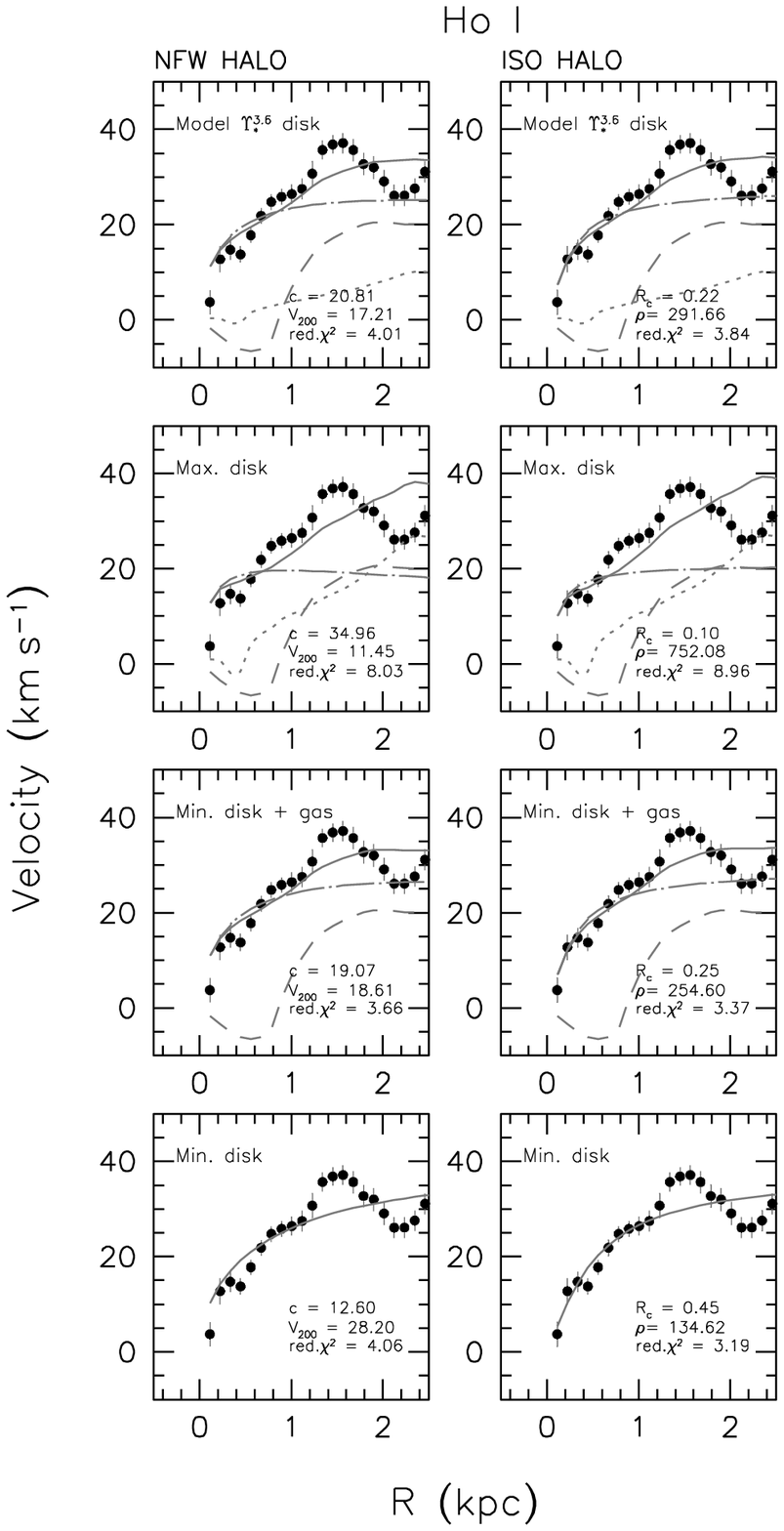}
\caption{{\bf Mass modeling results:} Disk-halo decomposition of the Ho I rotation curve 
under various \ML\ assumptions (\MLsps, maximum disk, $\rm minimum disk + gas$ and minimum disks).
The black dots indicate the bulk rotation curve, and the short and long dashed lines show the
rotation velocities of the stellar and gas components, respectively. The fitted parameters
of NFW and pseudo-isothermal halo models (long dash-dotted lines) are denoted on each panel.
\label{HoI_MD}}
\end{figure}
{\clearpage}

\begin{table*}
\scriptsize
\caption{Parameters of dark halo models for Ho I}
\label{NFWHALO_HoI}
\begin{center}
\begin{tabular}{@{}lcccc}
\hline
\hline
\noalign{\vskip 2pt}
& \multicolumn{3}{c}{NFW halo} \\
\cline{2-4} \\
\multicolumn{1}{c}{\ML\ assumption}   &  \multicolumn{1}{c}{$c$}   &  \multicolumn{1}{c}{$V_{200}$}  &  \multicolumn{1}{c}{$\chi_{red.}^2$} \\
\multicolumn{1}{c}{(1)} & \multicolumn{1}{c}{(2)}   & \multicolumn{1}{c}{(3)}      & \multicolumn{1}{c}{(4)}  \\
\hline
Min. disk        & 12.6 $\pm$ 2.8 ({\bf 9.0})      & 28.2 $\pm$ 4.3 ({\bf 36.9 $\pm$ 1.9})  & 4.05  ({\bf 4.56})         \\
Min. disk+gas    & 19.0 $\pm$ 4.1 ({\bf 9.0})     & 18.6 $\pm$ 2.1 ({\bf 35.6 $\pm$ 2.0})  & 3.66  ({\bf 4.90})         \\
Max. disk        & 34.9 $\pm$ 17.7 ({\bf 9.0})     & 11.4 $\pm$ 2.2 ({\bf 11.4 $\pm$ 2.2})  & 8.03  ({\bf 8.03})         \\
Model $\Upsilon_{*}^{3.6}$ disk & 20.8 $\pm$ 4.9 ({\bf 9.0}) & 17.2 $\pm$ 2.1 ({\bf 29.3 $\pm$ 2.3}) & 4.00 ({\bf 6.14}) \\
\hline\\
\noalign{\vskip 2pt}
 & \multicolumn{3}{c}{Pseudo-isothermal halo} \\
\cline{2-4} \\
\multicolumn{1}{c}{\ML\ assumption}    & \multicolumn{1}{c}{$R_C$} &   \multicolumn{1}{c}{$\rho_0$}    & \multicolumn{1}{c}{$\chi_{red.}^2$} \\
\multicolumn{1}{c}{(5)} & \multicolumn{1}{c}{(6)}   & \multicolumn{1}{c}{(7)}      & \multicolumn{1}{c}{(8)} \\
\hline
Min. disk        & 0.44 $\pm$ 0.07       & 134.6 $\pm$ 35.4         & 3.19       \\
Min. disk+gas    & 0.25 $\pm$ 0.06       & 254.6 $\pm$ 111.2         & 3.37 \\
Max. disk        & 0.10 $\pm$ 0.09       & 752.0 $\pm$ 1232.7         & 8.96  \\ 
Model $\Upsilon_{*}^{3.6}$ disk   & 0.22 $\pm$ 0.07 	   & 291.6 $\pm$ 152.9         & 3.84 \\
\hline
\end{tabular}
\medskip\noindent
\begin{minipage}{107mm}
\noindent
\\
{\bf Note.$\--$}
{\bf (1)(5):} The stellar mass-to-light ratio \ML\ assumptions. ``Model \MLsps\ disk'' uses the values derived from the population synthesis models in Section~\ref{Stars}.
{\bf (2):} Concentration parameter c of NFW halo model (NFW 1996, 1997). We also fit the NFW model to
the rotation curves with only $V_{200}$ as a free parameter after fixing $c$ to 9. The corresponding
best-fit $V_{200}$ and $\chi^{2}_{red}$ values are given in the brackets in (3) and (4), respectively.
{\bf (3):} The rotation velocity (\kms)\,at radius $R_{200}$ where the density constrast exceeds 200 (Navarro \etal\ 1996).
{\bf (4)(8):} Reduced $\chi^{2}$ value.
{\bf (6):} Fitted core-radius of pseudo-isothermal halo model (kpc).
{\bf (7):} Fitted core-density of pseudo-isothermal halo model ($10^{-3}$ \cubedens).
{\bf ($...$):} blank due to unphysically large value or not well-constrained uncertainties.
\end{minipage}
\end{center}
\end{table*}

\begin{figure}
\epsscale{1.0}
\figurenum{A.13}
\includegraphics[angle=0,width=1.0\textwidth,bb=60 80 540 680,clip=]{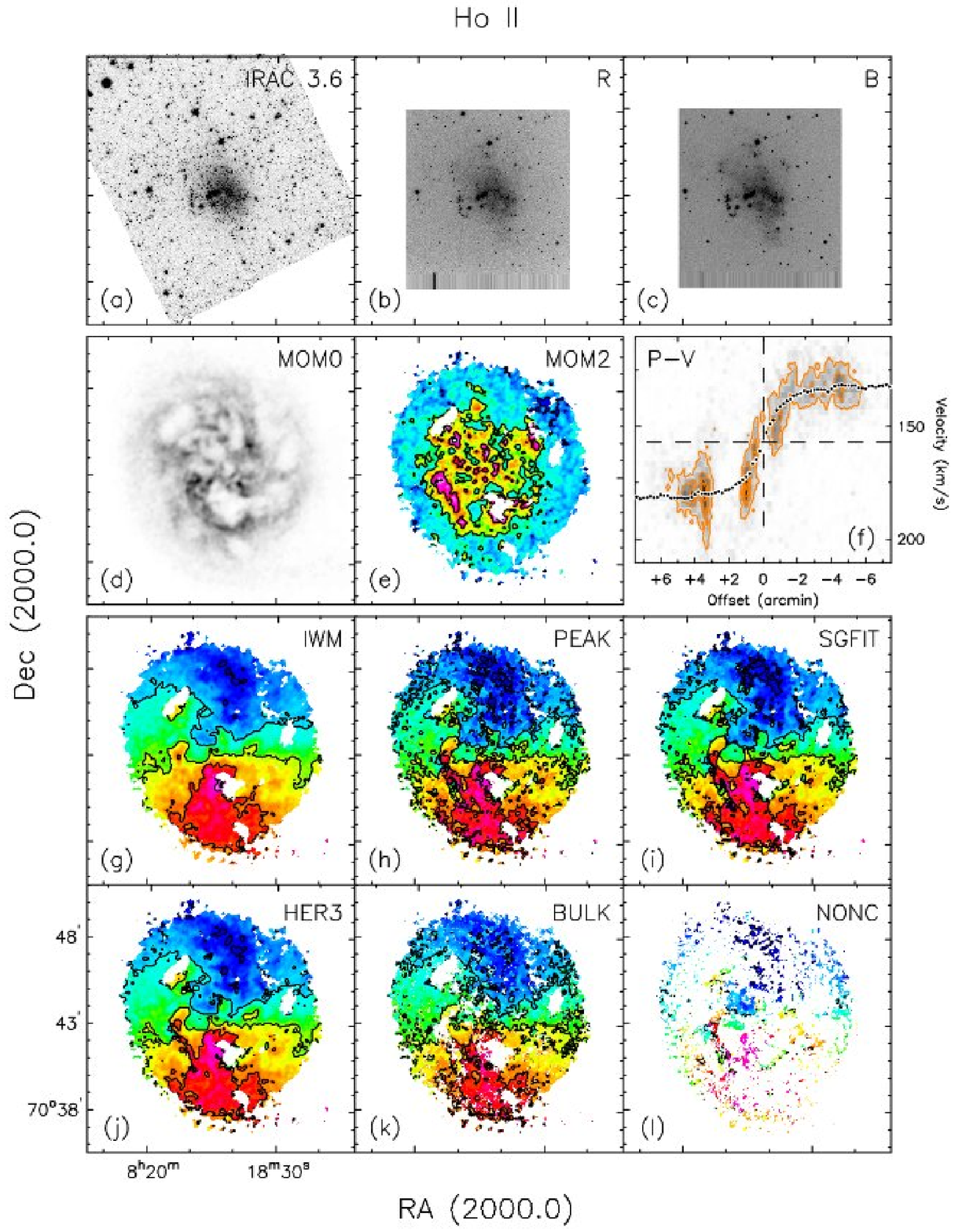}
\caption{{\bf Data:} Total intensity maps and velocity fields of Ho II. 
{\bf (a)(b)(c):}: Total intensity maps in {\it Spitzer} IRAC 3.6 $\mu$m, optical {\it R} and {\it B} bands.
{\bf (d):} Integrated H{\sc i} map (moment 0). The gray-scale levels run from 0 to 600 $\rm mJy\,beam^{-1}\,\kms$.
{\bf (e):} Velocity dispersion map (moment 2). Velocity contours run from $0$ to $25$\,\kms\,with a spacing of $5$\,\kms. 
{\bf (f):} Position-velocity diagram taken along the average position angle of the major axis as listed in Table 1. 
Contours start at $+2\sigma$ in steps of $5\sigma$. The dashed lines indicate the systemic velocity and position
of the kinematic center derived in this paper. Overplotted is the bulk rotation curve corrected for the average
inclination from the tilted-ring analysis as listed in Table 1. 
{\bf (g)(h)(i)(j)(k)(l):} Velocity fields. Contours run from $100$\,\kms\,to $200$\,\kms\,with a spacing of $15$\,\kms.
\label{HoII_MAPS}}
\end{figure}
{\clearpage}

\begin{figure}
\epsscale{1.0}
\figurenum{A.14}
\includegraphics[angle=0,width=1.0\textwidth,bb=45 140 570 720,clip=]{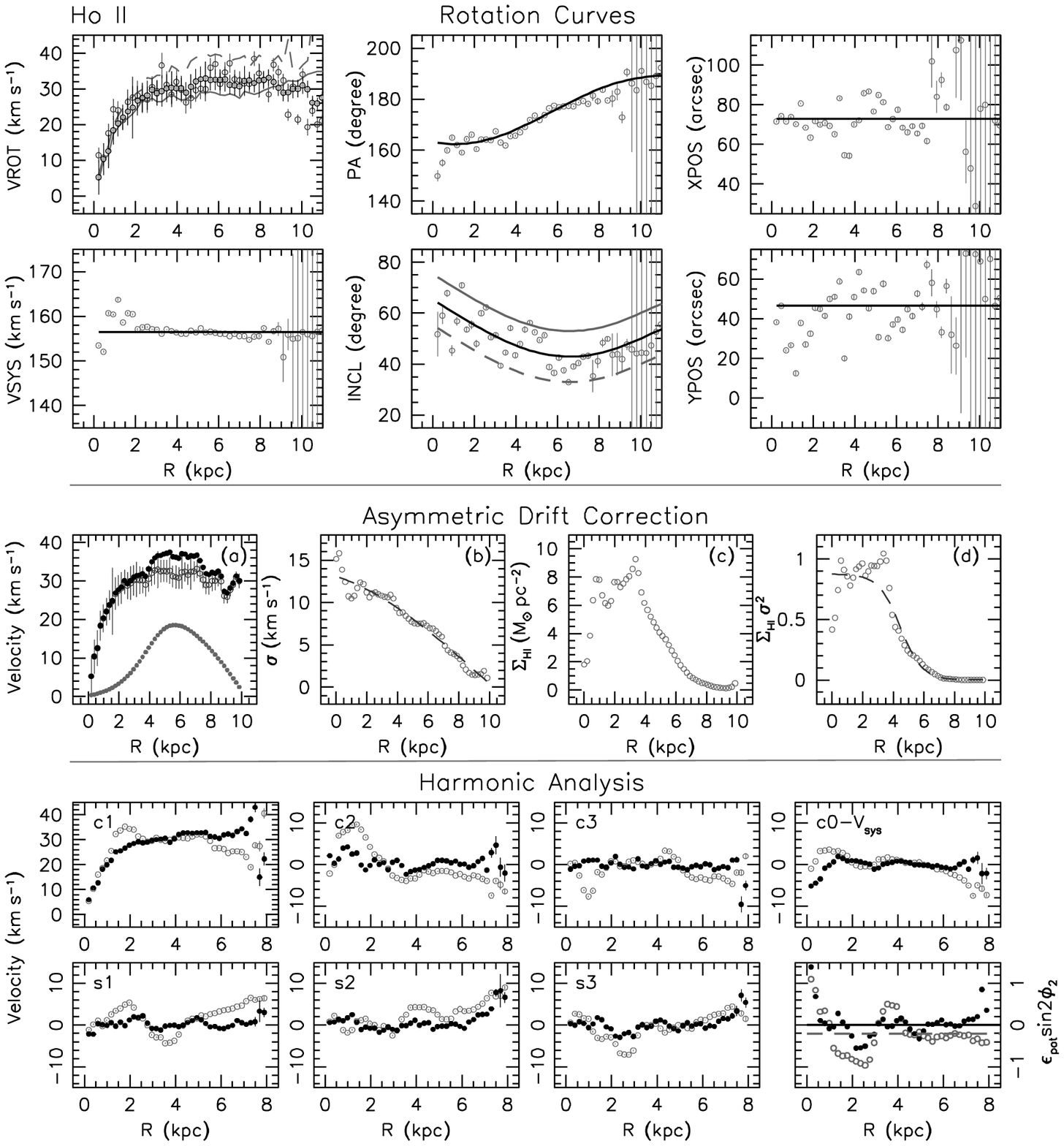}
\caption{{\bf Rotation curves:} The tilted ring model derived from the bulk velocity field of Ho II.
The open gray circles in all panels indicate the fit made with all ring parameters free. 
The gray dots in the VROT panel were derived using the entire velocity field after fixing other ring parameters
to the values (black solid lines) as shown in the panels. To examine the sensitivity of the rotation curve
to the inclination, we vary the inclination by $+10$ and $-10\,^{\circ}$ as indicated by the gray
solid and dashed lines, respectively, in the right-middle panel. 
We derive the rotation curves using these inclinations while keeping other ring
parameters the same. The resulting rotation curves are indicated
by gray solid (for $+10\,^{\circ}$ inclination) and dashed (for $-10\,^{\circ}$ inclination) lines in the VROT panel.
{\bf Asymmetric drift correction:}
{\bf a:} Gray filled dots indicate the derived radial velocity
correction for the asymmetric drift $\sigma_{\rm D}$. Black open and filled
dots represent the uncorrected and corrected curves for the asymmetric drift, respectively.
{\bf b:} Azimuthally averaged H{\sc i} velocity dispersion.
{\bf c:} Azimuthally averaged H{\sc i} surface density.
{\bf d:} The dashed line indicates a fit to $\Sigma\sigma^{2}$ with an analytical function.
{\bf Harmonic analysis:} Harmonic expansion of the velocity fields for Ho II.
The black dots and gray open circles indicate the results from the bulk and hermite $h_{3}$ velocity fields, respectively.
In the bottom-rightmost panel, the solid and dashed lines indicate global elongations of the potential
measured using the bulk and hermite $h_{3}$ velocity fields.
\label{HoII_TR_ADC_HD}}
\end{figure}
{\clearpage}

\begin{figure}
\epsscale{1.0}
\figurenum{A.15}
\includegraphics[angle=0,width=1.0\textwidth,bb=43 175 543 695,clip=]{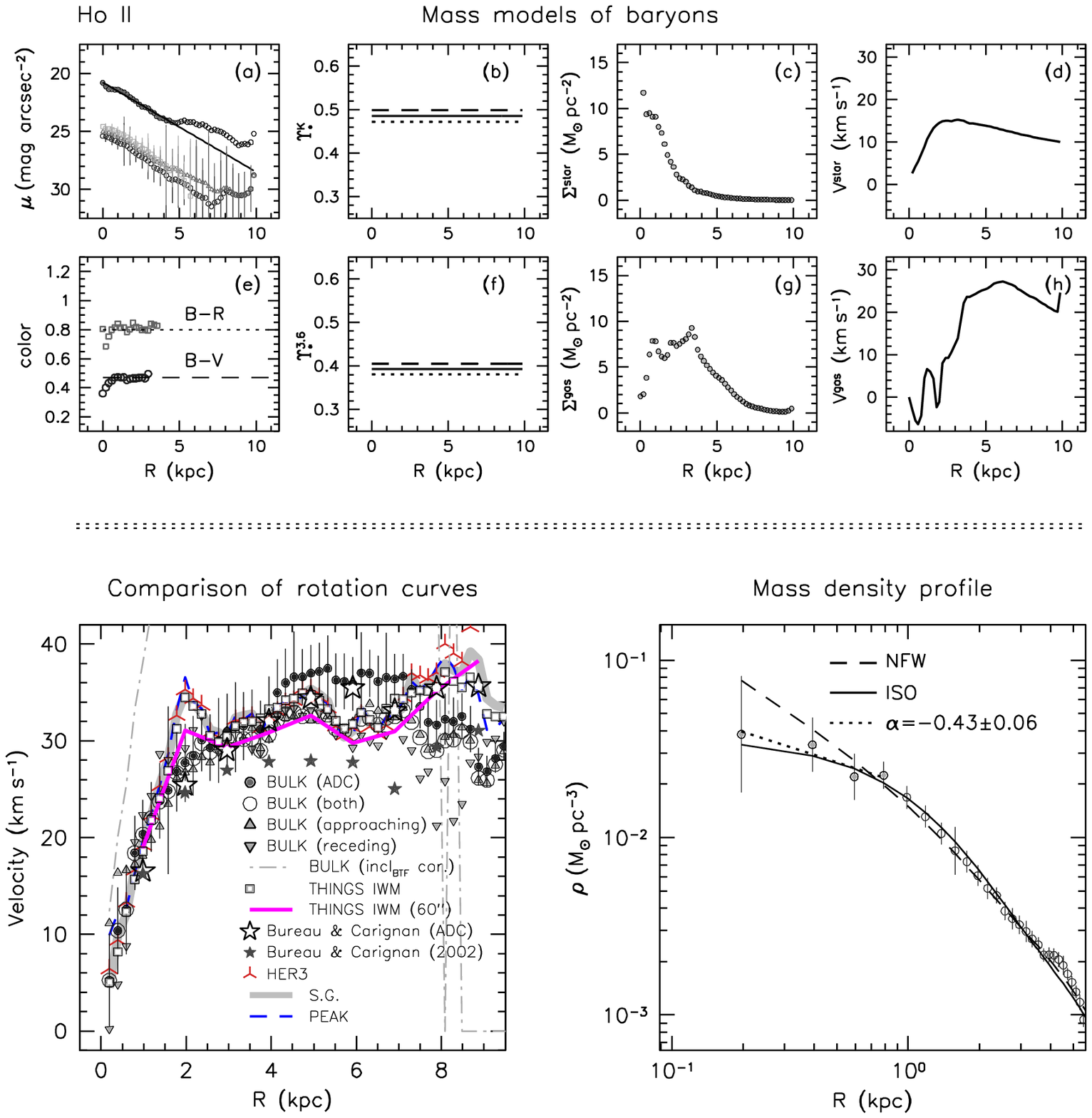}
\caption{
{\bf Mass models of baryons:} Mass models for the gas and stellar components of Ho II.
{\bf (a):} Azimuthally averaged surface brightness profiles in the 3.6$\mu$m, $R$, $V$ and $B$ bands (top to bottom).
{\bf (b)(f):} The stellar mass-to-light values in the $K$ and 3.6$\mu$m bands derived from stellar population synthesis models.
{\bf (c)(d):} The mass surface density and the resulting rotation velocity for the stellar component.
{\bf (e):} Optical colors.
{\bf (g)(h):} The mass surface density (scaled by 1.4 to account for He and metals) and the resulting
rotation velocity for the gas component.
{\bf Comparison of rotation curves:} Comparison of the H{\sc i} rotation curves derived from different types
of velocity fields (i.e., bulk, IWM, hermite $h_{3}$, single Gaussian and peak velocity fields as denoted
in the panel) and literature for Ho II. See Section~\ref{Deriving_Rotcurs} for more information.
{\bf Mass density profile:} The derived mass density profile of Ho II.
The open circles represent the mass density profile derived from the bulk rotation curve
assuming minimum disk. The inner density slope $\alpha$ is measured by a least squares fit (dotted line)
to the data points indicated by gray dots, and shown in the panel.
\label{HoII_VROT_BARYONS_ALPHA}}
\end{figure}
{\clearpage}

\begin{figure}
\epsscale{1.0}
\figurenum{A.16}
\includegraphics[angle=0,width=1.0\textwidth,bb=20 150 580 720,clip=]{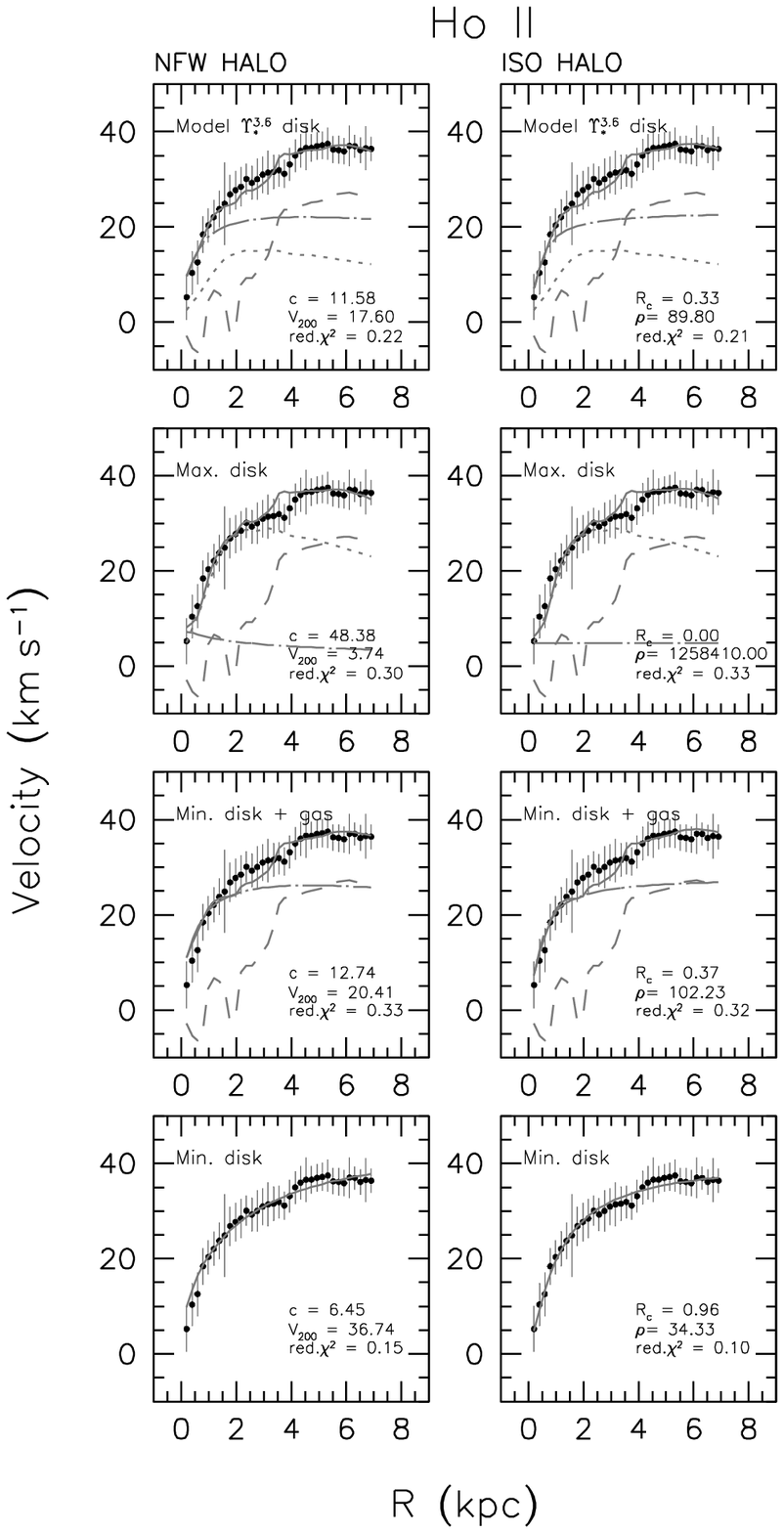}
\caption{{\bf Mass modeling results:} Disk-halo decomposition of the Ho II rotation curve 
under various \ML\ assumptions (\MLsps, maximum disk, $\rm minimum disk + gas$ and minimum disks).
The black dots indicate the bulk rotation curve, and the short and long dashed lines show the
rotation velocities of the stellar and gas components, respectively. The fitted parameters
of NFW and pseudo-isothermal halo models (long dash-dotted lines) are denoted on each panel.
\label{HoII_MD}}
\end{figure}
{\clearpage}

\begin{table*}
\scriptsize
\caption{Parameters of dark halo models for Ho II}
\label{NFWHALO_HoII}
\begin{center}
\begin{tabular}{@{}lcccc}
\hline
\hline
\noalign{\vskip 2pt}
& \multicolumn{3}{c}{NFW halo} \\
\cline{2-4} \\
\multicolumn{1}{c}{\ML\ assumption}   &  \multicolumn{1}{c}{$c$}   &  \multicolumn{1}{c}{$V_{200}$}  &  \multicolumn{1}{c}{$\chi_{red.}^2$} \\
\multicolumn{1}{c}{(1)} & \multicolumn{1}{c}{(2)}   & \multicolumn{1}{c}{(3)}      & \multicolumn{1}{c}{(4)}    \\
\hline
Min. disk        & 6.4 $\pm$ 0.4 ({\bf 9.0})     & 36.7 $\pm$ 1.3 ({\bf 31.2 $\pm$ 0.3})  & 0.15  ({\bf 0.28})         \\
Min. disk+gas    & 12.7 $\pm$ 1.4 ({\bf 9.0})      & 20.4 $\pm$ 0.7 ({\bf 22.9 $\pm$ 0.5})  & 0.32  ({\bf 0.42})         \\
Max. disk        & 48.3 $\pm$ 86.8 ({\bf 9.0})      & 3.7 $\pm$ 1.4 ({\bf 4.02 $\pm$ 1.7})  & 0.30  ({\bf 0.32})         \\
Model $\Upsilon_{*}^{3.6}$ disk & 11.5 $\pm$ 1.3 ({\bf 9.0}) & 17.6 $\pm$ 0.7 ({\bf 19.0 $\pm$ 0.4}) & 0.22 ({\bf 0.24}) \\
\hline\\
\noalign{\vskip 2pt}
 & \multicolumn{3}{c}{Pseudo-isothermal halo} \\
\cline{2-4} \\
\multicolumn{1}{c}{\ML\ assumption}    & \multicolumn{1}{c}{$R_C$} &   \multicolumn{1}{c}{$\rho_0$}    & \multicolumn{1}{c}{$\chi_{red.}^2$} \\
\multicolumn{1}{c}{(5)} & \multicolumn{1}{c}{(6)}   & \multicolumn{1}{c}{(7)}      & \multicolumn{1}{c}{(8)} \\
\hline
Min. disk        & 0.95 $\pm$ 0.04       & 34.3 $\pm$ 2.6         & 0.10       \\
Min. disk+gas    & 0.37 $\pm$ 0.05       & 102.2 $\pm$ 28.2         & 0.31 \\
Max. disk        & 0.00 $\pm$ 0.24       & $...$ $\pm$ $...$         & 0.33  \\ 
Model $\Upsilon_{*}^{3.6}$ disk   & 0.33 $\pm$ 0.05 	   & 89.8 $\pm$ 27.2         & 0.20 \\
\hline
\end{tabular}
\medskip\noindent
\begin{minipage}{107mm}
\noindent
\\
{\bf Note.$\--$}
{\bf (1)(5):} The stellar mass-to-light ratio \ML\ assumptions. ``Model \MLsps\ disk'' uses the values derived from the population synthesis models in Section~\ref{Stars}.
{\bf (2):} Concentration parameter c of NFW halo model (NFW 1996, 1997). We also fit the NFW model to
the rotation curves with only $V_{200}$ as a free parameter after fixing $c$ to 9. The corresponding
best-fit $V_{200}$ and $\chi^{2}_{red}$ values are given in the brackets in (3) and (4), respectively.
{\bf (3):} The rotation velocity (\kms)\,at radius $R_{200}$ where the density constrast exceeds 200 (Navarro \etal\ 1996).
{\bf (4)(8):} Reduced $\chi^{2}$ value.
{\bf (6):} Fitted core-radius of pseudo-isothermal halo model (kpc).
{\bf (7):} Fitted core-density of pseudo-isothermal halo model ($10^{-3}$ \cubedens).
{\bf ($...$):} blank due to unphysically large value or not well-constrained uncertainties.
\end{minipage}
\end{center}
\end{table*}

\begin{figure}
\figurenum{A.17}
\epsscale{1.0}
\includegraphics[angle=0,width=1.0\textwidth,bb=60 80 540 680,clip=]{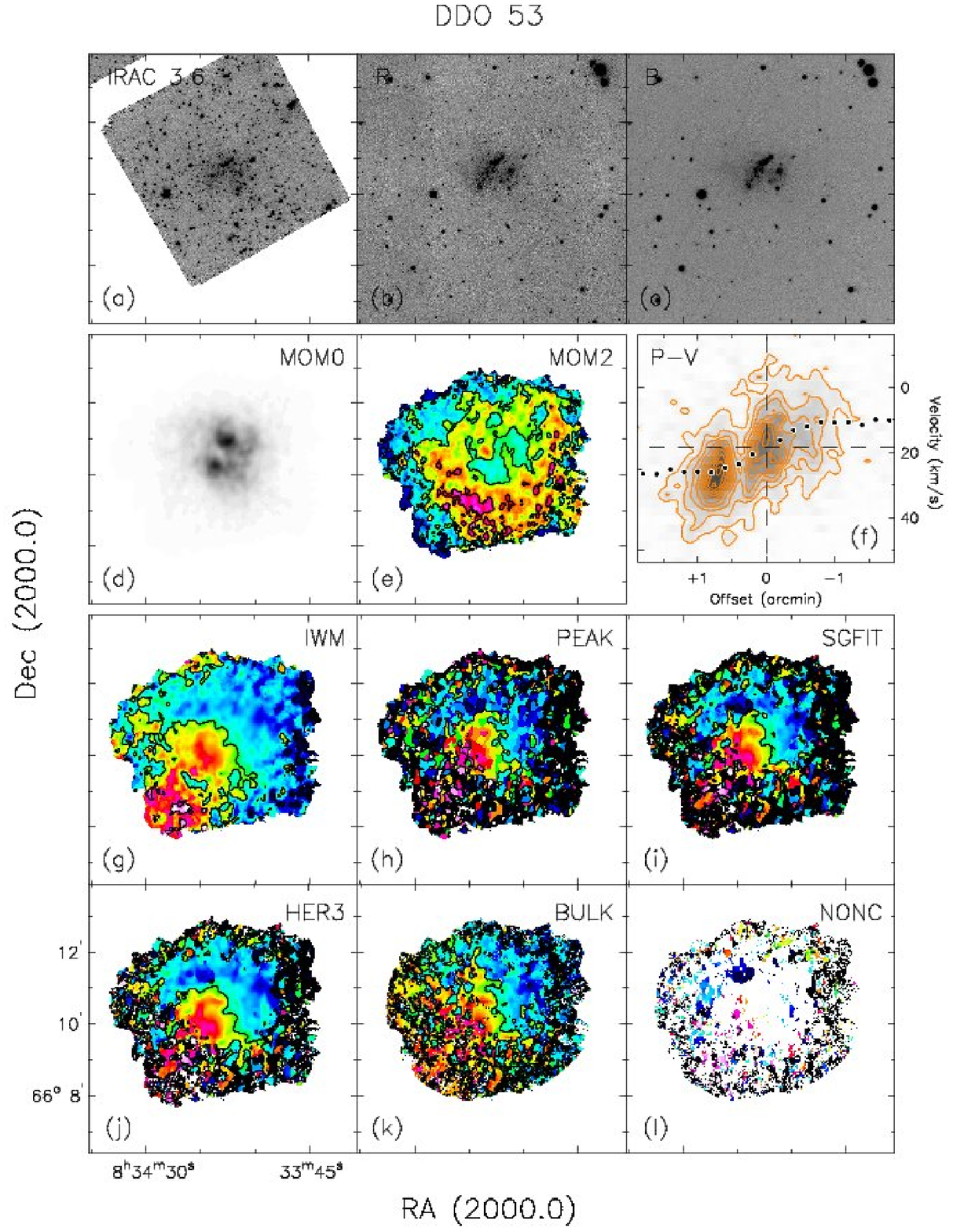}
\caption{{\bf Data:} Total intensity maps and velocity fields of DDO 53.
{\bf (a)(b)(c):}: Total intensity maps in {\it Spitzer} IRAC 3.6 $\mu$m, optical {\it R} and {\it B} bands.
{\bf (d):} Integrated H{\sc i} map (moment 0). The gray-scale levels run from 0 to 350 $\rm mJy\,beam^{-1}\,\kms$.
{\bf (e):} Velocity dispersion map (moment 2). Velocity contours run from $0$ to $25$\,\kms\,with a spacing of $5$\,\kms. 
{\bf (f):} Position-velocity diagram taken along the average position angle of the major axis as listed in Table 1. 
Contours start at $+2\sigma$ in steps of $3\sigma$. The dashed lines indicate the systemic velocity and position
of the kinematic center derived in this paper. Overplotted is the bulk rotation curve corrected for the average
inclination from the tilted-ring analysis as listed in Table 1. 
{\bf (g)(h)(i)(j)(k)(l):} Velocity fields. Contours run from $-10$\,\kms\,to $40$\,\kms\,with a spacing of $15$\,\kms.
\label{DDO53_MAPS}}
\end{figure}
{\clearpage}

\begin{figure}
\epsscale{1.0}
\figurenum{A.18}
\includegraphics[angle=0,width=1.0\textwidth,bb=45 140 570 720,clip=]{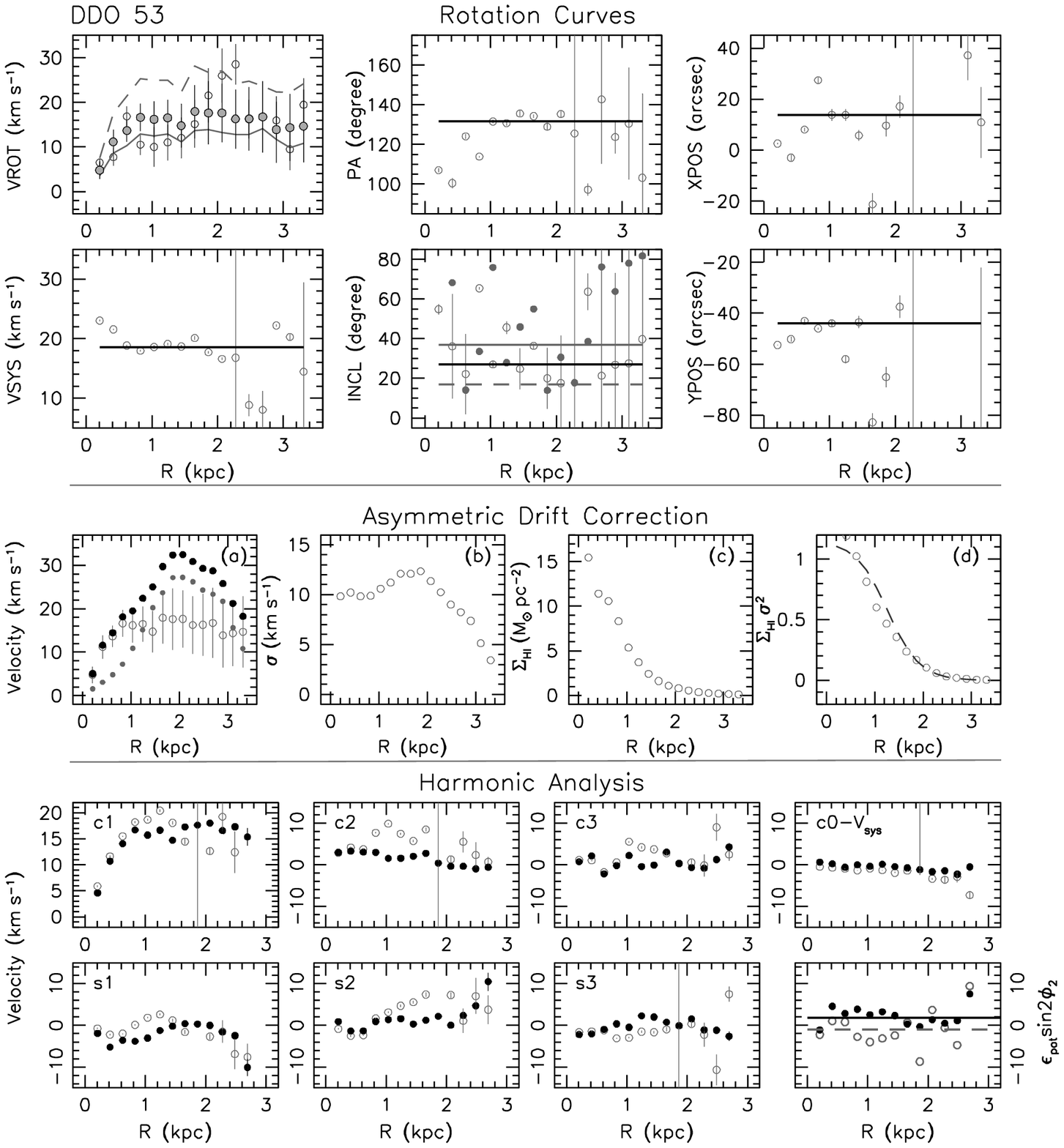}
\caption{{\bf Rotation curves:} The tilted ring model derived from the bulk velocity field of DDO 53.
The open gray circles in all panels indicate the fit made with all ring parameters free. 
The gray dots in the VROT panel were derived using the entire velocity field after fixing other ring parameters
to the values (black solid lines) as shown in the panels. To examine the sensitivity of the rotation curve
to the inclination, we vary the inclination by $+10$ and $-10\,^{\circ}$ as indicated by the gray
solid and dashed lines, respectively, in the right-middle panel. 
We derive the rotation curves using these inclinations while keeping other ring
parameters the same. The resulting rotation curves are indicated
by gray solid (for $+10\,^{\circ}$ inclination) and dashed (for $-10\,^{\circ}$ inclination) lines in the VROT panel.
{\bf Asymmetric drift correction:}
{\bf a:} Gray filled dots indicate the derived radial velocity
correction for the asymmetric drift $\sigma_{\rm D}$. Black open and filled
dots represent the uncorrected and corrected curves for the asymmetric drift, respectively.
{\bf b:} Azimuthally averaged H{\sc i} velocity dispersion.
{\bf c:} Azimuthally averaged H{\sc i} surface density.
{\bf d:} The dashed line indicates a fit to $\Sigma\sigma^{2}$ with an analytical function.
{\bf Harmonic analysis:} Harmonic expansion of the velocity fields for DDO 53.
The black dots and gray open circles indicate the results from the bulk and hermite $h_{3}$ velocity fields, respectively.
In the bottom-rightmost panel, the solid and dashed lines indicate global elongations of the potential
measured using the bulk and hermite $h_{3}$ velocity fields.
\label{DDO53_TR_ADC_HD}}
\end{figure}
{\clearpage}

\begin{figure}
\epsscale{1.0}
\figurenum{A.19}
\includegraphics[angle=0,width=1.0\textwidth,bb=43 175 543 695,clip=]{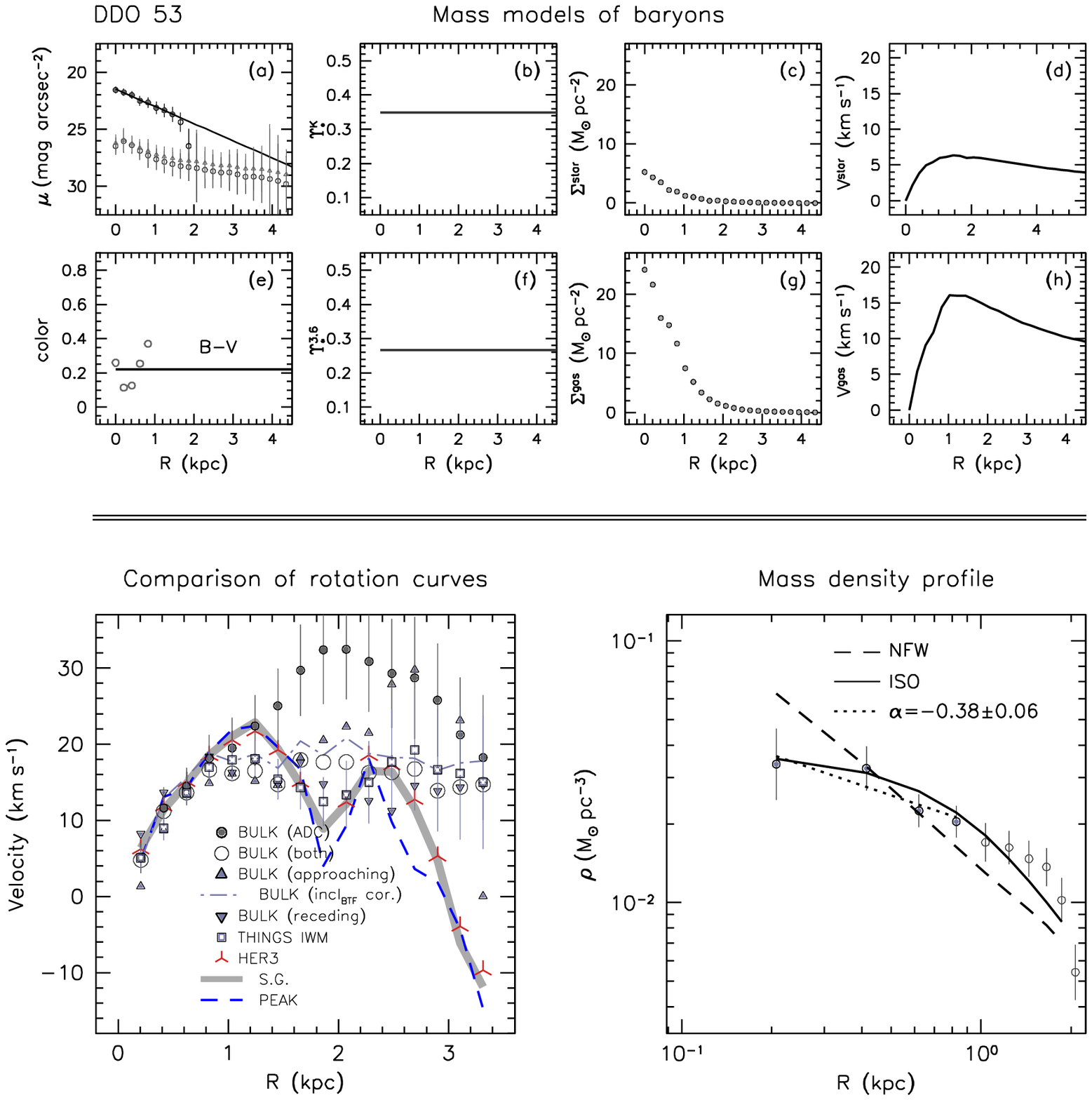}
\caption{
{\bf Mass models of baryons:} Mass models for the gas and stellar components of DDO 53.
{\bf (a):} Azimuthally averaged surface brightness profiles in the 3.6$\mu$m, $V$ and $B$ bands (top to bottom).
{\bf (b)(f):} The stellar mass-to-light values in the $K$ and 3.6$\mu$m bands derived from stellar population synthesis models.
{\bf (c)(d):} The mass surface density and the resulting rotation velocity for the stellar component.
{\bf (e):} Optical color.
{\bf (g)(h):} The mass surface density (scaled by 1.4 to account for He and metals) and the resulting
rotation velocity for the gas component.
{\bf Comparison of rotation curves:} Comparison of the H{\sc i} rotation curves derived using different types
of velocity fields (i.e., bulk, IWM, hermite $h_{3}$, single Gaussian and peak velocity fields as denoted
in the panel) for DDO 53. See Section~\ref{Deriving_Rotcurs} for more information.
{\bf Mass density profile:} The derived mass density profile of DDO 53.
The open circles represent the mass density profile derived from the bulk rotation curve
assuming minimum disk. The inner density slope $\alpha$ is measured by a least squares fit (dotted line)
to the data points indicated by gray dots, and shown in the panel.
\label{DDO53_VROT_BARYONS_ALPHA}}
\end{figure}
{\clearpage}

\begin{figure}
\epsscale{1.0}
\figurenum{A.20}
\includegraphics[angle=0,width=1.0\textwidth,bb=20 150 580 720,clip=]{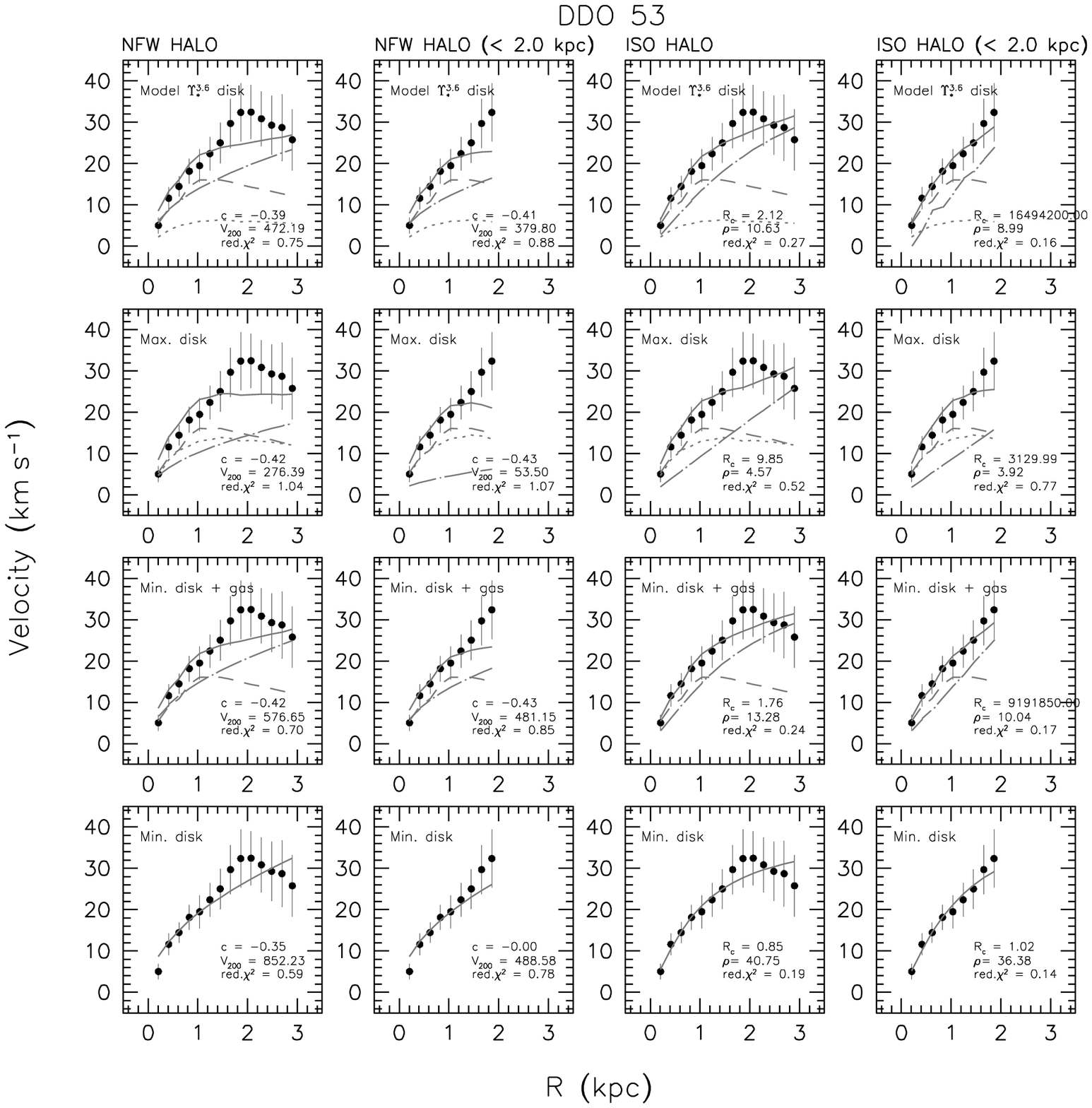}
\caption{{\bf Mass modeling results:} Disk-halo decomposition of the DDO 53 rotation curve
under various \ML\ assumptions (\MLsps, maximum disk, $\rm minimum disk + gas$ and minimum disks).
The black dots indicate the bulk rotation curve, and the short and long dashed lines show the
rotation velocities of the stellar and gas components, respectively. The fitted parameters
of NFW and pseudo-isothermal halo models (long dash-dotted lines) are denoted on each panel.
\label{DDO53_MD}}
\end{figure}
{\clearpage}

\begin{table*}
\scriptsize
\caption{Parameters of dark halo models for DDO 53}
\label{NFWHALO_DDO53}
\begin{center}
\begin{tabular}{@{}lccccccrc}
\hline
\hline
\noalign{\vskip 2pt}
& \multicolumn{3}{c}{NFW halo (entire region)} && \multicolumn{3}{c}{NFW halo ($<$ 2.0 kpc)}\\
\cline{2-4} \cline{6-8}\\
\multicolumn{1}{c}{\ML\ assumption}   &  \multicolumn{1}{c}{$c$}   &  \multicolumn{1}{c}{$V_{200}$}  &  \multicolumn{1}{c}{$\chi_{red.}^2$}    &&  \multicolumn{1}{c}{$c$}   &  \multicolumn{1}{c}{$V_{200}$}   &  \multicolumn{1}{c}{$\chi_{red.}^2$}\\
\multicolumn{1}{c}{(1)} & \multicolumn{1}{c}{(2)}   & \multicolumn{1}{c}{(3)}      & \multicolumn{1}{c}{(4)}  && \multicolumn{1}{c}{(5)}  & \multicolumn{1}{c}{(6)} & \multicolumn{1}{c}{(7)}  \\
\hline
Min. disk        & $<0.1$ ({\bf 9.0})      & 852.2 $\pm$ $...$ ({\bf 24.2 $\pm$ 2.3})  & 0.59  ({\bf 0.86})      &&  $<0.1$       & 488.5 $\pm$ $...$    & 0.78\\
Min. disk+gas    & $<0.1$ ({\bf 9.0})      & 576.6 $\pm$ $...$ ({\bf 13.9 $\pm$ 2.5})  & 0.70  ({\bf 0.88})      &&  $<0.1$        & 481.1 $\pm$ $...$    & 0.84\\
Max. disk        & $<0.1$ ({\bf 9.0})     & 276.4 $\pm$ $...$ ({\bf 0.04 $\pm$ $...$})  & 1.04  ({\bf 6.51})      &&  $<0.1$        & 53.5 $\pm$ $...$    & 1.06\\
Model $\Upsilon_{*}^{3.6}$ disk & $<0.1$ ({\bf 9.0})  & 472.1 $\pm$ $...$ ({\bf 12.1 $\pm$ 2.6}) & 0.75 ({\bf 7.12}) &&  $<0.1$   & 379.7 $\pm$ $...$    & 0.88\\
\hline\\
\noalign{\vskip 2pt}
 & \multicolumn{3}{c}{Pseudo-isothermal halo (entire region)} && \multicolumn{3}{c}{Pseudo-isothermal halo ($<$ 6.0 kpc)}\\
\cline{2-4} \cline{6-8} \\
\multicolumn{1}{c}{\ML\ assumption}    & \multicolumn{1}{c}{$R_C$} &   \multicolumn{1}{c}{$\rho_0$}    & \multicolumn{1}{c}{$\chi_{red.}^2$}     && \multicolumn{1}{c}{$R_C$}      & \multicolumn{1}{c}{$\rho_0$}   & \multicolumn{1}{c}{$\chi_{red.}^2$}\\
\multicolumn{1}{c}{(8)} & \multicolumn{1}{c}{(9)}   & \multicolumn{1}{c}{(10)}      & \multicolumn{1}{c}{(11)}    && \multicolumn{1}{c}{(12)}  & \multicolumn{1}{c}{(13)} & \multicolumn{1}{c}{(14)} \\
\hline
Min. disk        & 0.85 $\pm$ 0.10       & 40.7 $\pm$ 5.1         & 0.18         && 1.02 $\pm$ 0.16       & 36.3 $\pm$ 4.4         & 0.13\\
Min. disk+gas    & 1.75 $\pm$ 0.54       & 13.2 $\pm$ 3.1         & 0.24         && $...$       & 10.0 $\pm$ 0.9         & 0.17\\
Max. disk        & 9.85 $\pm$ 81.9       & 4.5 $\pm$ 2.6         & 0.52          && $...$       & 3.9 $\pm$ 5.4         & 0.76\\
Model $\Upsilon_{*}^{3.6}$ disk & 2.11 $\pm$ 0.88 & 10.6 $\pm$ 2.9 & 0.26         && $...$       & 8.9 $\pm$ 1.0         & 0.15\\
\hline
\end{tabular}
\medskip\noindent
\begin{minipage}{154mm}
\noindent
\\
{\bf Note.$\--$}
{\bf (1)(8):} The stellar mass-to-light ratio \ML\ assumptions. ``Model \MLsps\ disk'' uses the values derived from the population synthesis models in Section~\ref{Stars}.
{\bf (2)(5):} Concentration parameter c of NFW halo model (NFW 1996, 1997). We also fit the NFW model to
the rotation curves with only $V_{200}$ as a free parameter after fixing $c$ to 9. The corresponding
best-fit $V_{200}$ and $\chi^{2}_{red}$ values are given in the brackets in (3) and (4), respectively.
{\bf (3)(6):} The rotation velocity (\kms)\,at radius $R_{200}$ where the density constrast exceeds 200 (Navarro \etal\ 1996).
{\bf (4)(7)(11)(14):} Reduced $\chi^{2}$ value.
{\bf (9)(12):} Fitted core-radius of pseudo-isothermal halo model (kpc).
{\bf (10)(13):} Fitted core-density of pseudo-isothermal halo model ($10^{-3}$ \cubedens).
{\bf ($...$):} blank due to unphysically large value or not well-constrained uncertainties.
\end{minipage}
\end{center}
\end{table*}

\begin{figure}
\epsscale{1.0}
\figurenum{A.21}
\includegraphics[angle=0,width=1.0\textwidth,bb=60 80 540 680,clip=]{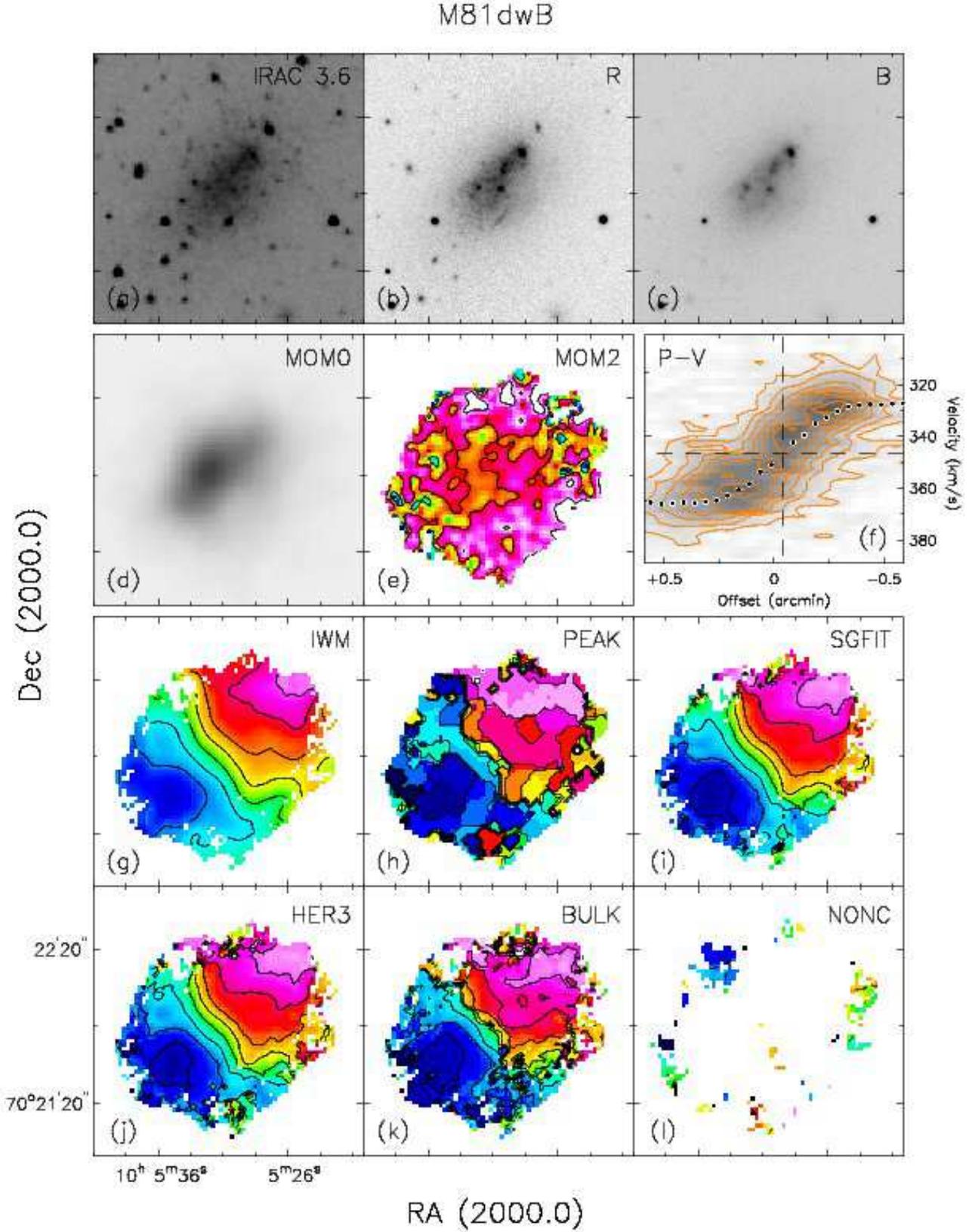}
\caption{{\bf Data:} Total intensity maps and velocity fields of M81dwB.
{\bf (a)(b)(c):}: Total intensity maps in {\it Spitzer} IRAC 3.6 $\mu$m, optical {\it R} and {\it B} bands.
{\bf (d):} Integrated H{\sc i} map (moment 0). The gray-scale levels run from 0 to 500 $\rm mJy\,beam^{-1}\,\kms$.
{\bf (e):} Velocity dispersion map (moment 2). Velocity contours run from $0$ to $25$\,\kms\,with a spacing of $5$\,\kms. 
{\bf (f):} Position-velocity diagram taken along the average position angle of the major axis as listed in Table 1. 
Contours start at $+2\sigma$ in steps of $3\sigma$. The dashed lines indicate the systemic velocity and position
of the kinematic center derived in this paper. Overplotted is the bulk rotation curve corrected for the average
inclination from the tilted-ring analysis as listed in Table 1. 
{\bf (g)(h)(i)(j)(k)(l):} Velocity fields. Contours run from $320$\,\kms\,to $380$\,\kms\,with a spacing of $5$\,\kms.
\label{M81DWB_MAPS}}
\end{figure}
{\clearpage}

\begin{figure}
\epsscale{1.0}
\figurenum{A.22}
\includegraphics[angle=0,width=1.0\textwidth,bb=45 140 570 720,clip=]{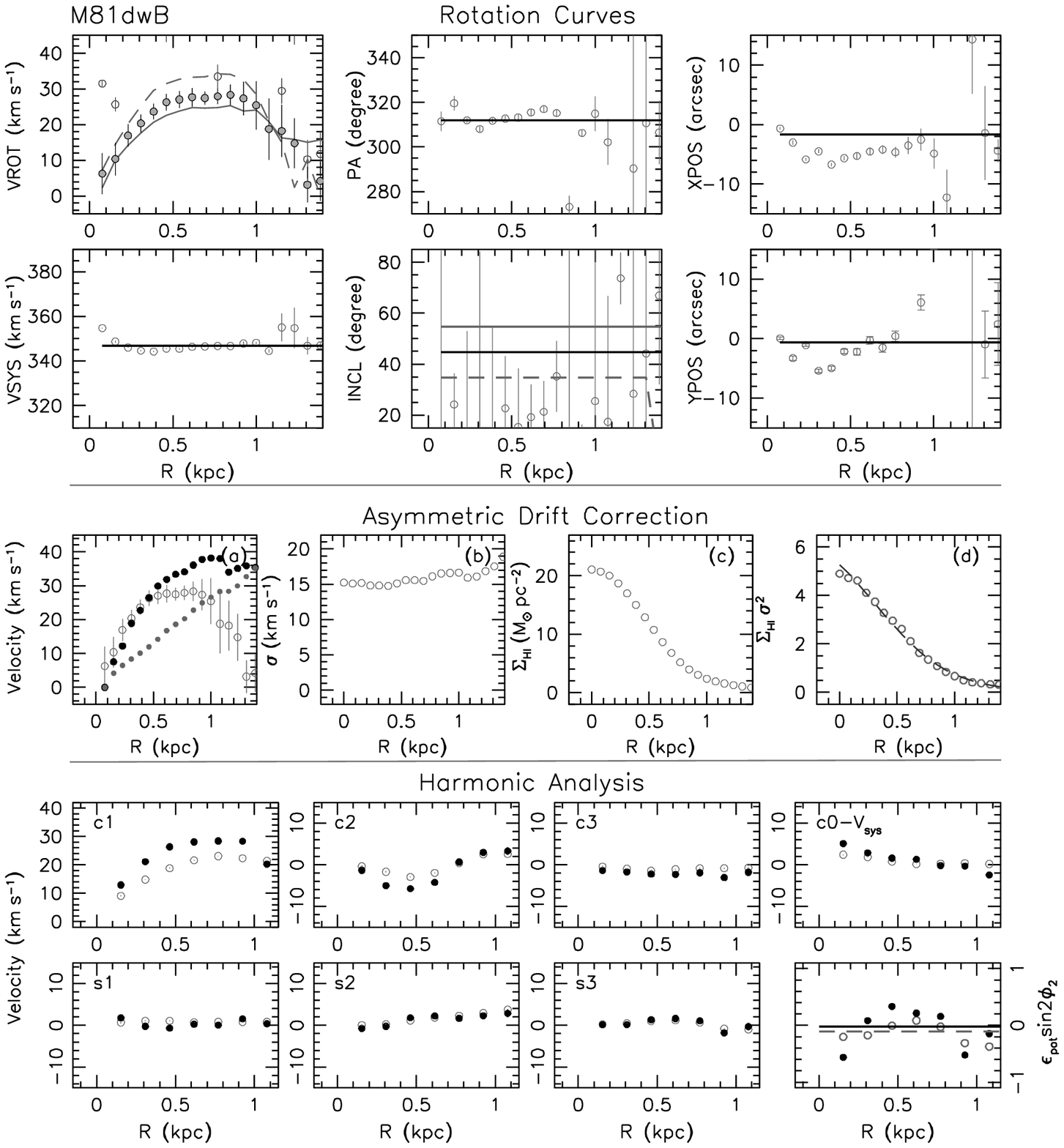}
\caption{{\bf Rotation curves:} The tilted ring model derived from the bulk velocity field of M81dwB.
The open gray circles in all panels indicate the fit made with all ring parameters free. 
The gray dots in the VROT panel were derived using the entire velocity field after fixing other ring parameters
to the values (black solid lines) as shown in the panels. To examine the sensitivity of the rotation curve
to the inclination, we vary the inclination by $+10$ and $-10\,^{\circ}$ as indicated by the gray
solid and dashed lines, respectively, in the right-middle panel. 
We derive the rotation curves using these inclinations while keeping other ring
parameters the same. The resulting rotation curves are indicated
by gray solid (for $+10\,^{\circ}$ inclination) and dashed (for $-10\,^{\circ}$ inclination) lines in the VROT panel.
{\bf Asymmetric drift correction:}
{\bf a:} Gray filled dots indicate the derived radial velocity
correction for the asymmetric drift $\sigma_{\rm D}$. Black open and filled
dots represent the uncorrected and corrected curves for the asymmetric drift, respectively.
{\bf b:} Azimuthally averaged H{\sc i} velocity dispersion.
{\bf c:} Azimuthally averaged H{\sc i} surface density.
{\bf d:} The dashed line indicates a fit to $\Sigma\sigma^{2}$ with an analytical function.
{\bf Harmonic analysis:} Harmonic expansion of the velocity fields for M81dwB.
The black dots and gray open circles indicate the results from the hermite $h_{3}$ and IWM velocity fields, respectively.
In the bottom-rightmost panel, the solid and dashed lines indicate global elongations of the potential
measured using the hermite $h_{3}$ and IWM velocity fields.
\label{M81DWB_TR_ADC_HD}}
\end{figure}
{\clearpage}

\begin{figure}
\epsscale{1.0}
\figurenum{A.23}
\includegraphics[angle=0,width=1.0\textwidth,bb=43 175 543 695,clip=]{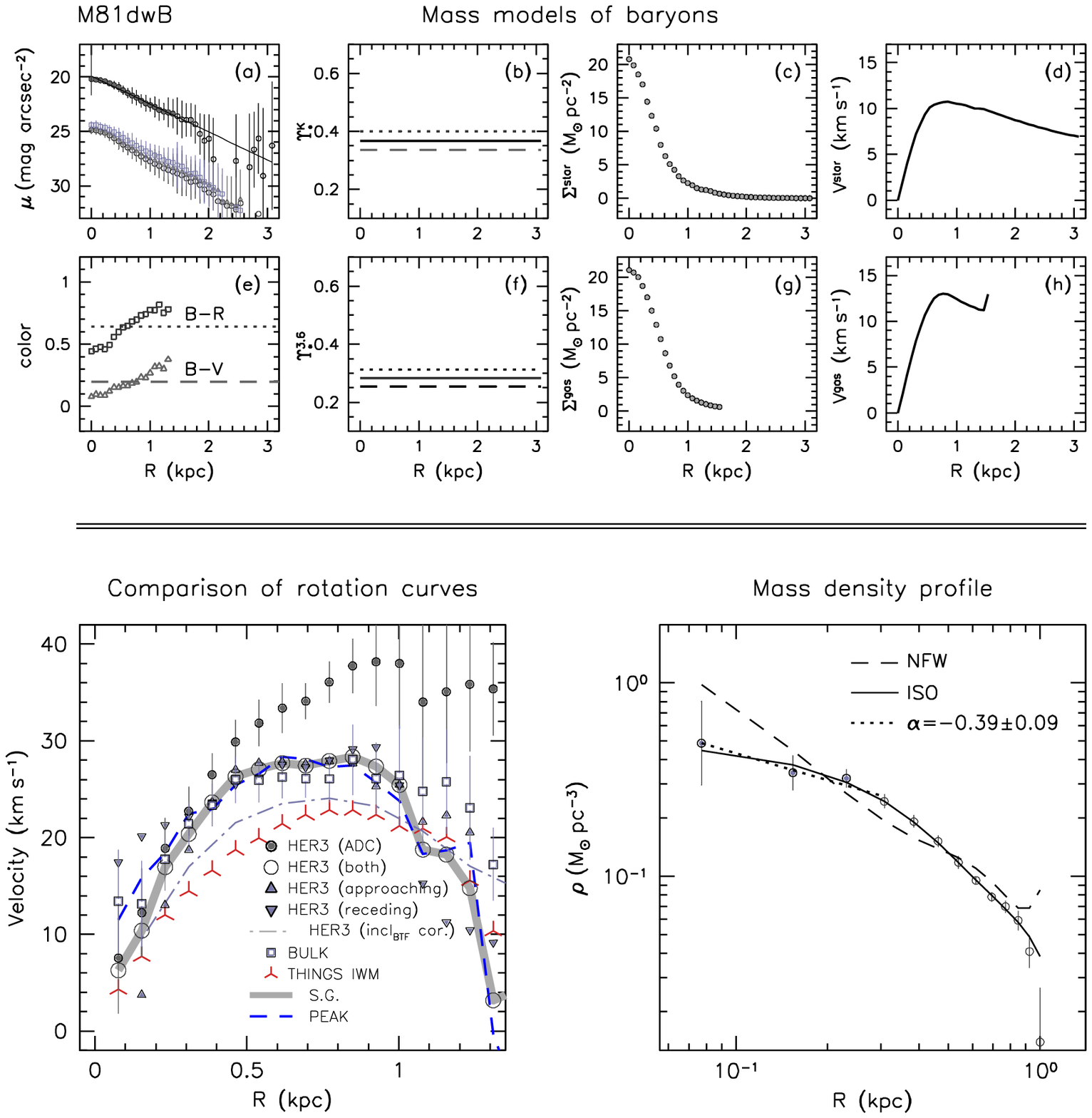}
\caption{
{\bf Mass models of baryons:} Mass models for the gas and stellar components of M81dwB.
{\bf (a):} Azimuthally averaged surface brightness profiles in the 3.6$\mu$m, $R$, $V$ and $B$ bands (top to bottom).
{\bf (b)(f):} The stellar mass-to-light values in the $K$ and 3.6$\mu$m bands derived from stellar population synthesis models.
{\bf (c)(d):} The mass surface density and the resulting rotation velocity for the stellar component.
{\bf (e):} Optical colors.
{\bf (g)(h):} The mass surface density (scaled by 1.4 to account for He and metals) and the resulting
rotation velocity for the gas component.
{\bf Comparison of rotation curves:} Comparison of the H{\sc i} rotation curves derived using different types
of velocity fields (i.e., bulk, IWM, hermite $h_{3}$, single Gaussian and peak velocity fields as denoted in
the panel) for M81dwB. See Section~\ref{Deriving_Rotcurs} for more information.
{\bf Mass density profile:} The derived mass density profile of M81dwB.
The open circles represent the mass density profile derived from the hermite $h_{3}$ rotation curve
assuming minimum disk. The inner density slope $\alpha$ is measured by a least squares fit (dotted line)
to the data points indicated by gray dots, and shown in the panel.
\label{M81DWB_VROT_BARYONS_ALPHA}}
\end{figure}
{\clearpage}

\begin{figure}
\figurenum{A.24}
\epsscale{1.0}
\includegraphics[angle=0,width=1.0\textwidth,bb=20 150 580 720,clip=]{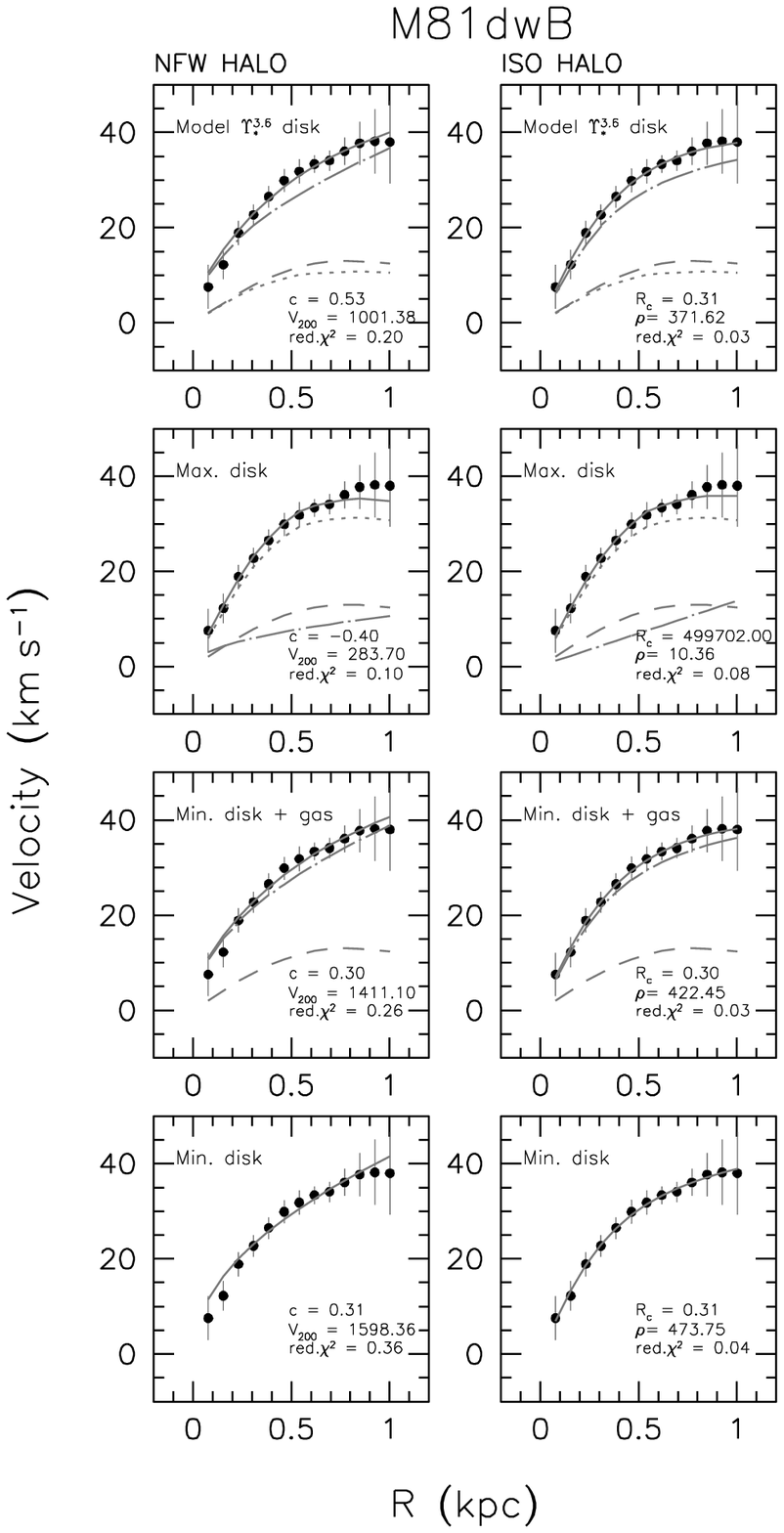}
\caption{{\bf Mass modeling results:} Disk-halo decomposition of the M81dwB rotation curve 
under various \ML\ assumptions (\MLsps, maximum disk, $\rm minimum disk + gas$ and minimum disks).
The black dots indicate the hermite $h_{3}$ rotation curve, and the short and long dashed lines show the
rotation velocities of the stellar and gas components, respectively. The fitted parameters
of NFW and pseudo-isothermal halo models (long dash-dotted lines) are denoted on each panel.
\label{M81DWB_MD}}
\end{figure}
{\clearpage}

\begin{table*}
\scriptsize
\caption{Parameters of dark halo models for M81 dwB}
\label{NFWHALO_M81dwB}
\begin{center}
\begin{tabular}{@{}lrcccc}
\hline
\hline
\noalign{\vskip 2pt}
& \multicolumn{3}{c}{NFW halo} \\
\cline{2-4} \\
\multicolumn{1}{c}{\ML\ assumption}   &  \multicolumn{1}{c}{$c$}   &  \multicolumn{1}{c}{$V_{200}$}  &  \multicolumn{1}{c}{$\chi_{red.}^2$} \\
\multicolumn{1}{c}{(1)} & \multicolumn{1}{c}{(2)}   & \multicolumn{1}{c}{(3)}      & \multicolumn{1}{c}{(4)}    \\
\hline
Min. disk        & 0.83 $\pm$ 101.6 ({\bf 9.0})      & 1003.0 $\pm$ $...$ ({\bf 78.7 $\pm$ 3.4})  & 0.36  ({\bf 0.91})         \\
Min. disk+gas    & 0.29 $\pm$ 171.5 ({\bf 9.0})     & 1411.0 $\pm$ $...$ ({\bf 69.1 $\pm$ 2.8})  & 0.25  ({\bf 0.63})         \\
Max. disk        & $<0.1$ ({\bf 9.0})      & 283.7 $\pm$ $...$ ({\bf 9.91 $\pm$ 1.5})  & 0.10  ({\bf 0.14})         \\
Model $\Upsilon_{*}^{3.6}$ disk & 0.52 $\pm$ 105.6 ({\bf 9.0}) & 1001.3 $\pm$ $...$ ({\bf 62.1 $\pm$ 2.5}) & 0.19 ({\bf 0.48}) \\
\hline\\
\noalign{\vskip 2pt}
 & \multicolumn{3}{c}{Pseudo-isothermal halo} \\
\cline{2-4} \\
\multicolumn{1}{c}{\ML\ assumption}    &  \multicolumn{1}{c}{$R_C$} &   \multicolumn{1}{c}{$\rho_0$}    & \multicolumn{1}{c}{$\chi_{red.}^2$} \\
\multicolumn{1}{c}{(5)} & \multicolumn{1}{c}{(6)}   & \multicolumn{1}{c}{(7)}      & \multicolumn{1}{c}{(8)} \\
\hline
Min. disk        & 0.31 $\pm$ 0.01       & 473.7 $\pm$ 19.9         & 0.03       \\
Min. disk+gas    & 0.30 $\pm$ 0.01       & 422.4 $\pm$ 19.9         & 0.03 \\
Max. disk        & $...$       	   & 10.36 $\pm$ 2.2         & 0.07  \\ 
Model $\Upsilon_{*}^{3.6}$ disk   & 0.30 $\pm$ 0.01 	   & 371.6 $\pm$ 19.3         & 0.03 \\
\hline
\end{tabular}
\medskip\noindent
\begin{minipage}{109mm}
\noindent
\\
{\bf Note.$\--$}
{\bf (1)(5):} The stellar mass-to-light ratio \ML\ assumptions. ``Model \MLsps\ disk'' uses the values derived from the population synthesis models in Section~\ref{Stars}.
{\bf (2):} Concentration parameter c of NFW halo model (NFW 1996, 1997). We also fit the NFW model to
the rotation curves with only $V_{200}$ as a free parameter after fixing $c$ to 9. The corresponding
best-fit $V_{200}$ and $\chi^{2}_{red}$ values are given in the brackets in (3) and (4), respectively.
{\bf (3):} The rotation velocity (\kms)\,at radius $R_{200}$ where the density constrast exceeds 200 (Navarro \etal\ 1996).
{\bf (4)(8):} Reduced $\chi^{2}$ value.
{\bf (6):} Fitted core-radius of pseudo-isothermal halo model (kpc).
{\bf (7):} Fitted core-density of pseudo-isothermal halo model ($10^{-3}$ \cubedens).
{\bf ($...$):} blank due to unphysically large value or not well-constrained uncertainties.
\end{minipage}
\end{center}
\end{table*}

\begin{figure}
\figurenum{A.25}
\epsscale{1.0}
\includegraphics[angle=0,width=1.0\textwidth,bb=20 50 360 480,clip=]{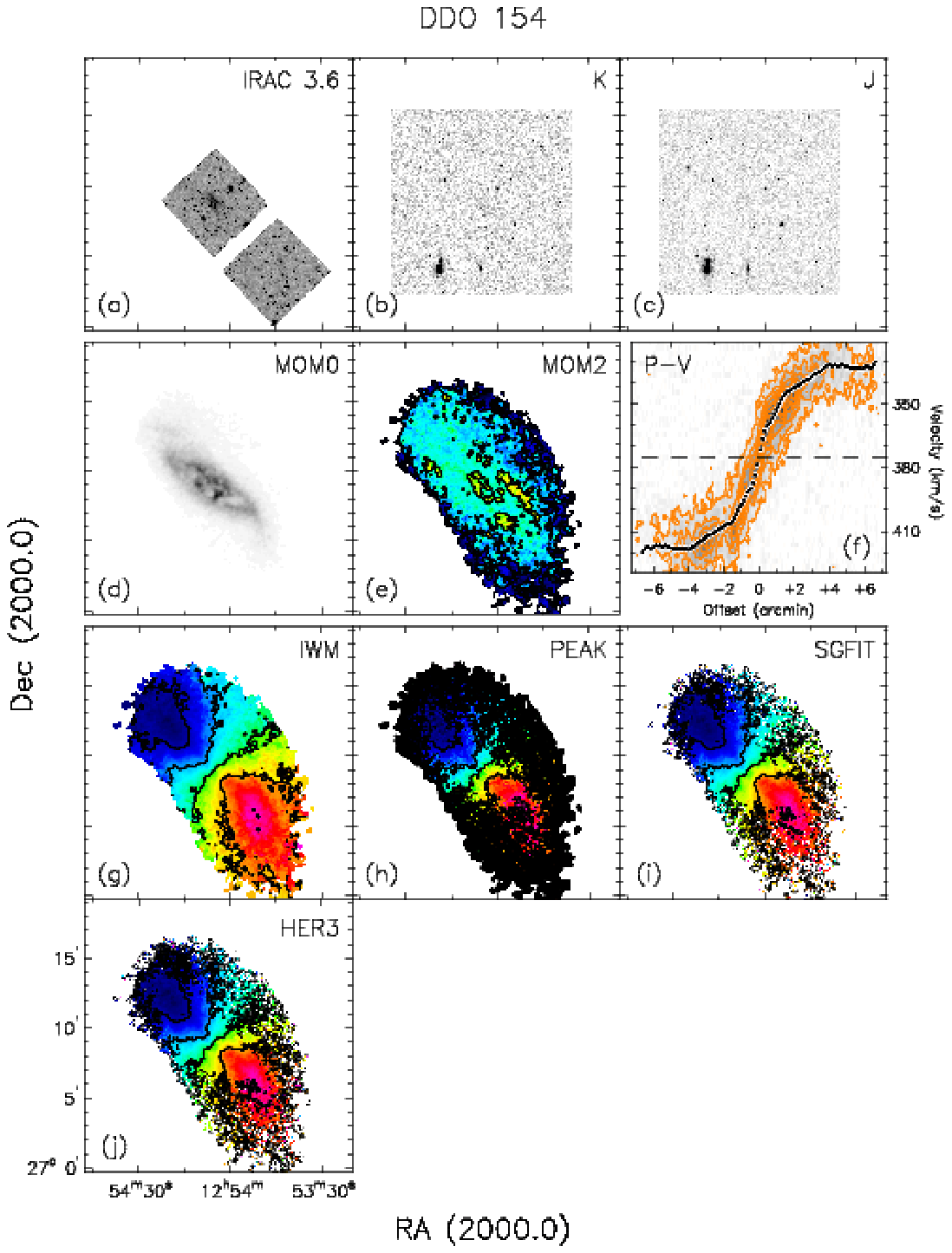}
\caption{{\bf Data:} Total intensity maps and velocity fields of DDO 154.
{\bf (a)(b)(c):}: Total intensity maps in {\it Spitzer} IRAC 3.6 $\mu$m, {\it K} and {\it J} bands.
{\bf (d):} Integrated H{\sc i} map (moment 0). The gray-scale levels run from 0 to 450 $\rm mJy\,beam^{-1}\,\kms$.
{\bf (e):} Velocity dispersion map (moment 2). Velocity contours run from $0$ to $15$\,\kms\,with a spacing of $5$\,\kms. 
{\bf (f):} Position-velocity diagram taken along the average position angle of the major axis as listed in Table 1. 
Contours start at $+3\sigma$ in steps of $5\sigma$. The dashed lines indicate the systemic velocity and position
of the kinematic center derived in this paper. Overplotted is the bulk rotation curve corrected for the average
inclination from the tilted-ring analysis as listed in Table 1. 
{\bf (g)(h)(i)(j):} Velocity fields. Contours run from $320$\,\kms\,to $440$\,\kms\,with a spacing of $20$\,\kms.
\label{DDO154_MAPS}}
\end{figure}
{\clearpage}

\begin{figure}
\figurenum{A.26}
\epsscale{1.0}
\includegraphics[angle=0,width=1.0\textwidth,bb=45 165 570 695,clip=]{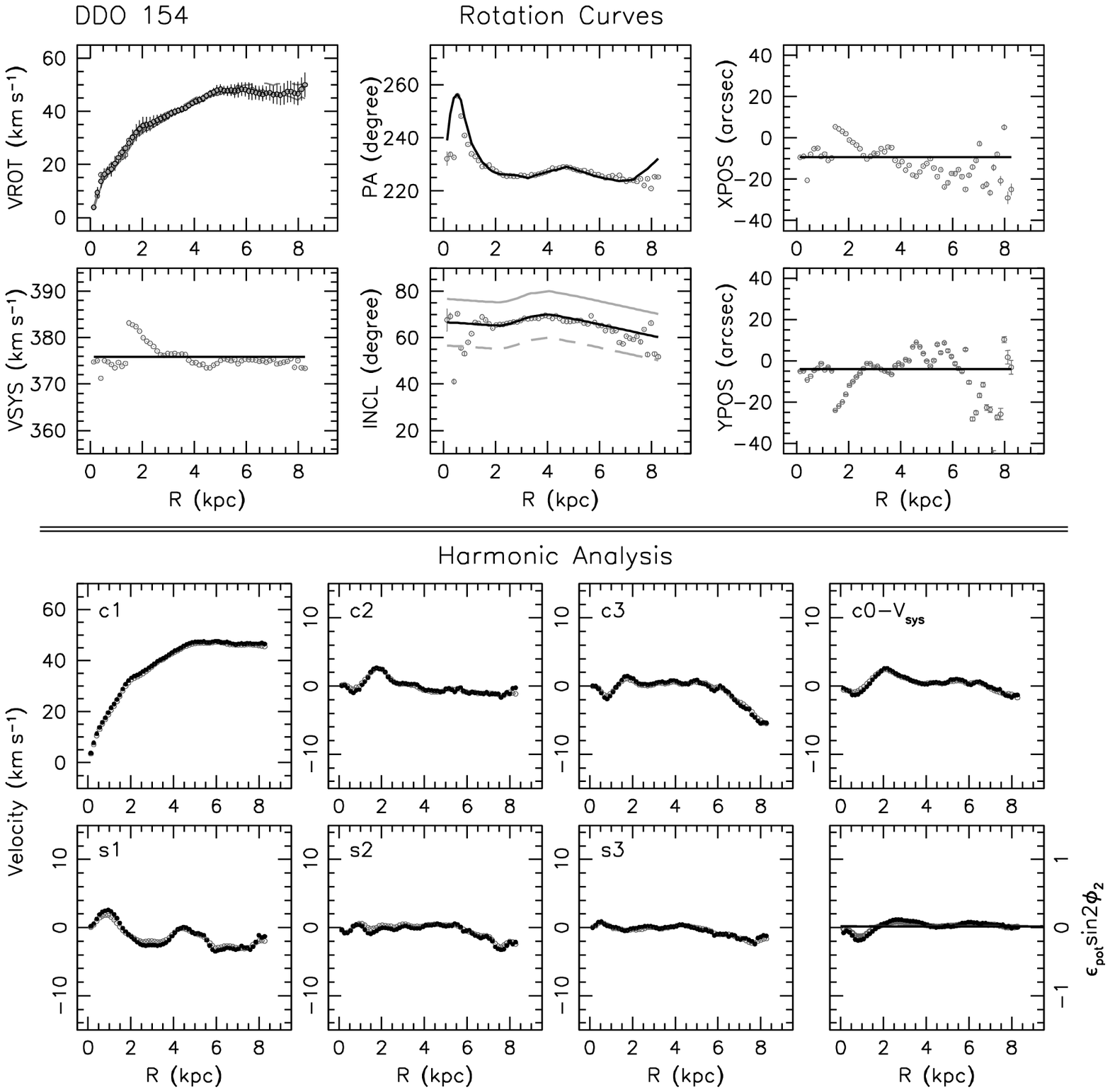}
\caption{{\bf Rotation curves:} The tilted ring model derived from the bulk velocity field of DDO 154.
The open gray circles in all panels indicate the fit made with all ring parameters free. 
The gray dots in the VROT panel were derived using the entire velocity field after fixing other ring parameters
to the values (black solid lines) as shown in the panels. To examine the sensitivity of the rotation curve
to the inclination, we vary the inclination by $+10$ and $-10\,^{\circ}$ as indicated by the gray
solid and dashed lines, respectively, in the right-middle panel. 
We derive the rotation curves using these inclinations while keeping other ring
parameters the same. The resulting rotation curves are indicated
by gray solid (for $+10\,^{\circ}$ inclination) and dashed (for $-10\,^{\circ}$ inclination) lines in the VROT panel.
{\bf Harmonic analysis:} Harmonic expansion of the velocity fields for DDO 154.
The black dots and gray open circles indicate the results from the hermite $h_{3}$ and IWM velocity fields, respectively.
In the bottom-rightmost panel, the solid and dashed lines indicate global elongations of the potential
measured using the hermite $h_{3}$ and IWM velocity fields.
\label{DDO154_TR_HD}}
\end{figure}
{\clearpage}

\begin{figure}
\epsscale{1.0}
\figurenum{A.27}
\includegraphics[angle=0,width=1.0\textwidth,bb=43 175 543 695,clip=]{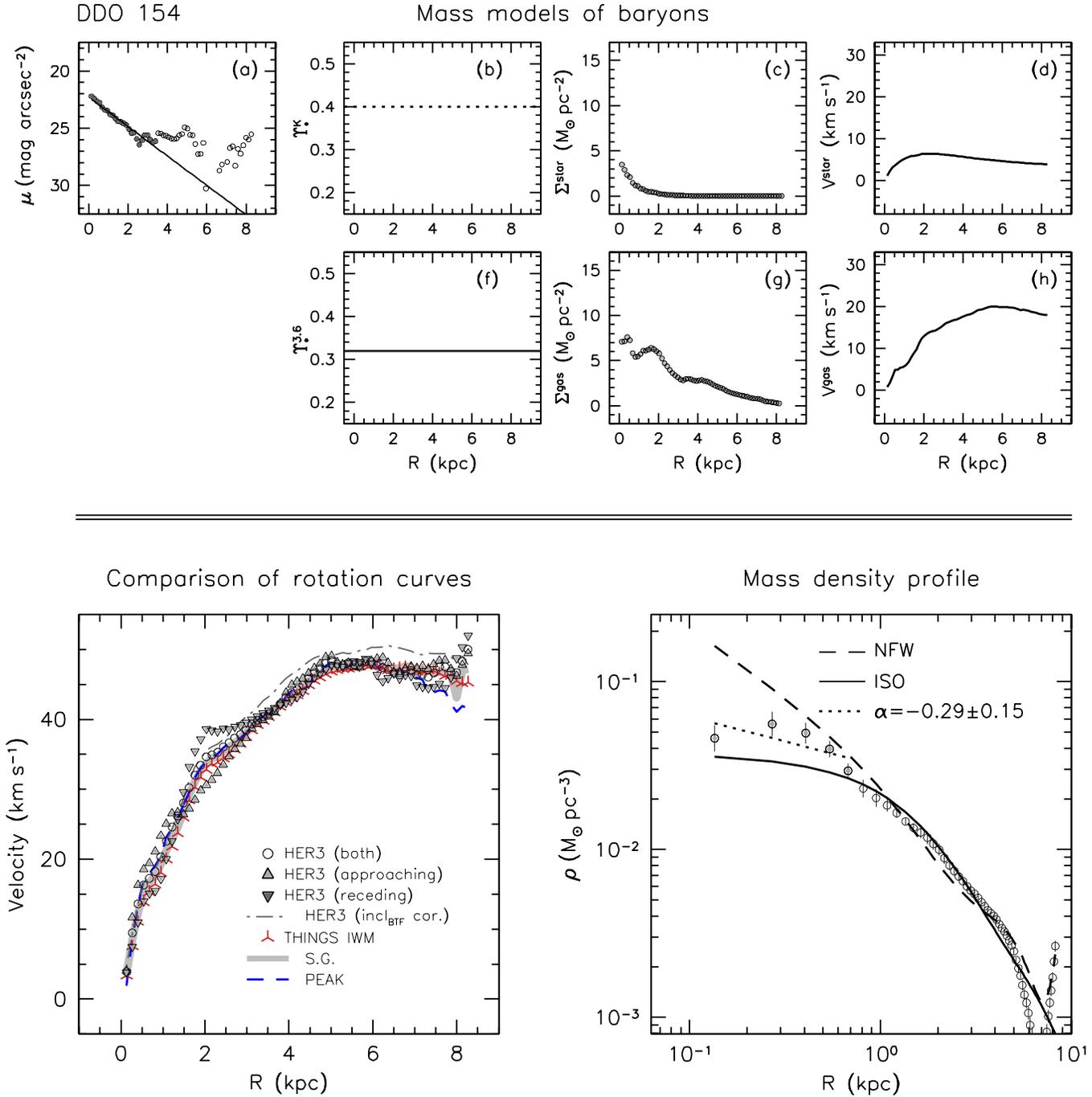}
\caption{{\bf Mass models of baryons:} Mass models for the gas and stellar components of DDO 154. 
{\bf (a):} The azimuthally averaged 3.6$\mu$m surface brightness profile.
{\bf (b)(f):} The stellar mass-to-light values in the $K$ and 3.6$\mu$m bands derived from stellar population synthesis models.
{\bf (c)(d):} The mass surface density and the resulting rotation velocity for the stellar component.
{\bf (g)(h):} The mass surface density (scaled by 1.4 to account for He and metals) and the resulting
rotation velocity for the gas component.
{\bf Comparison of rotation curves:} Comparison of the H{\sc i} rotation curves for DDO 154. See Section~\ref{Deriving_Rotcurs}
for a detailed discussion. These rotation curves have also been discussed in detail \cite{deBlok_2008}. 
{\bf Mass density profile:} The derived mass density profile of DDO 154. 
The open circles represent the mass density profile derived from the bulk rotation curve
assuming minimum disk. The inner density slope $\alpha$ is measured by a least squares fit (dotted line)
to the data points indicated by gray dots, and shown in the panel.
\label{DDO154_VROT_BARYONS_ALPHA}}
\end{figure}
{\clearpage}

\begin{figure}
\figurenum{A.28}
\epsscale{1.0}
\includegraphics[angle=0,width=1.0\textwidth,bb=20 150 580 720,clip=]{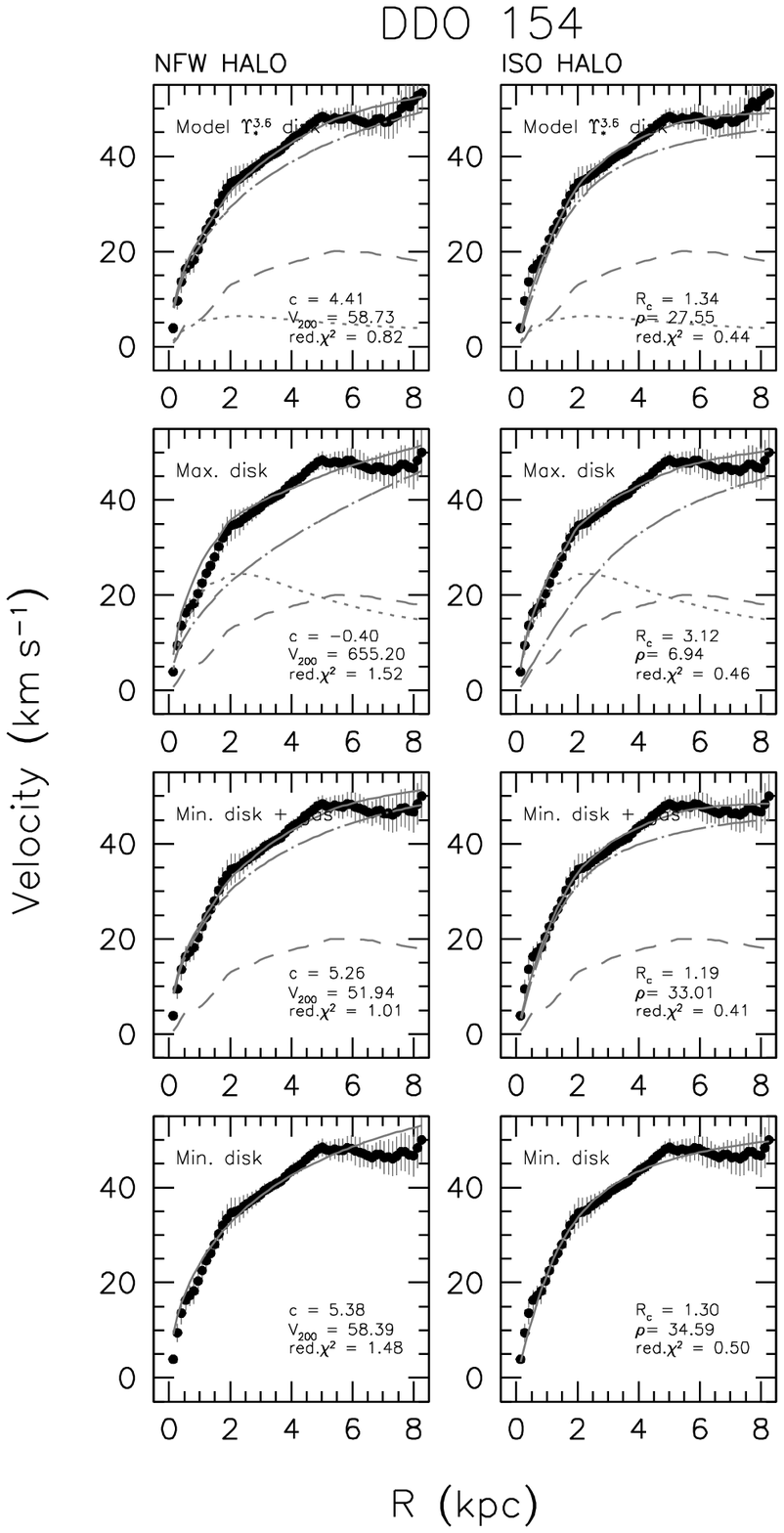}
\caption{{\bf Mass modeling results:} Disk-halo decomposition of the DDO 154 rotation curve 
under various \ML\ assumptions (\MLsps, maximum disk, $\rm minimum disk + gas$ and minimum disks).
The black dots indicate the hermite $h_{3}$ rotation curve, and the short and long dashed lines show the
rotation velocities of the stellar and gas components, respectively. The fitted parameters
of NFW and pseudo-isothermal halo models (long dash-dotted lines) are denoted on each panel.
See \cite{deBlok_2008} for more details. 
\label{DDO154_MD}}
\end{figure}
{\clearpage}

\begin{table*}
\scriptsize
\caption{Parameters of dark halo models for DDO 154}
\label{NFWHALO_M81dwB}
\begin{center}
\begin{tabular}{@{}lrcccc}
\hline
\hline
\noalign{\vskip 2pt}
& \multicolumn{3}{c}{NFW halo} \\
\cline{2-4} \\
\multicolumn{1}{c}{\ML\ assumption}   &  \multicolumn{1}{c}{$c$}   &  \multicolumn{1}{c}{$V_{200}$}  &  \multicolumn{1}{c}{$\chi_{red.}^2$} \\
\multicolumn{1}{c}{(1)} & \multicolumn{1}{c}{(2)}   & \multicolumn{1}{c}{(3)}      & \multicolumn{1}{c}{(4)}    \\
\hline
Min. disk        & 5.3 $\pm$ 0.5 ({\bf 9.0})     & 58.3 $\pm$ 4.2 ({\bf 41.7 $\pm$ 0.5})  & 1.48  ({\bf 2.71})         \\
Min. disk+gas    & 5.2 $\pm$ 0.4 ({\bf 9.0})     & 51.9 $\pm$ 3.3 ({\bf 37.2 $\pm$ 0.5})  & 1.01  ({\bf 2.06})         \\
Max. disk        & $<0.1$ ({\bf 9.0})      & 655.1 $\pm$ $...$ ({\bf 30.0 $\pm$ 0.9})  & 1.52  ({\bf 5.66})         \\
Model $\Upsilon_{*}^{3.6}$ disk & 4.4 $\pm$ 0.4 ({\bf 9.0}) & 58.7 $\pm$ 4.2 ({\bf 36.7 $\pm$ 0.5}) & 0.82 ({\bf 2.22}) \\
\hline\\
\noalign{\vskip 2pt}
 & \multicolumn{3}{c}{Pseudo-isothermal halo} \\
\cline{2-4} \\
\multicolumn{1}{c}{\ML\ assumption}    &  \multicolumn{1}{c}{$R_C$} &   \multicolumn{1}{c}{$\rho_0$}    & \multicolumn{1}{c}{$\chi_{red.}^2$} \\
\multicolumn{1}{c}{(5)} & \multicolumn{1}{c}{(6)}   & \multicolumn{1}{c}{(7)}      & \multicolumn{1}{c}{(8)} \\
\hline
Min. disk        & 1.30 $\pm$ 0.04       & 34.5 $\pm$ 1.8         & 0.50       \\
Min. disk+gas    & 1.19 $\pm$ 0.04       & 33.0 $\pm$ 1.9         & 0.40 \\
Max. disk        & 3.11 $\pm$ 0.19       & 6.9 $\pm$ 0.4         & 0.45  \\ 
Model $\Upsilon_{*}^{3.6}$ disk   & 1.33 $\pm$ 0.05 	   & 27.5 $\pm$ 1.6         & 0.43 \\
\hline
\end{tabular}
\medskip\noindent
\begin{minipage}{109mm}
\noindent
\\
{\bf Note.$\--$}
{\bf (1)(5):} The stellar mass-to-light ratio \ML\ assumptions. ``Model \MLsps\ disk'' uses the values derived from the population synthesis models in Section~\ref{Stars}.
{\bf (2):} Concentration parameter c of NFW halo model (NFW 1996, 1997). We also fit the NFW model to
the rotation curves with only $V_{200}$ as a free parameter after fixing $c$ to 9. The corresponding
best-fit $V_{200}$ and $\chi^{2}_{red}$ values are given in the brackets in (3) and (4), respectively.
{\bf (3):} The rotation velocity (\kms)\,at radius $R_{200}$ where the density constrast exceeds 200 (Navarro \etal\ 1996).
{\bf (4)(8):} Reduced $\chi^{2}$ value.
{\bf (6):} Fitted core-radius of pseudo-isothermal halo model (kpc).
{\bf (7):} Fitted core-density of pseudo-isothermal halo model ($10^{-3}$ \cubedens).
{\bf ($...$):} blank due to unphysically large value or not well-constrained uncertainties.
\end{minipage}
\end{center}
\end{table*}

\end{document}